\documentclass{ws-rv975x65}
\usepackage[square]{ws-rv-van}     % numbered citation/references (default)
\makeindex
\usepackage{stmaryrd} % for using the symbol \oblong (small square)
\usepackage{tikz} % for drawing figures
\usepackage{enumitem} % for enumerate items in alphabetic order, etc.

\providecommand\phantomsection{} % to use \phantomsection with or without hyperref without an error

\usetikzlibrary{decorations.markings}
\tikzset{arrow data/.style 2 args={
      decoration={
         markings,
         mark=at position #1 with \arrow{#2}},
         postaction=decorate}
      }
\tikzset{mylabel/.style  args={at #1 #2  with #3}{
    postaction={decorate,
    decoration={
      markings,
      mark= at position #1
      with  \node [#2] {#3};
 } } } }

\newcommand{\Pl}{\ell_\mathrm{Pl}} % Planck length
\newcommand{\mubar}{{\bar{\mu}}} % mu bar
\newcommand{\abs}[1]{{\left|{#1}\right|}} % abs (variant delimiters)
\newcommand{\Abs}[1]{{\big|{#1}\big|}} % abs (fixed delimiters)
\newcommand{\norm}[1]{{\Vert{#1}\Vert}} % norm
\newcommand{\inner}[2]{{\langle {#1}\vert {#2} \rangle}} % inner product
\newcommand{\oinner}[2]{{( {#1}\vert {#2} \rangle}} % round inner product
\newcommand{\ooinner}[2]{{( {#1}\vert {#2} )}} % double-round inner product
\newcommand{\ket}[1]{\vert{#1}\rangle} % ket
\newcommand{\bra}[1]{\langle{#1}\vert} % bra
\newcommand{\oket}[1]{\vert{#1})} % round ket
\newcommand{\obra}[1]{({#1}\vert} % round bra

\newcommand{\sgn}{{\mathrm{sgn}}} % sign
\newcommand{\Tr}{{\mathrm{Tr}}} % trace

\newcommand{\grav}{\mathrm{grav}} % gravitational part
\newcommand{\matt}{\mathrm{matt}} % matt part
\newcommand{\Eucl}{\mathrm{Eucl}} % Euclidean part
\newcommand{\kin}{\mathrm{kin}} % kinematic (Hilbert space, states)
\newcommand{\phys}{\mathrm{phys}} % physical (Hilbert space, states)
\newcommand{\inv}{\mathrm{inv}} % invariant (Hilbert space, states)
\newcommand{\Diff}{\mathrm{Diff}} % diffeomorphism group
\newcommand{\hil}{\mathcal{H}} % Hilbert space
\newcommand{\Cyl}{\mathrm{Cyl}} % space of cylindrical functions
\newcommand{\calC}{\mathcal{C}} % for constraints

\newcommand{\tE}{\mbox{$\tilde{E}$}} % densitized triad

\newcommand{\bibnote}{\bibitem[$\,*\,$]{}} % note mark in the bibliography

\begin{document}

\chapter{Loop Quantum Gravity}\label{ra_ch1}

\author{Dah-Wei Chiou%\footnote{Author footnote.}
}
%\index[aindx]{Author, F.} % or \aindx{Author, F.}
%\index[aindx]{Author, S.} % or \aindx{Author, S.}

\address{Department of Physics,\\
National Taiwan Normal University, Taipei 11677, Taiwan}

\address{Department of Physics and Center for Condensed Matter Sciences,\\
National Taiwan University, Taipei 10617, Taiwan%\footnote{Affiliation footnote.}
}

\begin{abstract}
This article presents an ``in-a-nutshell'' yet self-contained introductory review on loop quantum gravity (LQG)---a background-independent, nonperturbative approach to a consistent quantum theory of gravity. Instead of rigorous and systematic derivations, it aims to provide a general picture of LQG, placing emphasis on the fundamental ideas and their significance. The canonical formulation of LQG, as the central topic of the article, is presented in a logically orderly fashion with moderate details, while the spin foam theory, black hole thermodynamics, and loop quantum cosmology are covered briefly. Current directions and open issues are also summarized.

\vspace{1in}

\noindent
*This article is to appear in International Journal of Modern Physics D and in the book ``One Hundred Years of General Relativity:
Cosmology and Gravity'', edited by Wei-Tou Ni (World Scientific, Singapore, 2015).
The related chapters in the book include:
\begin{enumerate}[label=\textit{\Alph*}]
\item \textbf{Cosmic microwave background}, M.~Bucher\label{ch:Bucher}
\item \textbf{Cosmic structure}, M.~Davis\label{ch:Davis}
\item \textbf{Cosmic inflation}, K.~Sato and J.~Yokoyama\label{ch:Sato}
\item \textbf{Inflation, string theory and cosmic strings}, D.~Chernoff and S.-H.~Tye\label{ch:Chernoff}
\item \textbf{Quantum gravity in curved spacetime}, S.~P.~Kim\label{ch:Kim}
\item \textbf{Perturbative quantum gravity comes of age}, R.~Woodard\label{ch:Woodard}
\item \textbf{Black hole thermodynamics}, S.~Carlip\label{ch:Carlip}
\end{enumerate}

\end{abstract}
%\markright{Customized Running Head for Odd Page} % default is Chapter Title.

\body

\newpage

\setcounter{tocdepth}{4}

\tableofcontents

\newpage

\section{Introduction}\label{sec:introduction}
Quantum gravity (QG) is the research in theoretical physics that seeks a consistent quantum theory of gravity. It is considered by many as the open problem of paramount importance in fundamental physics, as its task is to unify quantum mechanics (more specifically, quantum field theory, QFT) and general relativity (GR), which are the two greatest theories discovered in the twentieth century and have become the cornerstones of modern physics.

Ever since it was recognized that the gravitational field needs to be quantized, the quest for a satisfactory quantum description of spacetime has never stopped, and gradually QG has grown into a vast area of research along many different paths with various doctrines. The two most developed approaches are string theory (see \cref{ch:Chernoff} and references therein) and loop quantum gravity (LQG). Other directions include causal sets, dynamical triangulation, emergent gravity, $\mathcal{H}$ space theory, noncommutative geometry, supergravity, thermogravity, twistor theory, and many more. (See Refs.~\cite{Rovelli:1997qj,Carlip:2001wq,Woodard:2009ns,Kiefer:2012boa} for surveys and more references on different theories of QG.)

Among other approaches, the strength of LQG is that it provides a compelling description of quantum spacetime in a nonperturbative, background-independent fashion. (On the other hand, see Chaps.~\ref{ch:Kim} and \ref{ch:Woodard} for the perturbative approach of QG.) The beauty lies in its faithful attempt to establish a conceptual framework that consolidates the apparently conflicting tenets of QFT and GR. LQG deliberately adopts the ``minimalist'' approach in the sense that it focuses solely on the search for a consistent quantum theory of gravity without requiring any extraordinary ingredients such as extra dimensions, supersymmetry, and so on (although many of these can be incorporated compatibly). Unlike string theory, the aim of which is much more ambitious, LQG does not intend to address the unification problem of finding a ``theory of everything'' in which all forces, including gravity, are unified. The central problem confronted by LQG simply is: What is the consistent quantum field theory of which the low-energy limit is classical GR, if it exists after all?

Having been developed about 25 years (see Ref.~\cite{Rovelli:2010bf} for its history and current status), LQG has faced many challenges and achieved encouraging progress. Many long-standing open problems within the approach have been solved, and LQG now provides a rigorous mathematical foundation for QG. One of the main results is the discovery that the spectra of area and volume are discrete and the corresponding quantum states, called \emph{spin networks}, reveal the microscopic structure of space that is granular in Planck scale. The discreteness of space is not postulated \textit{ad hoc} but a direct consequence of quantization in the same nature as the discrete levels of energy in an atom. Because of the discreteness, LQG enables one to derive the Bekenstein-Hawking entropy of black holes from first principles, offering a microscopic explanation of its proportionality to the area of the horizon. (Also see \cref{ch:Carlip} for other aspects of black hole thermodynamics.) Furthermore, the construction of LQG relies heavily on diffeomorphism invariance. Once the gauge degrees of diffeomorphism are factored away, the resulting quantum states, called \emph{s-knots}, are not to be interpreted as quantum excitations \emph{on} a space but, rather, \emph{of} the space itself, because any external reference to localization of quantum states becomes irrelevant and only their contiguous relation remains significant. In terms of this relational structure, LQG profoundly manifests \emph{background independence}---the key principle of classical GR---in the context of QFT.
With many successes achieved, the major weakness of LQG is in our limited understanding about its quantum dynamics and low-energy physics, but these two aspects have been intensively investigated and some promising solutions have emerged.

LQG is closely related to the spin foam theory, an alternative approach of QG. While LQG is based on the canonical (Hamiltonian) formalism, the spin foam theory can be viewed as the covariant (sun-over-histories) approach of LQG. The history of a spin network (more precisely, an s-knot) evolving over time is called a \emph{spin foam}, which represents a quantized spacetime in the same sense that a spin network represents a quantized space. The transition amplitude is given by a discrete version of path integral, which is the sum, with proper weight factors, over all possible spin foams sending the initial spin network to the final one. Extensive research on the spin foam theory has made rapid advances, suggesting that a specific class of spin foam models might provide a covariant definition of the LQG dynamics.

Applying principles of LQG to cosmological settings, one obtains loop quantum cosmology (LQC). LQC is a symmetry-reduced model of LQG with only a finite number of degrees of freedom. The simplification of symmetry reduction makes possible a detailed investigation on the ramifications of loop quantum effects, many of which remain obscure in the full theory of LQG. One of the striking results of LQC is that the big bang singularity is resolved by the loop quantum effects and replaced by a \emph{quantum bounce}, which bridges the present universe with a preexistent one. The bouncing scenario opens up a new paradigm of cosmology. Research of LQC has led to many successes and represents a very active research field in recent years.

There are many books and review articles available on LQG\cite{Ashtekar:1991hf,Thiemann:2002nj,Ashtekar:2004eh,Rovelli:2004tv, Thiemann:2007zz,Rovelli:2008zza,Gambini:2011zz,Rovelli:2014book}, the spin foam theory\cite{Perez:2003vx,Oriti:2003wf,Perez:2004hj,Perez:2012wv}, and LQC\cite{Bojowald:2008zzb,Ashtekar:2011ni,Bojowald:2011zzb,Bojowald:2011book}.
This article serves as an ``in-a-nutshell'' yet self-contained introductory review on LQG. It is centered on the canonical formulation of LQG with moderate details but covers the spin foam theory, black hole thermodynamics, and LQC only very briefly.

The article is organized as follows. After a quick review on the problems of QG in \sref{sec:motivations}, the main topic of canonical approach of LQG is presented from \sref{sec:connection theories} to \sref{sec:low-energy physics} in a logically orderly fashion. Additionally, the spin foam theory, black hole thermodynamics, and LQC are briefly described in Secs.\ \ref{sec:spin foam theory}, \ref{sec:black hole thermodynamics}, and \ref{sec:LQC}, respectively. Finally, \sref{sec:current directions} outlines the current directions and open issues.

A large part of this article is heavily based on Refs.~\cite{Ashtekar:2004eh,Rovelli:2004tv} without delving into all technicalities.
Conventions are given as follows. The signature of the spacetime metric $g_{\alpha\beta}$ is given by $(-,+,+,+)$. The Greek letters in lower case are used for the coordinates of the 4-dimensional spacetime manifold, while the Latin letters in lower case starting $a,b,\dots$ are for the coordinates of the 3-dimensional space manifold. The Latin letters in upper case starting $I,J,K,\dots=0,1,2,3$ are used for the 4-dimensional ``internal space'' (the vector space on which the Lorentz group $SO(3,1)$ is represented), while the Latin letters in lower case starting $i,j,k,\dots=1,2,3$ are for the 3-dimensional internal space (the vector space on which $SU(2)$, the covering group of $SO(3)$, is represented). In case the Euclidean GR, in place of the Lorentzain GR, the metric signature is given by $(+,+,+,+)$ and the Lorentz group $SO(3,1)$ is replaced by the Euclidean rotational group $SO(4)$. The speed of light is set to $c=1$, but we keep the (reduced) Planck constant $\hbar$ and the gravitational constant $G$ explicit.

\section{Motivations}\label{sec:motivations}
Despite the fact that enormous efforts in various approaches have been devoted for QG, it remains the Holy Grail of theocratical physics. In the section, we briefly discuss the conceptual problems in QG. (See Ref.~\cite{Isham:1991mm,Kiefer:2012boa,Woodard:2009ns} for full elaboration.)

\subsection{Why quantum gravity?}
The very first question about QG is: Why do we even bother to quantize gravity at all? Apart from many aesthetic considerations for an elegant unification of known fundamental physics, the logical necessity of a quantum description of gravity follows from the conflicts between the two fundamental theories of GR and QFT. The fundamental theories collide in the classical Einstein field equation
\begin{equation}\label{Einstein equation}
R_{\mu\nu}-\frac{1}{2}R\,g_{\mu\nu}=8\pi G T_{\mu\nu},
\end{equation}
which relates the dynamics of nongravitational matter (in terms of $T_{\mu\nu}$) described by the theory of QFT to the dynamics of geometry (in terms of $g_{\mu\nu}$) described by the theory of GR. Note that the metric $g_{\mu\nu}$ enters the definition of the energy-momentum tensor via $T_{\mu\nu}\equiv T_{\mu\nu}(g)=\delta S/\delta g^{\mu\nu}$ ($S$ is the action) and that QFT is defined only on a given (flat or curved) background.

As matter fields are quantized in QFT, the only way to make sense of \eref{Einstein equation} without quantizing $g_{\mu\nu}$ is to start with a given background $g_0$ and replace the right-hand side with the expectation value $\langle\,\hat{T}_{\mu\nu}(g_0)\rangle$. Once we solve the classical field $g_{\mu\nu}$ by this prescription, we treat the solution as the new background and iterate the procedure. However, the iteration is not guaranteed to yield convergent solutions and thus leads to inconsistency. In fact, one cannot consistently couple a classical system to a quantum one\cite{Woodard:2009ns,Wald:1984rg}, even though QFT in curved spacetime provides an adequate effective theory in low-energy regime.

The conflicts force one to quantize both geometry and matter fields simultaneously or to invoke other more radical modifications.

\subsection{Difficulties of quantum gravity}
What after all are the difficulties that prevent one from quantizing both geometry and matter easily at the same time? It turns out the difficulties lie at the very contradictory difference between the fundamental principles of GR and QFT.

In GR, spacetime is dynamical as well as matter fields. Matter and geometry are essentially on the equal footing in this regard. In QFT, by contrast, all dynamical variables are quantized except spacetime. Spacetime is treated as a given fixed background, which parameterizes all dynamical fields and provides the \textit{a priori} causal structure needed for field quantization. Therefore, one encounters the stark paradox when joining GR and QFT together.

There are various approaches trying to circumvent the paradox, differing on which features of GR and QFT are viewed as fundamental and unchangeable and which features as inessential and modifiable, as surveyed in Refs.~\cite{Rovelli:1997qj,Carlip:2001wq,Woodard:2009ns,Kiefer:2012boa}.

\subsection{Background-independent approach}\label{sec:background-independent approach}
In conventional QFT, dynamics of dynamical fields takes place on the ``stage'' of a fixed spacetime with given metric. By analogy, dynamical fields are animals that roam around and chase one another upon the stage of an island of spacetime.

GR, on the other hand, has taught us a very different paradigm of \emph{background independence}, according to which spacetime is itself a dynamical entity (gravitational field) in many respects the same as other nongravitational entities. That said, physical entities, both gravitational and nongravitational, are not residing and moving in spacetime but, rather, they reside and move on top of one another. In the words of Rovelli's metaphor\cite{Rovelli:2004tv}:
``It is as if we had observed in the ocean many animals living on an island: animals on the island. Then we discover that the island itself is in fact a great whale. So the animals are no longer on the island, just animals on animals."

Among various theories of QG, LQG is special in the sense that it reveres the paradigm of background independence earnestly and aims to formulate a nonconventional QFT of GR in genuine conformity with the paradigm, essentially in a nonperturbative fashion, so as to reconcile the aforementioned paradox. This of course is a very challenging task and the first step is to begin with the ``proper'' mathematical framework, which turns out to be the canonical description of connection dynamics of GR in terms of \emph{Ashtekar connection}. The shift from metric to Ashtekar connection opens the possibility of employing nonperturbative techniques in gauge theories (especially, lattice gauge theories) to QG.

\section{Connection theories of general relativity}\label{sec:connection theories}
Classical GR is usually presented as \emph{geometrodynamics}, i.e., a dynamical theory of metrics, but it can also be recast as \emph{connection dynamics}, i.e., a dynamical theory of connections. The merit of this reformulation is to cast GR in a language closer to that of gauge theories, for which quantization is better understood. In this section, we present a brief introduction to connection theories of GR, largely excerpted from Sec.~2 of Ref.~\cite{Ashtekar:2004eh}, and address some remarks in \sref{sec:remarks on connection theories}. For introductory purpose, we will focus only on the gravitational part of the action and phase space. For inclusion of matter and the cosmological constant, see e.g.\ Ref.~\cite{Ashtekar:1989ju} (also see \sref{sec:inclusion of matter fields}); for extension to supergravity and other spacetime dimensions, see e.g.\ Refs.~\cite{Jacobson:1987cj,Matschull:1993hy,Freidel:1999rr} (also see \sref{sec:SUSY and other dimensions}).

\subsection{Connection dynamics}
The most widely known formulation of classical GR is the dynamical theory of metric $g_{\mu\nu}$, which is based on the Riemannian geometry and governed by the Einstein-Hilbert action
\begin{equation}\label{EH action}
S[g]=\frac{1}{16\pi G}\int d^4x \sqrt{\abs{g}}\,R,
\end{equation}
where $R$ is the Ricci scalar and $g$ is the determinant of $g_{\mu\nu}$. Classical GR also admits many other formulations, which are equivalent to one another to various extent.

In the Palatini formulation\cite{Capovilla:1989ac,Capovilla:1991qb} based on the Riemann-Cartan geometry, instead of $g_{\mu\nu}$, the basic gravitational variables are taken to be the coframe field ${e_\mu}^I$, which gives the orthonormal cotetrad (or vierbein), and the $so(1,3)$-valued connection 1-form field ${{\omega_\mu}^I}_J$, which corresponds to the gauge group $SO(3,1)$ of local Lorentz transformation. The Palatini action is given by
\begin{equation}\label{Palatini action}
S[e,\omega]=\frac{1}{32\pi G}\int \epsilon_{IJKL}e^I\wedge e^J\wedge\Omega^{KL},
\end{equation}
where
\begin{equation}
\Omega:=d\omega+\omega\wedge\omega
\end{equation}
is the curvature of the connection ${{\omega_\mu}^I}_J$. The coframe ${e_\mu}^I$ determines the spacetime metric via
\begin{equation}\label{g and e}
g_{\mu\nu}=\eta_{IJ}{e_\mu}^I{e_\nu}^J.
\end{equation}
Meanwhile, the connection ${{\omega_\mu}^I}_J$ is completely determined by the coframe via
\begin{equation}\label{e and omega}
de+\omega\wedge e=0,
\end{equation}
which is the equation of motion obtained by varying the Palatini action with respect to $\omega_\mu$. As far as equations of motion are concerned, the Palatini action \eref{Palatini action}, imposed by \eref{e and omega}, reduces to the familiar Einstein-Hilbert action \eref{EH action}. Thus, these two actions lead to the same theory at least at the classical level.

It is tedious yet straightforward to perform the Legendre transform on the Palatini action to obtain the Hamiltonian theory\cite{Ashtekar:1988sw}, but the resulting theory has certain second-class constraints. When the second-class constraints are solved, one is led to the standard Hamiltonian description of geometrodynamics (in terms of cotriads) and thus loses all reference to connection dynamics; meanwhile, the form of geometrodynamics is rather complicated and seems insurmountable for quantum theory. This problem can be avoided by modifying the Palatini action with Holst's augmentation\cite{Holst:1995pc} as
\begin{equation}\label{Palatini-Holst action}
S[e,\omega]=\frac{1}{32\pi G}\int \epsilon_{IJKL}e^I\wedge e^J\wedge\Omega^{KL}
-\frac{1}{16\pi G\gamma}\int e^I\wedge e^J\wedge\Omega_{IJ},
\end{equation}
where $\gamma$ is an arbitrary but fixed number, called the \emph{Barbero-Immirzi} parameter.
The celebrated feature is that this augmentation does not change the equations of motion of \eref{Palatini action}. Inclusion of Holst's modification renders it more elegant to perform the Legendre transform in the sense that all constraints are first class (as we will see shortly) and consequently makes the quantum theory founded on the Hamiltonian description more manageable as significance of connection dynamics is retained.
(While different numerical values of $\gamma$ yield equivalent theories at the classical level, it should be noted that, in the quantum theory, they give rise to different sectors that are not unitarily equivalent to one another as will be shown in \sref{sec:operators and quantum geometry}.)

In the non-abelian gauge theory, the Yang-Mills action can also be augmented with a topological term called the $\theta$-term as
\begin{equation}\label{YM action}
S=\frac{1}{2\alpha^2}\int \Tr\, ^\star\!F \wedge F
+\frac{\theta}{8\pi^2}\int \Tr\, F\wedge F,
\end{equation}
where $F$ is the field strength (curvature) of the Yang-Maills connection, $\star$ is the Hodge dual, and $\alpha$ and $\theta$ are known as the coupling constant and the $\theta$-angle respectively.\footnote{Analogous to the Barbero-Immirze parameter $\gamma$, different numerical values of $\theta$ yield equivalent theories at the classical level but give rise to inequivalent $\theta$-sectors in the quantum theory.} It is noticeable that \eref{Palatini-Holst action} bears close resemblance to \eref{YM action}. Whether this intriguing resemblance carries any profound meaning arouses a lot of curiosity. LQG, in particular, exploits the resemblance to the extent that we can adopt the nonperturbative techniques (such as the quantization scheme in terms of Wilson loops) well developed in the context of gauge theories.

\subsection{Canonical (Hamiltonian) formulation}
We have reformulated the classical theory of GR in a fashion very close to non-abelian gauge theories. Prior to quantizing the theory, we have to take a further step to translate the covariant description into the canonical (or Hamiltonian) description in order to apply the standard scheme of \emph{equal-time quantization} used in QFT.

First, we apply the ADM foliation, which foliates the 4-dimensional spacetime manifold $\mathcal{M}$ into a family of spacelike surfaces  (called ``leaves'') $\Sigma_t$, labelled by a time coordinate $t$ and with spatial coordinates on each slice given by $x^i$. Under the ADM foliation, variables on the spacetime manifold are split into 3+1 decomposition. The coframe (cotetrad) ${e_\mu}^I$ is split into the cotriad $e_a^i$ plus the \emph{lapse} function $N$ and the \emph{shift} vector field $N^a$. The lapse and shift describe how each of the leaves $\Sigma_t$ are welded together in the foliation.\footnote{More precisely, let $n^\alpha$ be the unit vector normal to the slice $\Sigma_t$, we have $t^\alpha=Nn^\alpha+N^\alpha$ with $N^\alpha n_\alpha=0$.} Correspondingly, the 4-dimensional metric $g_{\mu\nu}$ of $\mathcal{M}$ is split into the 3-dimensional metric $q_{ab}$ of $\Sigma$ plus $N$ and $N^a$. Similarly, apart from $(\omega^i\cdot t):=-\frac{1}{2}\epsilon^{ijk}\omega_{jk}\cdot t$, the $so(3,1)$-valued connection 1-form $\omega_\mu^{IJ}$ on $\mathcal{M}$ is naturally decomposed into two $su(2)$-valued 1-forms on $\Sigma$: $\Gamma=\Gamma_a^i\tau_idx^a$ and $K=K_a^i\tau_idx^a$, where $\tau_i=\tau^i=\sigma_i/(2i)$ are the generators of $SU(2)$ with $\sigma_i$ being the Pauli matrices. As a consequence of \eref{e and omega}, $\Gamma_a^i$ is the $su(2)\cong so(3)$ spin connection on $\Sigma_t$ associated with the cotriad $e^i_a$, i.e.,
\begin{equation}
de^i+{\epsilon^i}_{jk}\Gamma^i\wedge e^k=0,
\end{equation}
and $K_a^i$ is the extrinsic curvature of $\Sigma_t$ imbedded in $\mathcal{M}$. The 3+1 splitting is summarized in \tref{tab:ADM}.
\begin{table}[ht]
\tbl{Decomposition under the ADM foliation.\label{tab:ADM}}
{\begin{tabular}{ccc}
\toprule
$\mathcal{M}$ & $\rightarrow$ & $\Sigma_t$ \\
$\mu,\nu,\dots$ & $\rightarrow$ & $a,b,\dots$ \\
$I,J,\dots=(0,1,2,3)$ & $\rightarrow$ & $i,j,\dots=(1,2,3)$ \\
$so(3,1)$ & $\rightarrow$ & $so(3)\cong su(2)$ \\
\colrule
${e_\mu}^I$ & $\rightarrow$ & $e_a^i,\ N,\ N^a$ \\
$g_{\mu\nu}$ & $\rightarrow$ & $q_{ab},\ N,\ N^a$ \\
${{\omega_\mu}^I}_J$ & $\rightarrow$ & $\Gamma^i_a,\ K^i_a,\ (\omega^i\cdot t)$ \\
$de^I+{\omega^I}_J\wedge e^J=0$ & $\rightarrow$ & $de^i+{\epsilon^i}_{jk}\Gamma^i\wedge e^k=0$ \\
\botrule
\end{tabular}
}
%\begin{tabnote}
%$^{\text a}$Sample table footnote.
%\end{tabnote}
%\label{ra_tbl1}
\end{table}

Performing the Legendre transformation on the Palatini-Holst action \eref{Palatini-Holst action}, we obtain
\begin{equation}
S=\frac{1}{8\pi G\gamma}\int dt\int_{\Sigma_t} d^3x \left(\tE^a_i\mathcal{L}_tA^i_a
-h(\tE^a_i,A^i_a,N,N^a,(\omega^i\cdot t))\right),
\end{equation}
where $\mathcal{L}$ is the Lie derivative, and the Hamiltonian density $h$ is given by
\begin{equation}
h=(\omega^i\cdot t)G_i+N^aC_a+NC.
\end{equation}
The canonical pair $(A_a^i,\tE^b_j)$ is given by the \emph{Ashtekar connection} (1-form)
\begin{equation}
A_a^i(x):=\Gamma_a^i(x)+\gamma K_a^i(x)
\end{equation}
and the \emph{densitized inverse triad} (vector density of weight 1)
\begin{equation}
\tE_i^a:=\frac{1}{2}e_b^je_c^k\epsilon^{abc}\epsilon_{ijk},
\end{equation}
which gives the 3-metric $q_{ab}$ via
\begin{equation}
q\,q^{ab}=\tE^a_i\tE^b_j\delta^{ij}
\end{equation}
with
\begin{equation}
q\equiv\abs{\det q_{ab}}=\det \tE^a_i.
\end{equation}
The Hamiltonian density $h$ is given by three (local) constraints $G_i=0$ (Gauss constraint), $C_a=0$ (diffeomorphism constraint), and $C=0$ (scalar constraint) associated with the Lagrange multipliers $(\omega^i\cdot t)$, $N^a$, and $N$, respectively.
The three constraints are given by
\begin{subequations}\label{three constraints}
\begin{eqnarray}
G_i &=& \mathcal{D}_a\tE^a_i: =\partial_a\tE^a_i
+{\epsilon_{ij}}^k A_a^j\tE^a_k,\\
C_a &=& \tE^b_i F_{ab}^i
-\frac{\sigma-\gamma^2}{\sigma\gamma}K_a^iG_i,\\
\label{constraint C}
C &=& \frac{\gamma}{2\sqrt{\abs{q}}}\tE^a_i\tE^b_j
\left[
{\epsilon^{ij}}_k F_{ab}^k+(\sigma-\gamma^2)2K^i_{[a}K^j_{b]}\right]\nonumber\\
&&\quad\mbox{}
+8\pi G(\gamma^2-\sigma)\partial_a\left(\frac{\tE_i^a}{\sqrt{\abs{q}}}\right)G^i,
\end{eqnarray}
\end{subequations}
where $F_{ab}^k$ is the curvature of the connection $A_a^i$ and $\sigma=-1$.\footnote{Had we considered the Euclidean GR instead of the Lorentzain GR, the Lorentz group $SO(3,1)$ would have been replaced by the Euclidean rotational group $SO(4)$. Correspondingly, $\sigma=+1$ and $A_a^i:=\Gamma_a^i-\sigma\gamma K_a^i =\Gamma_a^i-\gamma K_a^i$ for the Euclidean case.}

The fundamental variables are $A_a^i$ and $\tE^a_i$, which satisfy the canonical relation
\begin{equation}\label{PB of A and E}
\{A_a^i(x),\tE^b_j(y)\}
=8\pi G\gamma\,\delta^i_j\delta_a^b\delta^3(x-y),
\end{equation}
and all three constraints can be expressed in terms of $A^i_a$ and $\tE^a_i$.\footnote{Note that $K_a^i=(\Gamma_a^i-A_a^i)/(\sigma\gamma)$ and $\Gamma_a^i$ is a non-polynomial function of $\tE^a_i$.}
For any smooth $su(2)$-valued field $\lambda=\lambda^i\tau_i$ on $\Sigma$, the functional
\begin{equation}\label{C Gauss}
\calC_\mathrm{G}[\lambda]:= \frac{1}{8\pi G\gamma}\int_\Sigma d^3x \lambda^i G_i
\end{equation}
generates precisely the infinitesimal internal $SU(2)$ gauge rotations by $\lambda$:
\begin{equation}\label{under C Gauss}
\left\{A_a^i,\calC_\mathrm{G}[\lambda]\right\} = -\mathcal{D}_a\lambda^i
\quad\text{and}\quad
\left\{\tE^a_i,\calC_\mathrm{G}[\lambda]\right\} = {\epsilon_{ij}}^k\lambda^j\tE^a_k.
\end{equation}
Consequently, the Gauss constraint $G_i=0$ is the ``Gauss law'' that ensures the gauge invariance under the internal $SU(2)$ rotations. Furthermore, removing from $C^a$ and $C$ a suitable multiple of the Gauss constraint, for any smooth vector field $N^a$ and scalar field $N$, we can show that the functional
\begin{equation}
\calC_\Diff[\vec{N}]:= \frac{1}{8\pi G\gamma} \int_\Sigma d^3x \left(N^a\tE_b^iF_{ab}^i-(N^aA_a^i)G_i\right)
\end{equation}
generates the infinitesimal 3d diffeomorphism ($\Diff$) on $\Sigma$ along the direction $\vec{N}$:
\begin{equation}\label{under C Diff}
\left\{A_a^i,\calC_\Diff[\vec{N}]\right\} = \mathcal{L}_{\vec{N}}A_a^i
\quad\text{and}\quad
\left\{\tE^a_i,\calC_\Diff[\vec{N}]\right\} = \mathcal{L}_{\vec{N}}\tE^a_i,
\end{equation}
and the functional
\begin{equation}\label{C[N]}
\calC[N]:=\frac{1}{16\pi G} \int_\Sigma d^3x N
\frac{\tE^a_i\tE^b_j}{\sqrt{\abs{q}}}
\left[
{\epsilon^{ij}}_k F_{ab}^k+(\sigma-\gamma^2)2K^i_{[a}K^j_{b]}\right]
\end{equation}
generates the infinitesimal translation in time along the direct $Nn^\alpha$:
\begin{equation}\label{under C scalar}
\left\{A_a^i,\calC[N]\right\} = \mathcal{L}_{Nn^\alpha}A_a^i
\quad\text{and}\quad
\left\{\tE^a_i,\calC[N]\right\} = \mathcal{L}_{Nn^\alpha}\tE^a_i,
\end{equation}
where $\mathcal{L}$ is the Lie derivative.\footnote{It should be noted that Eqs.~(\ref{under C Diff}) and (\ref{under C scalar}) are valid only if $A_a^i$ and $\tE^a_i$ already satisfy the Gauss constraint, because we have removed a multiple of the Gaussian constraint from $C^a$ and $C$.}
Altogether, the three constraints ensure invariance under the $SU(2)$ gauge rotation, the diffeomorphism on $\Sigma$, and the translation in time.

The three constraints are \emph{first class} in Dirac's terminology; that is, the Poisson bracket of any two of the constraints vanishes in the restricted phase space imposed by the three constraints. Explicitly, the Poisson brackets between the constraints, which give the so-called ``constraint algebra'', read as\footnote{Note that the constraint algebra \eref{PB bw constraints} is not closed in the usual sense, as the arguments of $\calC_\Diff$ and $\calC_\mathrm{G}$ on the right-hand side of \eref{PB C C} involve the phase space variables $A_a$ and $\tE^a$.}
\begin{subequations}\label{PB bw constraints}
\begin{eqnarray}
\left\{\calC_\mathrm{G}[\lambda],\calC_\mathrm{G}[\lambda']\right\} &=& \calC_\mathrm{G}[[\lambda,\lambda']],\\
\left\{\calC_\mathrm{G}[\lambda],\calC_\Diff[\vec{N}]\right\} &=& -\calC_\mathrm{G}[\mathcal{L}_{\vec{N}}\lambda],\\
\left\{\calC_\mathrm{G}[\lambda],\calC[N]\right\} &=& 0,\\
\left\{\calC_\Diff[\vec{N}],\calC_\Diff[\vec{N}']\right\} &=& \calC_\Diff[[\vec{N},\vec{N}']] \equiv \calC[\mathcal{L}_{\vec{N}}\vec{N}'] \equiv -\calC[\mathcal{L}_{\vec{N}'}\vec{N}],\\
\label{PB CDiff C}
\left\{\calC_\Diff[\vec{N}],\calC[M]\right\} &=& \calC[\mathcal{L}_{\vec{N}}M],\\
\label{PB C C}
\left\{\calC[N],\calC[M]\right\} &=& \sigma\left(\calC_\Diff[\vec{S}] + \calC_\mathrm{G}[S^aA_a]\right)\nonumber\\
&& \quad\mbox{} + \frac{\sigma-\gamma^2}{(8\pi G\gamma)^2}\, \calC_\mathrm{G}\left[\abs{q}^{-1} [\tE^a\partial_a N,\tE^b\partial_bM]\right],
\end{eqnarray}
\end{subequations}
where in the last equation the vector field $S^a$ is given by
\begin{equation}
S^a = \left(N\partial_bM-M\partial_bN\right)\frac{\delta^{ij}\tE^b_i\tE^a_j}{\abs{q}}
\equiv \sigma\,q^{ab}\left(N\partial_bM-M\partial_bN\right).
\end{equation}

While the three constraints have to be satisfied, the evolution of the canonical pair is dictated by the Hamilton's equations
\begin{equation}\label{Adot Edot}
\frac{\partial A_a^i}{\partial t} = \big\{A_a^i,H\big\}
\quad\text{and}\quad
\frac{\partial{\tE}^a_i}{\partial t} = \big\{\tE^a_i,H\big\},
\end{equation}
where the (total) Hamiltonian is $H:=(8\pi G\gamma)^{-1}\int_\Sigma d^3x\, h$.\footnote{This Hamiltonian is peculiar and sometimes referred to as ``super-Hamiltonian'' in the sense that it vanishes identically when the three constraints are imposed. The fact that the Hamiltonian vanishes is a characteristic of reparametrization-invariant theories.} Because all three constraints are first class, once all of them are satisfied for an initial state of $(A_a^i,\tE^b_j)$, they will continue to be satisfied under the evolution of \eref{Adot Edot}. The two equations of motion in \eref{Adot Edot} together with the three constraints $G_i=0$, $C_a=0$, $C=0$ are completely equivalent to the Einstein field equation.

\subsection{Remarks on connection theories}\label{sec:remarks on connection theories}
Historically, the Ashtekar connection was not obtained by starting from the Palatini-Holst action. Rather, the original idea\cite{Ashtekar:1986yd} is to replace the Lorentz connection ${\omega_\mu}^{IJ}$ in the Palatini theory with the \emph{complex} Lorentz connection ${A_\mu}^{IJ}$ defined as
\begin{equation}\label{self-dual A}
{A_\mu}^{IJ}:=
\frac{1}{2}\left({\omega_\mu}^{IJ}-\frac{i}{2}\,{\epsilon^{IJ}}_{KL}{\omega_\mu}^{KL}\right),
\end{equation}
which satisfies the \emph{self-dual} condition: $\frac{1}{2}{\epsilon^{IJ}}_{KL}{A_\mu}^{IJ}=i{A_\mu}^{IJ}$. Correspondingly, $\Omega^{KL}$ in \eref{Palatini action} is replaced by $F^{IJ}$, which is the curvature of the connection $A^{IJ}$ and can also be viewed as the self-dual part of $\Omega^{IJ}$, i.e., $F^{IJ}:= \frac{1}{2}\Omega^{IJ}-\frac{1}{4}{\epsilon^{IJ}}_{KL}{\Omega}^{KL}$. It turns out, in terms of the complex connection, all equations in the classical theory are simplified dramatically\cite{Ashtekar:1987gu}, as the new variable manifests the fact that $so(3,1)_\mathbb{C}$ is isomorphic to $su(2)_\mathbb{C}\oplus su(2)_\mathbb{C}$ (where $g_\mathbb{C}$ denotes complexification of $g$). The essence of using the new variable is closely related to the constructions of the twistor theory\cite{Penrose:1999cw} and the $\mathcal{H}$ space theory\cite{Lidvigsen:1981}.

However, the complex connection takes values in the Lie algebra of a noncompact group, posing an obstacle to constructing the corresponding quantum theory. It is Holst\cite{Holst:1995pc} who first realized that the imaginary number $i$ in \eref{self-dual A} can be replaced by any (real or complex) parameter $\gamma$ and this prescription corresponds to adding the second term in \eref{Palatini-Holst action}. Since then, most progress towards the quantum theory has been made with real $\gamma$.
Taking $\gamma=\pm i$ in the Lorentzian GR ($\sigma=-1$) and $\gamma=\pm1$ in the Euclidean GR ($\sigma=+1$), one encounters the special cases, in which the (classical) theory has a richer geometrical structure---particularly, the three constraints in \eref{three constraints} are drastically simplified (note $\sigma-\gamma^2=0$) and the Poisson brackets in \eref{PB bw constraints} form a closed Lie algebra.

Rigorously speaking, the formulation we adopted is not a theory of ``connection'' dynamics but of ``coframe-connection'' dynamics, as we also include the coframe ${e_\mu}^I$ in addition to the connection ${\omega_\mu}^{IJ}$ as the fundamental variables. Connection and coframe are the gauge variables in the Poincar\'{e} gauge theory of gravity (see Ref.~\cite{Blagojevic:2013xpa} for a review). If one postulates a Lagrangian quadratic in the gauge variables for the Poincar\'{e} gauge theory, the most general Lagrangian is given in Eq.~(5.13) of Ref.~\cite{Blagojevic:2013xpa}:
\begin{eqnarray}\label{PG Lagrangian}
S &=& \frac{1}{16\pi G}\int\left(a_0 R + b_0X -2\Lambda\right)\eta\nonumber\\
 && \; \mbox{} + \text{12 more terms quadratic in curvature, torsion, and $X$},
\end{eqnarray}
where $R:={R_{\alpha\beta}}^{\beta\alpha}$ is the Ricci scalar, $X:=\frac{1}{4!}\eta_{\alpha\beta\gamma\delta}R^{\alpha\beta\gamma\delta}$, $\Lambda$ is the cosmological constant, and $\eta$ is the volume 4-form. The Palatini-Holst action \eref{Palatini-Holst action} only takes the first line of \eref{PG Lagrangian} with $a_0=1$, $b_0=\gamma^{-1}$ (and usually $\Lambda=0)$.\footnote{The Lagrangian with the first two terms $\sim a_0R+b_0X$ was first discussed by Hojman \textit{et al.}\cite{Hojman:1980kv}, but the second term $\sim b_0X$ is referred to as Holst's modification in the literature of LQG, mainly because Holst was the first to relate it to the Ashtekar variable. Likewise, the coframe-connection formulation of the first term $\sim a_0R$ is known as the Einstein-Cartan or the Einstein-Cartan-Sciama-Kibble theory in the literature of connection theories (see Ref.~\cite{Blagojevic:2013xpa} for a historical account) but often referred to as the Palatini formulation in LQG, partly because Palatini's particular idea was underscored in Ashtekar's seminal paper Ref.~\cite{Ashtekar:1986yd}. Although the eponyms adopted in LQG are not historically faithful, we adhere to them in conformity with the LQG convention.
%===% which may be justified by \href{http://en.wikipedia.org/wiki/Stigler's_law_of_eponymy}{Stigler's law}---``No scientific discovery is named after its original discoverer''.
} From the standpoint of the Poincar\'{e} gauge theory, it seems rather \textit{ad hoc} to include Holst's term but neglect the second line completely, since Holst's term is equivalent to a certain torsion square term via Nieh-Yan identity. Indeed, it would be theoretically more compelling if one could construct the quantum theory based on the generic Lagrangian \eref{PG Lagrangian}, but inclusion of any terms in the second line unfortunately makes the constraints in the canonical description very complicated and no longer first class, rendering the Hamiltonian formulation inadequate for quantization. The fact that the Palatini-Holst action \textit{per se} is privileged seems to suggest that, even with real $\gamma$, the $su(2)_\mathbb{C}\oplus su(2)_\mathbb{C}$ structure of the complex Lorentz algebra still plays a profound role in QG (as advocated by the twistor theory and the $\mathcal{H}$ space theory), although its direct relevance remains obscure.\footnote{\label{foot:other dimensions}Accordingly, LQG is most naturally constructed for 4-dimensional spacetime. Nevertheless, generalization to other dimensions is still possible, depending on what principles of LQG are to be upheld. (See \sref{sec:SUSY and other dimensions}.)}

Finally, it should be noted that in the case of vacuum the Barbero-Immirzi parameter $\gamma$ does not appear in the equations of motion and thus have no physical effect at the classical level, but this is not true in general. In the presence of minimally coupled fermions, the parameter appears in the equations of motion, giving rise to a four-fermion interaction\cite{Perez:2005pm}. LQG nevertheless takes the Palatini-Holst action (with or without matter) as the starting point, adopting the attitude that the four-fermion interaction is very likely to be real on account of the fact that the effect of gravity on fermions is very difficult to measure.

\section{Quantum kinematics}\label{sec:quantum kinematics}
Departing from the canonical description of GR described in \sref{sec:connection theories}, it is time to construct the quantum theory.

\subsection{Quantization scheme}
Following the standard strategy for quantization in gauge theories, as $A=A_a^i\tau_i dx^a$ is considered as the connection potential and $\tE=\tE^a_i\tau^i\partial_a$ is analogous to the $SU(2)$ electric field (note $\tau^i=\delta^{ij}\tau_j=\tau_i$), one would take functionals of $A$ as the kinematical quantum states to start with. However, this straight approach does not lead us far enough. In LQG, instead of the connection field $A(x)$, we consider the \emph{holonomy} (i.e.\ Wilson line) $h_\gamma$ defined as the path-ordered integral
\begin{equation}\label{h gamma}
h_\gamma := \mathcal{P} \exp \int_\gamma A_a^i\tau_i dx^a
\end{equation}
over an oriented 1-dimensional curve $\gamma$ in $\Sigma$. Correspondingly, instead of the electric field $\tE$, we consider the surface integral
\begin{equation}\label{E[S,f]}
E_{\mathcal{S},f}:=\int_\mathcal{S} \epsilon_{abc}\tE^c_i\, f^i dx^a\wedge dx^b
\end{equation}
over a 2-dimensional surface $\mathcal{S}$ in $\Sigma$ and with a smooth smearing $su(2)$-valued function $f(x)=f(x)^i\tau_i$. The Poisson bracket $\left\{h_\gamma,E_{\mathcal{S},f}\right\}$ can be straightforwardly computed by \eref{PB of A and E}. The bracket vanishes if $\gamma$ and $\mathcal{S}$ do not intersect or $\gamma$ lies within $\mathcal{S}$. Furthermore, $\{h_\gamma\}$ and $\{E_{\mathcal{S},f}\}$ are closed under the Poisson bracket and form the \emph{loop algebra}.

The reason why we should take $h_\gamma$ instead of $A(x)$ as the fundamental variable is essentially because $h_\gamma$ transforms more ``nicely'' than $A$ does under both the $SU(2)$ and the 3d diffeomorphism transformations, and consequently it is much more manageable to remove the gauge overcounting in the quantum theory.

Under the $SU(2)$ gauge transformation, $A(x)$ is transformed via
\begin{equation}\label{A under SU2}
A(x) \rightarrow A'(x)=\Lambda(x)A(x)\Lambda^\dag(x)+\Lambda(x)d\,\Lambda(x)^\dag,
\end{equation}
while $h_\gamma$ is transformed via
\begin{equation}\label{h under SU2}
h_\gamma \rightarrow h'_\gamma=\Lambda(x_f^\gamma)h_\gamma\, \Lambda(x_i^\gamma)^\dag,
\end{equation}
where $x_i^\gamma$ and $x_f^\gamma$ are the initial and final endpoints of $\gamma$. While \eref{A under SU2} involves every point $x$ for $A$, \eref{h under SU2} involves only the endpoints of $\gamma$ for $h_\gamma$.

On the other hand, consider a 3d diffeomorphism map $\varphi:\Sigma\rightarrow\Sigma$ that is smooth and invertible everywhere. Under $\varphi$, the connection $A$ transforms as a 1-form:
\begin{equation}\label{A under Diff}
A \rightarrow A'=\varphi^*A,
\end{equation}
where $\varphi^*$ is the pullback of $\varphi$. It follows that
\begin{equation}\label{h under Diff}
h_\gamma(A) \rightarrow h_\gamma(A')=h_\gamma(\varphi^*A)\equiv h_{\varphi\gamma}(A),
\end{equation}
where $\varphi\gamma\equiv\varphi(\gamma)$ is the image of $\gamma$ by $\varphi$.
Compared with \eref{A under Diff} for $A$, the transformation \eref{h under Diff} for $h_\gamma$ is more manageable.

We will first construct the kinematical Hilbert space $\hil$ using the functions of holonomies, and then implement the three constraints, one after another, to obtain the $SU(2)$-invariant Hilbert space $\hil_\mathrm{G}$, the $SU(2)$- and $\Diff$-invariant Hilbert space $\hil_\inv$, and finally the physical Hilbert space $\hil_\phys$. The scheme can be summarized as
\begin{equation}\label{quantization scheme}
\hil \quad\mathop{\longrightarrow}\limits_{SU(2)}\quad
\hil_\mathrm{G} \quad\mathop{\longrightarrow}\limits_{\Diff}\quad
\hil_\inv \quad\mathop{\longrightarrow}\limits_{\calC}\quad
\hil_\phys.
\end{equation}
The states of $\hil$ will be called \emph{cylindrical functions}; the states of $\hil_\mathrm{G}$ called \emph{spin networks}; and the states of $\hil_\inv$ called \emph{s-knots}. Finally, the physical Hilbert space $\hil_\phys$ is supposed to reveal the quantum dynamics.

\subsection{Cylindrical functions}
Let $\Gamma$ be an oriented \emph{graph}, which is defined as a collection of a finite number of oriented and (piecewise) smooth edges $\gamma_l$ with $l=1,\dots,L$ embedded in $\Sigma$. Consider a smooth function $f$ of $L$ $SU(2)$ elements. The couple $(\Gamma,f)$ defines a functional of $A$ as
\begin{equation}\label{cylindrical fun}
\Psi_{\Gamma,f}[A]\equiv\inner{A}{\Psi_{\Gamma,f}}:=f(h_{\gamma_1}(A),\dots,h_{\gamma_L}(A)).
\end{equation}
These functionals (of $A$) are called \emph{cylindrical functions} (of holonomies).
Let $\Cyl$ denote the linear space of cylindrical functions for all $\Gamma$ and $f$.
With a suitable topology (the detail is not important here), $\Cyl$ is dense in the space of all continuous functionals of $A$; in this sense, cylindrical functions grasp \emph{all} information of continuous functionals of $A$.

If two cylindrical functions are defined for the same graph $\Gamma$, we can define the inner product between them as
\begin{equation}\label{Gamma f Gamma g}
\inner{\Psi_{\Gamma,f}}{\Psi_{\Gamma,g}}:=
\int_{SU(2)^{L}}d\mu_1\dots d\mu_L\,
{f(h_{\gamma_1},\dots,h_{\gamma_L})}^*
\,g(h_{\gamma_1},\dots,h_{\gamma_L}),
\end{equation}
where $d\mu_l$ is the Haar measure on $SU(2)$. If two cylindrical functions are defined by two different couples $(\Gamma',f')$ and $(\Gamma'',g'')$, let $\Gamma=\{\gamma_1,\dots,\gamma_L\} =\{\gamma'_1,\dots,\gamma'_{L'}\} \cup \{\gamma''_{1},\dots,\gamma''_{L''}\}$ be the union of the two graphs $\Gamma'=\{\gamma'_1,\dots,\gamma'_{L'}\}$ and $\Gamma''=\{\gamma''_{1},\dots,\gamma''_{L''}\}$. We can extend the functions $f'$ and $g''$ to be defined in $\Gamma$ in the obvious way as $f(h_{\gamma_1},\dots,h_{\gamma_L}) :=f'(h_{\gamma'_1},\dots,h_{\gamma'_{L'}})$ and $g(h_{\gamma_1},\dots,h_{\gamma_L}) :=g''(h_{\gamma''_1},\dots,h_{\gamma''_{L''}})$. The inner product \eref{Gamma f Gamma g} can then be extended to any two given cylindrical functions as
\begin{equation}\label{Gamma' f' Gamma'' g''}
\inner{\Psi_{\Gamma'\!,\,f'}}{\Psi_{\Gamma''\!,\,g''}}:=\inner{\Psi_{\Gamma,f}}{\Psi_{\Gamma,g}}.
\end{equation}
This implements the inner product measure to the space $\Cyl$. We can then define the kinematical Hilbert space $\hil$ as the Cauchy completion of $\Cyl$ with respect to the norm of the inner product \eref{Gamma' f' Gamma'' g''}, and the dual space $\Cyl^*$ as the completion of $\Cyl$ with respect to the weak topology defined by \eref{Gamma' f' Gamma'' g''}.\footnote{More precisely, $\hil$ is the space of the Cauchy sequences $\ket{\Psi_n}$ in $\Cyl$ (i.e., $\norm{\Psi_m-\Psi_n}$ converges to zero); $\Cyl^*$ is the space of the sequences $\ket{\Psi_n}$ such that $\inner{\Psi_n}{\Psi}$ converges for all $\ket{\Psi}\in\Cyl$.} This complete the Gelfand triple $\Cyl\subset\hil\subset\Cyl^*$. (See Refs.~\cite{Ashtekar:1993wf,Marolf:1994cj,Ashtekar:1994mh} for the rigorous construction.)

The cylindrical functions with support on a given graph $\Gamma$ form a finite-dimensional subspace $\tilde{\hil}_\Gamma$ of $\hil$. Obviously, $\tilde{\hil}_\Gamma=L^2(SU(2)^L,d\mu^L)$ with $L$ being the number of edges in $\Gamma$. If $\Gamma\subset\Gamma'$, the Hilbert space $\tilde{\hil}_\Gamma$ is a subspace of the Hilbert space $\tilde{\hil}_{\Gamma'}$. This nested structure is called a projective family of Hilbert spaces and $\hil$ is (and can be defined as) the \emph{projective limit} of this family\cite{Marolf:1994cj,Ashtekar:1994mh}. It turns out $\hil$ can be viewed as the space of square integrable functionals in the Ashtekar-Lewandowski measure; i.e., $\hil=L^2[\mathcal{A},d\mu_\mathrm{AL}]$, where $\mathcal{A}$ is an extension of the space of smooth connections\cite{Ashtekar:1993wf}.

The Peter-Weyl theorem states that a basis for the Hilbert space of $L^2$ functions on $SU(2)$ is given by the matrix elements of the irreducible representations of the group. The irreducible representation of $SU(2)$ is labelled by an half-integer number $j$ and the matrix elements of the $j$ representation $R^{(j)}$ are denoted as ${R^{(j)\alpha}}_\beta(U)\in \mathbb{C}$ for $U\in SU(2)$. Then a basis of $\tilde{\hil}_\Gamma$ is composed of the states
\begin{equation}\label{basis states}
\ket{\Gamma,j_l,\alpha_l,\beta_l}
\equiv
\ket{\Gamma,j_1,\dots, j_L,\alpha_l,\dots,\alpha_L,\beta_l,\dots,\beta_L},
\end{equation}
which are defined via
\begin{equation}\label{basis states 2}
\inner{A}{\Gamma,j_l,\alpha_l,\beta_l}
={R^{(j_1)\alpha_1}}_{\beta_1}(h_{\gamma_1}(A))\cdots
{R^{(j_L)\alpha_L}}_{\beta_L}(h_{\gamma_L}(A)).
\end{equation}
For each graph $\Gamma$, the \emph{proper} subspace $\hil_\Gamma$ associated with $\Gamma$ is the subspace of $\tilde{\hil}_\Gamma$ spanned by $\ket{\Gamma,j_l,\alpha_l,\beta_l}$ with $j_l\neq0$. The proper subspaces $\hil_\Gamma$ are orthogonal to one another and span the whole $\hil$; i.e.,
\begin{equation}
\hil\cong \bigoplus_\Gamma\hil_\Gamma.
\end{equation}
The ``null' graph $\Gamma=\emptyset$ is included in the sum; it corresponds to the 1-dimensional Hilbert space spanned by the trivial state $\ket{\emptyset}$ defined as $\inner{A}{\emptyset}=1$.
An orthonormal basis of $\hil$ is simply composed of the states $\ket{\Gamma,j_l,\alpha_l,\beta_l}$ with all graphs $\Gamma$ (including $\emptyset$) and all spin labels $j_l=1/2,1,3/2,2,\dots$ (but $j_l\neq0$).

Among the basis states, an important special case is that the graph is given by a \emph{closed} curve (or a ``loop'') $\alpha$. As $\alpha$ has only one edge and no endpoints, the corresponding state $\ket{\Gamma,j_l,\alpha_l,\beta_l}$ becomes $\ket{\alpha,j}$, which is defined as
\begin{equation}\label{loop state}
\inner{A}{\alpha,j}=\Tr\, R^{(j)}\left(\mathcal{P}\exp \oint_\alpha A\right).
\end{equation}
These states $\ket{\alpha,j}$ are called \emph{loop states}. For $j=1/2$ particularly, $\inner{A}{\alpha,j}=\Tr\, h_\alpha(A)$ is the Wilson loop. Similarly, for a ``multiloop'' $[\alpha_l]=(\alpha_1,\dots,\alpha_n)$ given by a collection of a finite number of loops (without intersection), a \emph{multiloop state} $\ket{[\alpha_l],j_l}$ is defined as
\begin{equation}\label{multiloop state}
\inner{A}{[\alpha_l],j_l}=
\prod_{l=1}^n
\Tr\, R^{(j_l)}\left(\mathcal{P}\exp \oint_{\alpha_l} A\right).
\end{equation}

The kinematical Hilbert space $\hil$ carries a natural representation for both $SU(2)$ and $\Diff$ groups. Under the $SU(2)$ gauge transformation, the holonomy $h_\gamma$ transforms by \eref{h under SU2} and it follows that the basis states $\ket{\Gamma,j_l,\alpha_j,\beta_l}$ transform by
\begin{eqnarray}\label{basis state under SU2}
\ket{\Gamma,j_l,\alpha_j,\beta_l} &\rightarrow& \hat{U}_\Lambda\ket{\Gamma,j_l,\alpha_j,\beta_l}\\
&&
=\sum_{\alpha'_l,\beta'_l}
\left(
\prod_{l=1}^L
{R^{(j_l)\alpha_l}}_{\alpha'_l}(\Lambda(x_f^{\gamma_l})^\dag)
{R^{(j_l)\beta'_l}}_{\beta_l}(\Lambda(x_i^{\gamma_l}))\right)
\ket{\Gamma,j_l,\alpha'_l,\beta'_l}.\nonumber
\end{eqnarray}
On the other hand, under the $\Diff$ transformation, \eref{h under Diff} leads to
\begin{equation}\label{state under Diff}
\ket{\Psi_{\Gamma,f}} \rightarrow \hat{U}_\varphi\ket{\Psi_{\Gamma,f}}=\ket{\Psi_{\varphi\Gamma,f}}.
\end{equation}
That is, a cylindrical function $\Psi_{\Gamma,f}[A]$ with support on $\Gamma$ is sent to a new cylindrical function $\Psi_{\varphi\Gamma,f}[A]$ with support on the relocated graph $\varphi\Gamma$.
A moment of reflection tells that the inner product is $\Diff$-invariant:
\begin{equation}\label{inner product Diff-inv}
\inner{\hat{U}_\varphi\Psi_{\Gamma,f}}{\hat{U}_\varphi\Phi_{\Gamma',g}}
= \inner{\Psi_{\varphi\Gamma,f}}{\Phi_{\varphi\Gamma',g}}
= \inner{\Psi_{\Gamma,f}}{\Phi_{\Gamma',g}}.
\end{equation}
In fact, the uniqueness theorem tells that the Ashtekar-Lewandowski measure is the only unique measure that gives rise to the (${}^*$-algebra) representation of the kinematical algebra invariant under spatial diffeomorphisms\cite{Lewandowski:2005jk}.

\subsection{Spin networks}
We are now ready to implement the Gauss constraint upon $\hil$ to obtain the $SU(2)$-invariant space $\hil_\mathrm{G}$.

In the space of functionals $\Psi[A]$, we promote $A_a^i(x)$ to $\hat{A}_a^i(x)$ as the multiplicative operator:
\begin{equation}\label{hat A}
(\hat{A}_a^i(x)\Psi)[A]:=A_a^i(x)\Psi[A],
\end{equation}
and $\tE^a_i$ as the differential operator:
\begin{equation}\label{hat tE}
(\hat{\tE}^a_i\Psi)[A]:=-i8\pi G\hbar\gamma\frac{\delta\Psi[A]}{\delta A_a^i(x)},
\end{equation}
in accordance with \eref{PB of A and E}. The Gauss constraint $\calC_\mathrm{G}[\lambda]=0$ in \eref{C Gauss} is then promoted to the operator $\hat{\calC}_\mathrm{G}[\lambda]$:
\begin{equation}
(\hat{\calC}_\mathrm{G}[\lambda]\Psi)[A]=\Psi[A-\mathcal{D}\lambda],
\end{equation}
according to \eref{under C Gauss}. The $SU(2)$-invariant states are those lying in the kernel of $\hat{\calC}_\mathrm{G}[\lambda]$ for arbitrary $\lambda$.

In the quantum theory based on cylindrical functions of connections, in place of functionals of connections, it is the \emph{finite} gauge transformation that is of primary importance. Accordingly, the space $\hil_\mathrm{G}$ as the kernel of $\hat{\calC}_\mathrm{G}[\lambda]$ is obtained as the invariant subspace of $\hil$ under finite transformations $\hat{U}_\Lambda$, which act on $\ket{\Psi}$ as
\begin{equation}
(\hat{U}_\Lambda\Psi)[A]=\Psi[\Lambda A\,\Lambda^\dag+\Lambda d\Lambda^\dag]
\end{equation}
in accordance with \eref{A under SU2}.

For a given graph $\Gamma$, call the endpoints of the edges ``nodes'' and assume that, without loss of generality, the edges overlap, if they do at all, only at nodes. The oriented edges are then also referred to as ``links''.
Under a finite $SU(2)$ transformation $\hat{U}_\Lambda$, the basis states transform according to \eref{basis state under SU2}, which is invariant in the subspace $\hil_\Gamma$ and acts only on the nodes of $\Gamma$. At each node $v_n$, assume there are $n_\mathrm{in}$ ``ingoing'' links and $n_\mathrm{out}$ ``outgoing'' links. The task is to find the $SU(2)$-invariant (i.e., $j=0$) spin states within the tensor product $j_1\otimes\cdots j_{n_\mathrm{in}}\otimes \bar{j'}_{1}\otimes\cdots \bar{j'}_{n_\mathrm{out}} =0\oplus0\oplus\cdots$ at each node $v_n$. This can be achieved by finding appropriate linear superpositions:
\begin{eqnarray}
&&\ket{J=0,M=0} \nonumber\\
&=& \sum_{m_i,m'_j} {{i_n}^{m'_1\dots m'_{n_\mathrm{out}}}}_{m_1\dots m_{n_\mathrm{in}}}
\Big(\ket{j_1,m_1}\otimes\cdots\otimes\ket{j_{n_\mathrm{in}},m_{n_\mathrm{in}}}\Big)\\
&&\qquad \mbox{} \otimes
\Big(\ket{j'_{1},-m'_{1}}\otimes\cdots\otimes\ket{j'_{n_\mathrm{in}},-m'_{n_\mathrm{out}}}\Big),
\nonumber
\end{eqnarray}
where ${{i_n}^{\cdots}}_{\cdots}$ are called \emph{intertwiners} and regarded as generalized Clebsch-Gordan coefficients, which can be obtained by the standard method of the recoupling theory\cite{Kauffman:1994book} (also see Appendix~A of Ref.~\cite{Rovelli:2004tv}).

\begin{figure}

\center

\begin{tikzpicture}

%-------------------------------------------
%\draw[help lines] (0,0) grid (13,4);

\begin{scope}
%-------------------------------------------
% 1st link:
\draw [cyan,arrow data={0.5}{>},mylabel=at 0.5 above with {$j_1$}] (1,2) to [out=55,in=180] (2.8,3) to [out=0,in=140] (4.5,2);
% 2nd link:
\draw [cyan,arrow data={0.5}{>},mylabel=at 0.5 above with {$j_2$}] (1,2) to [out=40,in=165] (2.8,1.8) to [out=-15,in=195] (4.5,2);
% 3rd link:
\draw [cyan,arrow data={0.5}{>},mylabel=at 0.5 below with {$j_3$}] (4.5,2) to [out=225,in=5] (2.7,1) to [out=185,in=-45] (1,2);
% two nodes:
\draw [fill] (1,2) circle [radius=0.05] (4.5,2) circle [radius=0.05];
% labels:
\node at (1.2,2.4) {$\beta_1$};
\node at (1.4,2) {$\beta_2$};
\node at (1.2,1.65) {$\alpha_3$};
\node at (4.25,2.25) {$\alpha_1$};
\node at (4,1.9) {$\alpha_2$};
\node at (4.25,1.6) {$\beta_3$};
\end{scope}

\begin{scope}[shift={(6.5,0)},rotate=0]
%-------------------------------------------
% 1st link:
\draw [cyan,arrow data={0.5}{>},mylabel=at 0.5 above with {$j_1$}] (1,2) to [out=55,in=180] (2.8,3) to [out=0,in=140] (4.5,2);
% 2nd link:
\draw [cyan,arrow data={0.5}{>},mylabel=at 0.5 above with {$j_2$}] (1,2) to [out=40,in=165] (2.8,1.8) to [out=-15,in=195] (4.5,2);
% 3rd link:
\draw [cyan,arrow data={0.5}{>},mylabel=at 0.5 below with {$j_3$}] (4.5,2) to [out=225,in=5] (2.7,1) to [out=185,in=-45] (1,2);
% two nodes:
\draw [fill] (1,2) circle [radius=0.05] (4.5,2) circle [radius=0.05];
% two nodes:
\draw [fill,red] (1,2) node [left] {$i_1$} circle [radius=0.05] (4.5,2) node [right] {$i_2$} circle [radius=0.05];
\end{scope}

%-------------------------------------------
\begin{scope}
\node at (6,1.8) {{\Large $\Rightarrow$}};
\node at (6,2.3) {$\sum_{\alpha_l,\beta_l}\cdots$};
\end{scope}

\end{tikzpicture}

\caption{The left diagram represents the state $\ket{\Gamma,j_l,\alpha_l,\beta_l}$ while the right one represents the spin networks $\ket{\Gamma,j_l,i_n}$, both for the same simple graph $\Gamma$ composed of three oriented links and two nodes. The state on the right is a linear superposition of states on the left via the intertwiners.}
\label{fig:spin network}

\end{figure}
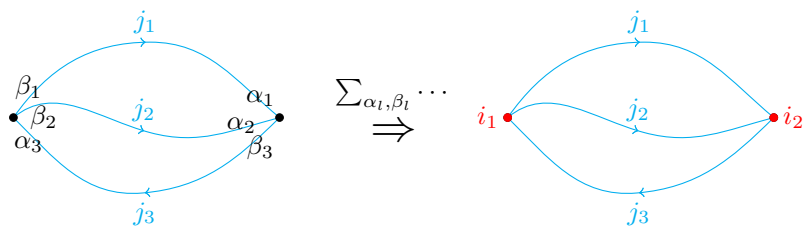

With every node specified with an intertwiner denoted as $i_n$, an $SU(2)$-invariant state $\ket{\Gamma,j_l,i_n}$ is given by
\begin{eqnarray}
\ket{\Gamma,j_l,i_n}
=
\sum_{\alpha_l,\beta_l}
\left(
\prod_{n=1}^N
{{i_n}^{\beta^{(n)}_1\dots\beta^{(n)}_{n_\mathrm{out}}}}_
{\alpha^{(n)}_1\dots\alpha^{(n)}_{n_\mathrm{in}}}
\right)
\ket{\Gamma,j_l,\alpha_l,\beta_l},
\end{eqnarray}
where $N$ is the number of nodes, $\alpha^{(n)}_i$ ($i=1,\dots,n_\mathrm{out}$) is one of $\alpha_l$ that is ``ingoing'' to the node $v_n$, and $\beta^{(n)}_i$ ($i=1,\dots,n_\mathrm{in}$) is one of $\beta_l$ that is ``outgoing'' from the node $v_n$. See \fref{fig:spin network} for a simple example. The states $\ket{\Gamma,j_l,i_n}$ are called \emph{spin networks}, and choice of $j_l$ and $i_n$ is called the ``coloring'' of the links and the nodes. Note that the loop states $\ket{\alpha,j}$ and multiloop states $\ket{[\alpha_l],j_l}$ are trivial spin networks without nodes.

It is clear that the $SU(2)$-invariant states of $\hil_\Gamma$ form a proper subspace of $\hil_\Gamma$, which inherits the same inner product structure of $\hil_\Gamma$. Consequently, we have the $SU(2)$-invariant Hilbert space $\hil_\mathrm{G}$ as the $SU(2)$-invariant subspace of $\hil$ and the corresponding Gelfand triple $\Cyl_\mathrm{G}\subset\hil_\mathrm{G}\subset\Cyl_\mathrm{G}^*$ in the obvious manner.

See Ref.~\cite{Rovelli:1995ac} for the original idea of spin networks and Refs.~\cite{Baez:1994hx,Baez:1995md} for the systematic implementation.

\subsection{S-knots}
The next step to implement the second constraint $\calC_\Diff(\vec{N})=0$ for the far more crucial invariance: the 3d diffeomorphism invariance. (We will follow the lines of Sec.~6.2 in \cite{Ashtekar:2004eh}.) To solve the constraint, we adopt the ``group averaging procedure'': the invariant states are obtained by averaging over the elements of $\Cyl$ that are transitive to one another by the invariance group\cite{Marolf:1995cn,Ashtekar:1995zh}. It turns out the $\Diff$-invariant states are not in $\hil$ but in the extended space $\Cyl^*$.

Given a graph $\Gamma$, denote by $\Diff_\Gamma$ the subgroup of $\Diff$ that maps $\Gamma$ to itself, and by $\mathrm{TDiff}_\Gamma$ the subgroup of $\Diff_\Gamma$ that preserves every edge of $\Gamma$ and its orientation. It is easy to see that the induced action of $\mathrm{TDiff}_\Gamma$ on $\Cyl_\Gamma$ is trivial. The quotient
\begin{equation}
\mathrm{GS}_\Gamma=\Diff_\Gamma/\mathrm{TDiff}_\Gamma
\end{equation}
is the group of \emph{graph symmetries} of $\Gamma$, which permutates the ordering and/or flips the orientations of the edges of $\Gamma$. $\mathrm{GS}_\Gamma$ is a finite and discrete group and it induces a nontrivial action $\widehat{\mathrm{GS}}_\Gamma$ on $\Cyl_\Gamma$ in the obvious way.

We construct the general solutions to the diffeomorphism constraint in two steps. First, given a state $\ket{\Psi_\Gamma}\in\hil_\Gamma$, we average it over the group of $\mathrm{GS}$ and obtain a projection map $\hat{P}_{\Diff,\Gamma}$ from $\hil_\Gamma$ to its subspace that is invariant under $\widehat{\mathrm{GS}}_\Gamma$:
\begin{equation}\label{Pdiff Gamma}
\hat{P}_{\Diff,\Gamma}\ket{\Psi_\Gamma} := \frac{1}{N_\Gamma} \sum_{\varphi\in\mathrm{GS}_\Gamma}\hat{U}_\varphi\ket{\Psi_\Gamma},
\end{equation}
where $N_\Gamma$ is the number of $\mathrm{GS}_\Gamma$ and $\hat{U}_\varphi$ is defined in \eref{state under Diff}. The state $\hat{P}_{\Diff,\Gamma}\ket{\Psi_\Gamma}$ is $\mathrm{GS}_\Gamma$-invariant, as for any $\varphi'\in\mathrm{GS}_\Gamma$ we have
\begin{eqnarray}
\hat{U}_{\varphi'}\hat{P}_{\Diff,\Gamma}\ket{\Psi_\Gamma} &=& \frac{1}{N_\Gamma} \sum_{\varphi\in\mathrm{GS}_\Gamma}\hat{U}_\varphi\hat{U}_\varphi\ket{\Psi_\Gamma}
=\sum_{\varphi\in\mathrm{GS}_\Gamma}\hat{U}_{\varphi'\circ\varphi}\ket{\Psi_\Gamma} \nonumber\\
&=& \sum_{\varphi'^{-1}\circ\varphi''\in\mathrm{GS}_\Gamma}\hat{U}_{\varphi''}\ket{\Psi_\Gamma}
=\hat{P}_{\Diff,\Gamma}\ket{\Psi_\Gamma}.
\end{eqnarray}
The map $\hat{P}_{\Diff,\Gamma}$ naturally extends to the projection $\hat{P}_\Diff$, which projects $\hil$ to its subspace that is invariant under $\widehat{\mathrm{GS}}_\Gamma$ for all $\Gamma$.

Next, we average over the remaining diffeomorphisms that move the graph $\Gamma$. This is a very huge group and the resulting group-averaged states are genuinely ``distributions'' and belong to $\Cyl^*$ instead of $\hil$. For any given state $\ket{\Psi_\Gamma}\in\hil_\Gamma$, we can define the group-averaged state $\obra{\eta(\Psi_\Gamma)}\in\Cyl^*$ by its linear action on arbitrary cylindrical functions $\ket{\Phi_{\Gamma'}}\in\Cyl$ as
\begin{equation}\label{eta Psi}
\oinner{\eta(\Psi_\Gamma)}{\Phi_{\Gamma'}}:=
\sum_{\varphi\in\Diff/\Diff_\Gamma} \inner{\hat{U}_\varphi\hat{P}_{\Diff,\Gamma}\Psi_\Gamma}{\Phi_{\Gamma'}},
\end{equation}
where $\inner{\cdot}{\cdot}$ is the inner product of $\hil$. This action is well defined because only a finite number of terms in the sum are nonzero when $\Gamma'=\varphi\Gamma$ for some $\varphi$. It follows from \eref{inner product Diff-inv} that the state $\obra{\eta(\Psi_\gamma)}$ is invariant under the action of $\Diff$:
\begin{equation}\label{Diff inv}
\oinner{\eta(\Psi_\Gamma)}{\hat{U}_\varphi\Phi_{\Gamma'}}
=\oinner{\eta(\Psi_\Gamma)}{\Phi_{\Gamma'}}
\end{equation}
for all $\varphi\in\Diff$. The space of these solutions to the diffeomorphism constraint is denoted by $\Cyl_\Diff^*$, and we have constructed a map
\begin{equation}\label{map eta}
\eta: \Cyl\rightarrow\Cyl_\Diff^*,
\end{equation}
which sends \emph{every} element of $\Cyl$, upon group averaging, to a $\Diff$-invariant state of $\Cyl_\Diff^*$.

It should be noted that $\eta$ is \emph{not} a projection as it maps $\Cyl$ onto a different space $\Cyl_\Diff^*$. Nevertheless, the group averaging procedure naturally endows $\Cyl_\Diff^*$ with the inner product
\begin{equation}\label{ooinner}
\ooinner{\eta(\Psi)}{\eta(\Phi)}:= \oinner{\eta(\Psi)}{\Phi},
\end{equation}
which is well defined as the right-hand side is independent of the specific choice of $\Psi$ and $\Phi$ for the same states $\obra{\eta(\Psi)}$ and $\obra{\eta(\Phi)}$. With respect to this inner product, the Hilbert space $\hil_\Diff$ of $\Diff$-invariant states is the Cauchy completion of $\Cyl_\Diff^*$. Finally, we can obtain the general solutions to both the Gauss and the diffeomorphism constraints by simply restricting the starting state $\ket{\Psi}\in\Cyl$ to be $SU(2)$ invariant; i.e., we start with $\ket{\Psi}\in\Cyl\cap\hil_\mathrm{G}$. The space $\Cyl_\inv^*$ of solutions to both the Gauss and diffeomorphism constraints is then given by
\begin{equation}
\Cyl_\inv^* = \eta\left(\Cyl\cap\hil_\mathrm{G}\right),
\end{equation}
and the Hilbert space $\hil_\inv$ of $SU(2)$- and $\Diff$-invariant states is the Cauchy completion of $\Cyl_\inv^*$.

What happens to operators? For a given ($SU(2)$ and) diffeomorphism invariant operator $\hat{\mathcal{O}}$ acting on $\hil$ (i.e., $\hat{\mathcal{O}}$ commutes with $\hat{U}_\varphi$; we will see examples of diffeomorphism invariant operators in \sref{sec:quantum geometry}), one can define the corresponding operator $\hat{\mathcal{O}}^*$ acting on $\Cyl_\Diff^*$ as
\begin{equation}\label{O star}
\oinner{\hat{\mathcal{O}}^*\eta(\Psi)}{\Phi}
:=\oinner{\eta(\Psi)}{\hat{\mathcal{O}}^\dag\Phi},
\end{equation}
which is well defined in the sense that $\hat{\mathcal{O}}^*$ maps from $\Cyl_\Diff^*$ into $\Cyl_\Diff^*$, since one can show that, by \eref{Diff inv}, $\oinner{\hat{\mathcal{O}}^*\eta(\Psi)}{\hat{U}_\varphi\Phi} =\oinner{\hat{\mathcal{O}}^*\eta(\Psi)}{\Phi}$ provided that $\hat{\mathcal{O}}^\dag$ commutes with $\hat{U}_\varphi$.
Furthermore, the operator $\hat{\mathcal{O}}^*$ is hermitian in $\hil_\Diff$ with respect to the inner product \eref{ooinner} if and only if $\hat{\mathcal{O}}$ is hermitian in $\hil$.

The fact that the ($SU(2)$ and) diffeomorphism invariant Hilbert space can be rigorously constructed is very significant, in contrast to the quantization program of geometrodynamics using metrics as fundamental variables, in which the precise construction remains elusive.

To understand the structure of $\hil_\inv$, we consider the action of $\eta$ on the basis states of spin networks $\ket{S}\equiv\ket{\Gamma,j_l,i_n}$ of $\hil_\mathrm{G}$. By \eref{eta Psi}, we have
\begin{eqnarray}\label{eta S' S}
&&\ooinner{\eta{(S')}}{\eta(S)}\equiv\oinner{\eta{(S')}}{S}\nonumber\\
&=&
\left\{
  \begin{array}{ll}
    0 & \quad\text{if}\ \varphi\Gamma\neq\Gamma'\ \text{for all}\ \varphi\in\Diff,\\
    \sum_{g_k\in\mathrm{GS}_\Gamma} \bra{S'}\hat{U}_{g_k}\ket{S} & \quad\text{if}\ \varphi\Gamma=\Gamma'\ \text{for some}\ \varphi\in\Diff,
  \end{array}
\right.
\end{eqnarray}
where $\hat{U}_{g_k}\in\widehat{\mathrm{GS}}_\Gamma$, and the state $\hat{U}_{g_k}\ket{S}$ is obtained from $\ket{S}$ by changing the ordering and/or the orientations of the edges of $\Gamma$. As equivalence classes of unoriented graphs under diffeomorphisms are called ``knots'' and classified by their knotting structures of edges, the first line of \eref{eta S' S} tells that two spin networks $\ket{S}$ and $\ket{S'}$ in $\hil_\mathrm{G}$ give rise to two orthogonal states $\obra{\eta(S)}$ and $\obra{\eta(S')}$ in $\hil_\inv$, unless $\Gamma$ and $\Gamma'$ are knotted in the same way. The states in $\hil_\inv$ are then distinguished by the knots, denoted as $K$, as well the colorings of links and nodes of $K$. The states with the same $K$ but different colorings, however, are not necessarily orthonormal to one another, due to the nontrivial action of the discrete symmetry group $\mathrm{GS}_\Gamma$ in the second line of \eref{eta S' S}. To obtain an orthonormal basis of $\hil_\inv$, we have to further diagonalize the quadratic form defined by the second line of \eref{eta S' S}. Denote $\obra{K,c}$ the resulting states, where the discrete label $c$ corresponds to the coloring of links and nodes of $\Gamma$ up to the complications caused by the discrete graph symmetry group $\mathrm{GS}_\Gamma$. We call the states $\obra{K,c}$ \emph{s-knots} and $c$ the coloring of the knot $K$.
It should be noted that diagonalizing the quadratic form in \eref{eta S' S} may yield degenerate eigenstates, as the projection map $\hat{P}_{\Diff,\Gamma}$ defined in \eref{Pdiff Gamma} might have a nontrivial kernel. As a result, the coloring $c$ for s-knots $\obra{K,c}$ in general has less choice than the coloring $j_l,i_n$ for spin networks $\ket{\Gamma,j_l,i_n}$.

Knots without nodes have been widely studied in the knot theory, while knots with nodes have also been studied but to a lesser extent\cite{Bruegmann:1992ak,Bruegmann:1992gp}. One peculiarity of knots with nodes is that there are graphs with 4-valent or higher valent nodes that cannot be mapped to one another by smooth diffeomorphisms.\footnote{Any 3d diffeomorphism $\varphi$ induces a linear map $J_p$ (pushforward) from the tangent space of a given point $p$ to the tangent space of $\varphi(p)$. In order for $\varphi$ to send two nodes into each other, we must have $J_p\vec{v}_i=\vec{v}'_i$, where $\vec{v}_i$, $i=1,\dots,n$, are the $n$ tangent vectors of the links at $p$ and $\vec{v}'_i$, $i=1,\dots,n$, are the $n$ tangent vectors of the links at $\varphi(p)$. In general, however, there is no pushforward map sending $n$ given vectors into $n$ other given directions when $n\geq4$, as the rank of $J_p$ is only 3.} Consequently, the knot classes are labelled by continuous parameters (moduli) and therefore the space $\hil_\inv$ remains nonseparable as the kinematical Hilbert space $\hil$\cite{Rovelli:1995ac,Grot:1996kj}. This problem can be resolved by extending the group $\Diff$ of \emph{smooth} diffeomorphisms to the group $\Diff^*$ of \emph{extended} diffeomorphisms\cite{Fairbairn:2004qe}. Extended diffeomorphisms are maps from $\Sigma$ to $\Sigma$ that are smooth and invertible everywhere in $\Sigma$ except at a finite number of isolated points. With $\Diff$ replaced by $\Diff^*$, the knot classes are discretely classified and consequently the Hilbert space $\hil_\inv$ is \emph{separable}. The ``excessive size'' of $\hil$ in terms of nonseparability turns out to be just a gauge artifact. It is argued in Ref.~\cite{Fairbairn:2004qe} (also see Sec.~6.7 of Ref.~\cite{Rovelli:2004tv}) that $\Diff^*$ in place of $\Diff$ is in fact more natural as the quantum theory is concerned, for diffeomorphisms in $\Diff$ are too ``rigid'' in the sense that they leave invariant the linear structures of tangent spaces at nodes, which however have no direct physical significance.

The physical interpretation of the s-knots states will be clear after we define the operators.

\section{Operators and quantum geometry}\label{sec:operators and quantum geometry}
The two fundamental variables in the canonical theory is the connection $A_a^i$ and its conjugate momentum $\tE^a_i$. In the space of functionals of $A$, the corresponding operators $\hat{A}_a^i$ and $\hat{\tE}^a_i$ are given by Eqs.~(\ref{hat A}) and (\ref{hat tE}), respectively. These two operators are however not good operators in the space $\Cyl$ of cylindrical functions. We will follow the lines of Sec.~6.6 of Ref.~\cite{Rovelli:2004tv} to construct the appropriate quantum operators.

\subsection{Holonomy operator}
In $\Cyl$, the operators of connections have to be replaced by the operators of holonomies defined in \eref{h gamma}. Let ${(h_\gamma)^A}_B$ be the matrix elements of the holonomy $h_\gamma$. Then the corresponding operator ${(\hat{h}_\gamma)^A}_B$ acting on $\ket{\Psi}\in\Cyl$ is simply the multiplicative operator defined as
\begin{equation}\label{hat h}
\left({(\hat{h}_\gamma)^A}_B\Psi\right)[A]:={(h_\gamma)^A}_B(A)\Psi[A].
\end{equation}
The right-hand side is clearly in $\Cyl$. In fact, any cylindrical function is immediately well defined as a multiplicative operator in $\Cyl$. For example, the operator $\hat{S}$ associated with the spin network $\ket{S}$ acts on another spin network $\ket{S'}$ as
\begin{equation}\label{hat S}
\hat{S}\ket{S'}=\ket{S\cup S'},
\end{equation}
where $\ket{S\cup S'}$ is the spin network formed by gluing $\mathcal{S}$ and $S'$ together in the obvious way:
\begin{equation}
\bra{A}\hat{S}\ket{S'}=\inner{A}{S}\inner{A}{S'}.
\end{equation}
Furthermore, it is clear that $\hat{S}$ is $SU(2)$ invariant (i.e., $\hat{S}$ commutes with $\hat{U}_\Lambda$) and leaves the Hilbert space $\hil_\mathrm{G}$ invariant. Spin networks $\ket{S}$ can be constructed as the action of $\hat{S}$ acting on the trivial state $\ket{\emptyset}$; i.e., $\hat{S}\ket{\emptyset}=\ket{S}$. In this sense, $\ket{\emptyset}$ is analogous to the Fock vacuum.

\subsection{Area operator}
On the other hand, when acting on cylindrical functions, the operator defined in \eref{hat tE} leads to
\begin{equation}\label{dh dA}
\frac{i}{8\pi G\hbar\gamma}\hat{\tE}^a_i(x)\,h_\gamma=
\frac{\delta}{\delta A_a^i(x)}h_\gamma = \int_\gamma ds\, \dot{\gamma}^a(s)\, \delta^3(\gamma(s),x)\, \left(h_{\gamma_1}\tau_i\, h_{\gamma_2}\right),
\end{equation}
where $s$ is an arbitrary parametrization of the curve $\gamma$, $\dot{\gamma}^a(s)\equiv d\gamma^a(s)/ds$ is the tangent to the curve at the point $\gamma(s)$, and $\gamma_1$ and $\gamma_2$ are the two segments into which $\gamma$ is separated by the point $\gamma(s)$. The right-hand side is a 2-dimensional Dirac distribution ($\delta^3(\cdots)$ is integrated over one dimension) and thus does not belong to $\Cyl$. As $\hat{\tE}^a_i$ is not a well-defined operator in $\hil$, we instead seek the desired operator by smearing $\tE$ over a 2-dimensional surface $\mathcal{S}$. In accordance with \eref{E[S,f]} (with $f=1$), we define the operator
\begin{equation}
\hat{E}_i(\mathcal{S}):=-8\pi i G\hbar\gamma \int_\mathcal{S} d\sigma^1d\sigma^2 n_a(\sigma)
\frac{\delta}{\delta A_a^i(x(\sigma))},
\end{equation}
where $\sigma=(\sigma^1,\sigma^2)$ are the parametrizations of the surface $\mathcal{S}$, and
\begin{equation}
n_a(\sigma)=\epsilon_{abc}
\frac{\partial x^b(\sigma)}{\partial\sigma^1}
\frac{\partial x^c(\sigma)}{\partial\sigma^a}
\end{equation}
is the 1-form normal to the surface $\mathcal{S}$ at the point $x(\sigma)$. When acting on holonomies, by \eref{dh dA}, $\hat{E}_i(\mathcal{S})$ yields
\begin{equation}
\hat{E}_i(\mathcal{S})h_\gamma = -8\pi i G\hbar\gamma\sum_{p\in \mathcal{S}\cap\gamma} \pm h_{\gamma_1^p}\tau_i\, h_{\gamma_2^p},
\end{equation}
where $p$ are intersection points (one, many, or none) between the curve $\gamma$ and the surface $\mathcal{S}$, $\gamma_1^p$ and $\gamma_2^p$ are the two segments of $\gamma$ separated by the point $p$, and the sign $\pm$ is given by $+$ if the curve $\gamma$ pierces $\mathcal{S}$ at $p$ ``upwards'' to the orientation of $\mathcal{S}$ and $-$ if ``downwards'' to the orientation of $\mathcal{S}$. The generalization to an arbitrary representation of the holonomy is obvious:
\begin{equation}\label{hat EiS}
\hat{E}_i(\mathcal{S})R^{(j)}(h_\gamma) = -8\pi i G\hbar\gamma\sum_{p\in \mathcal{S}\cap\gamma} \pm h_{\gamma_1^p}R^{(j)}(\tau_i) h_{\gamma_2^p}.
\end{equation}
Therefore, $\hat{E}_i(\mathcal{S})$ is a well-defined operator on $\hil$.

The operator $\hat{E}_i(\mathcal{S})$ is not $SU(2)$ invariant, and we cannot obtain an $SU(2)$-invariant operator by simply contracting the index $i$ as
\begin{equation}\label{E square}
\widehat{E^2}(\mathcal{S}):=\delta^{ij}\hat{E}_i(\mathcal{S})\hat{E}_j(\mathcal{S}),
\end{equation}
because the $SU(2)$ transformation of $\hat{E}_i(\mathcal{S})$ is complicated by the integral over $\mathcal{S}$. Instead, we first partition $\mathcal{S}$ into $N$ small surfaces $\mathcal{S}_I$, which become smaller and smaller as $N\rightarrow\infty$ (for each $N$, $\bigcup_{I=1}^N \mathcal{S}_I=\mathcal{S}$), and then define the \emph{area operator} associated with $\mathcal{S}$ as
\begin{equation}\label{area operator}
\hat{A}(\mathcal{S}) := \lim_{N\rightarrow\infty}\sum_I\sqrt{\widehat{E^2}(\mathcal{S}_I)},
\end{equation}
which corresponds to the classical area of $\mathcal{S}$,
\begin{equation}\label{A(S)}
A(\mathcal{S})=\int_\mathcal{S} \sqrt{n_a \tE^a_i n_b \tE^b_j \delta^{ij}}\,d^2\sigma.
\end{equation}

Let us study the action of $\hat{A}(\mathcal{S})$ on a spin network $\ket{\Gamma,j_l,i_n}$, assuming that no spin network nodes lies on $\mathcal{S}$. For sufficiently large $N$, none of $\mathcal{S}_I$ will contain more than one intersection point with $\Gamma$ and the sum over $I$ becomes a sum over the intersection points $p$ between $\mathcal{S}$ and $\Gamma$. Consequently, by \eref{hat EiS}, we have
\begin{equation}\label{area spectrum}
\hat{A}(\mathcal{S})\ket{\Gamma,j_l,i_n} = 8\pi G\hbar\gamma \sum_{p\in \mathcal{S}\cap\Gamma} \sqrt{j_p(j_p+1)}\ \ket{\Gamma,j_l,i_n},
\end{equation}
where $p$ are the intersection points between $\mathcal{S}$ and $\Gamma$, and $j_p$ is the color of the link that pierces $\mathcal{S}$ at $p$. Since the operator $\hat{A}(\mathcal{S})$ is diagonal on spin network states and its eigenvalues are real, the operator $\hat{A}(\mathcal{S})$ for an arbitrary $\mathcal{S}$ is well defined in $\hil_\mathrm{G}$ and is hermitian. For a complete and rigourous construction of the area operator, see Refs.~\cite{Thiemann:2007zz,Ashtekar:1996eg,Frittelli:1996cj}. The general result (with the possibility that the spin network nodes lie on $\mathcal{S}$) is given by
\begin{eqnarray}\label{area spectrum generic}
&&\hat{A}(\mathcal{S})\ket{\Gamma,\cdots}\nonumber\\
&=& 4\pi G\hbar\gamma \sum_{p\in \mathcal{S}\cap\Gamma}
\sqrt{2j^\mathrm{u}_p(j^\mathrm{u}_p+1) +2j^\mathrm{d}_p(j^\mathrm{d}_p+1) -j^\mathrm{t}_p(j^\mathrm{t}_p+1)}\ \ket{\Gamma,\cdots},
\end{eqnarray}
where $j^\mathrm{u}_p$, $j^\mathrm{d}_p$, and $j^\mathrm{t}_p$ are the colors of the links that emerge upwards (u), downwards (d), and tangentially (t) to the surface $\mathcal{S}$, respectively.
It is the key result of LQG that the spectrum of area is discrete. The smallest nonzero area eigenvalue is given by (with $j=1/2$)
\begin{equation}\label{Delta}
\Delta = 2\sqrt{3}\, \pi G\hbar\gamma \equiv 2\sqrt{3}\, \pi \gamma \Pl^2,
\end{equation}
which is of the order of the Planck area $\Pl^2\equiv G\hbar$ (assuming $\gamma$ is of the order of unity).
Intrinsic discreteness of space at the Planck scale has long been expected in QG. In the context of LQG, this discreteness is not postulated or imposed by hand but rather arises as a direct consequence of the quantization in the same sense that the energy spectrum of an harmonic oscillator or of an atom is quantized.
Also note that different choices of the numerical value of $\gamma$ give rise to nonequivalent quantum theories as the difference is reflected in the spectrum of the area operator.

\subsection{Volume operator}
The two operators $\hat{S}$ in \eref{hat S} and $\hat{A}(\mathcal{S})$ in \eref{area operator} are in principle sufficient to define the quantum theory. To better understand quantum states, we also define the volume operator, which plays a key role in the physical interpretation of quantum states.

Consider a 3-dimensional region $\mathcal{R}$. The classical volume of $\mathcal{R}$ is given by
\begin{equation}\label{V(R)}
V(\mathcal{R})
=\int_\mathcal{R} d^3 x
\sqrt{\frac{1}{3!} \epsilon_{abc}\epsilon^{ijk}
\abs{\tE^a_i\tE^b_j\tE^c_k}}\, ,
\end{equation}
To construct the quantum counterpart, we first have to regularize the quantity $\frac{1}{3!}\epsilon^{ijk}\tE^a_i(x)\tE^b_j(x)\tE^c_k(x)$ by the ``3-hand holonomy'' defined as
\begin{eqnarray}
T^{abc}(x;s,t,r) &:=& \frac{1}{3!}\, \epsilon_{ijk}
R^{(1)}(h_{\gamma_{xr}})^{il}\tE^a_l(r)\nonumber\\
&&\quad\mbox{}\times
R^{(1)}(h_{\gamma_{xs}})^{jm}\tE^b_m(s)
R^{(1)}(h_{\gamma_{xt}})^{kn}\tE^c_n(t),
\end{eqnarray}
where $s$, $t$, and $r$ are three points (close to but different from the point $x$), and $\gamma_{xx'}$ are paths from $x$ to $x'$. When $s$, $t$, and $r$ are very close to $x$, $T^{abc}(x;s,t,r)$ approximates $\frac{1}{3!}\epsilon^{ijk}\tE^a_i(x)\tE^b_j(x)\tE^c_k(x)$. As the ``3-hand generalization'' of $\widehat{E^2}(\mathcal{S})$, for a given \emph{closed} surface, we define
\begin{equation}\label{E3}
\widehat{E^3}(\mathcal{S}):=\int_\mathcal{S}d^2\sigma \int_\mathcal{S}d^2\sigma' \int_\mathcal{S}d^2\sigma''
{n_a(\sigma)n_b(\sigma')n_c(\sigma'')\, T^{abc}(x;\sigma,\sigma',\sigma'')},
\end{equation}
where $x$ is a point in the interior of $\mathcal{S}$ (the exact position of $x$ is irrelevant as we will always consider the limit of small $\mathcal{S}$).

Partition the region $\mathcal{R}$ into small cubes $\mathcal{R}_I$ (for each $N$, $\bigcup_{n=I}^N\mathcal{R}_I=\mathcal{R}$) such that the coordinate volume of each cube is smaller than $\epsilon^3$ as $N\rightarrow\infty$. In the same spirit of \eref{area operator}, we can now define the \emph{volume operator} associated with $\mathcal{R}$ as
\begin{equation}
\hat{V}(\mathcal{R}) := \frac{1}{\sqrt{3!}}\lim_{\epsilon\rightarrow0 \atop (N\rightarrow\infty)} \sum_I \sqrt{\abs{\widehat{E^3}(\partial\mathcal{R}_I)}}\,,
\end{equation}
where $\partial\mathcal{R}_I$ is the boundary surface of the cube $\mathcal{R}_I$.

When the operator $\hat{V}(\mathcal{R})$ acts on a spin network state $\ket{\Gamma,\dots}$, the three surface integrals over $\partial\mathcal{R}_I$ in \eref{E3} give three intersection points, as in the case of the area operator. For $\epsilon$ small enough, the only cubes whose surfaces have at least three intersections with the spin network are those containing a node of the spin network. Therefore, the sum over cubes $I$ reduces to the sum over the nodes $n\in \Gamma\cap\mathcal{R}$, and we have
\begin{equation}
\hat{V}(\mathcal{R})\ket{\Gamma,\dots}= \frac{1}{\sqrt{3!}}\lim_{\epsilon\rightarrow0 \atop (N\rightarrow\infty)} \sum_{n\in \Gamma\cap\mathcal{R}} \sqrt{\abs{\hat{W}_{I_n}}}\ \ket{\Gamma,\dots},
\end{equation}
where
\begin{equation}
\hat{W}_{I_n}\ket{\Gamma,\dots}\equiv \widehat{E^3}(\partial\mathcal{R}_{I_n})\ket{\Gamma,\dots}
=\sum_{r,s,t\in\Gamma\cap\partial\mathcal{R}_{I_n} \atop r\neq s\neq t\neq r}
\mathcal{T}(r,s,t)\ket{\Gamma,\dots}
\end{equation}
is the action of $\widehat{E^3}(\partial\mathcal{R}_{I_n})$ on $\ket{\Gamma,\dots}$, which is the sum over the triplets $(r,s,t)$ of distinct intersections between the spin network and the boundary of the cube $\mathcal{R}_{I_n}$ containing the node $n$. For each triplet $(r,s,t)$, the result of the action is denoted as $\mathcal{T}(s,t,r)\ket{\Gamma,\dots}$.
The key point is that, in the limit $\epsilon\rightarrow0$, the operator $\hat{W}_{I_n}$ does not change the graph of the spin network, nor the coloring of the links. Its only possible action is on the intertwiners. Consequently, we have
\begin{equation}\label{V on spin network}
\hat{V}(\mathcal{R})\ket{\Gamma,j_l,i_1\dotsi_N}
= (16\pi G\hbar\gamma)^{3/2}
\sum_{n\in\Gamma\cap\mathcal{R}}
{\mathcal{V}_{i_n}}^{i'_n} \ket{\Gamma,j_l,i_1\dots i'_n\dotsi_N},
\end{equation}
where the coefficients ${\mathcal{V}_{i_n}}^{i'_n}$ can be calculated by the recoupling theory.

A complete and rigorous construction of the volume operator can be found in Refs.~\cite{Ashtekar:1997fb,Lewandowski:1996gk}. The detailed calculation of ${\mathcal{V}_{i_n}}^{i'_n}$ is presented in Refs.~\cite{DePietri:1996pja,Thiemann:1996au}, where a list of eigenvalues is also given. It turns out that the node must be at least 4-valent in order to have a nonvanishing volume\cite{Loll:1995tp}.

The volume operator is well defined and hermitian with a discrete spectrum of nonnegative eigenvalues. Furthermore, we can choose a basis of intertwiners $i_n$ that diagonalize the matrices ${\mathcal{V}_{i_n}}^{i'_n}$ in \eref{V on spin network} so that the resulting spin networks are eigenstates of the area and the volume operators simultaneously. We denote $V_{i_n}$ the corresponding eigenvalues.

In addition to area and volume operators, the length operator can also be defined and also yields a discrete spectrum\cite{Thiemann:1996at}, but it is far more complicated and less understood than the area and volume operators.

\subsection{Quantum geometry}\label{sec:quantum geometry}
The volume operator essentially has contributions only from the nodes of a spin network, while the area operator has contributions from the links. Therefore, each node represents a quantum of volume and each link represents a quantum of area. That it, a spin network $\ket{\Gamma,j_1\dots j_L,i_1\dots,i_N}$ can be interpreted as an ensemble of $N$ quanta of volume, or $N$ ``chunks'' of space, which are separated from one another by the adjacent surfaces of $L$ quanta of area. Each chunk of space is located ``around'' the note $n$ with a quantized volume $V_{i_n}$. Two chunks are regarded as adjacent to each other if the two corresponding notes are connected by a link $j_l$, which corresponds to the adjacent surface with a quantized area $8\pi G\hbar\gamma\sqrt{j_l(j_l+1)}$. The graph $\Gamma$ dictates the adjacency relation among the chunks of space.

The physical picture is compelling that a spin network state determines a quantized 3d metric. However, it should be noted that both the area operator $\hat{A}(\mathcal{S})$ and the volume operator $\hat{V}(\mathcal{R})$ are \emph{not} diffeomorphism invariant (i.e., they do not commute with $\hat{U}_\varphi$), as the specification of $\mathcal{S}$ or $\mathcal{R}$ relies on spatial coordinates.\footnote{The only exception is the \emph{total} volume operator $\hat{V}\equiv\hat{V}(\Sigma)$, which is diffeomorphism invariant.} Nevertheless, we can specify the surface and region \emph{intrinsically} on the knot of the spin network itself: A ``region'' is simply as a collection of nodes $\{n_1,n_2,\dots\}_K:= \{n_1,\dots|n_i\in\text{notes of }K\}$ of the knot $K$ associated with a graph $\Gamma$; a ``surface'', as the boundary of a region, is a collection of knot links each of which connects to only one of $\{n_1,n_2,\dots\}_K$. In this way, the volume operator $\hat{V}(\{n_1,\dots\}_K)$ defined as
\begin{equation}\label{intrinsic V op}
\hat{V}(\{n_1,\dots\}_K)\ket{\Gamma,\dots}=
\left\{
  \begin{array}{ll}
    0 & \quad\text{if}\ [\Gamma]\neq K,\\
    \sum_{\{n_1,\dots\}} V_{i_{n_j}} & \quad\text{if}\ [\Gamma]= K,
  \end{array}
\right.
\end{equation}
where $[\Gamma]$ denotes the knot class represented by $\Gamma$,
and the area operator $\hat{A}(\partial\{n_1,\dots\}_K)$ defined as
\begin{equation}\label{intrinsic A op}
\hat{A}(\partial\{n_1,\dots\}_K)\ket{\Gamma,\dots}=
\left\{
  \begin{array}{ll}
    0 & \quad\text{if}\ [\Gamma]\neq K,\\
    \sum_{j_l\in\partial\{n_1,\dots\}_K} 8\pi G\hbar\gamma\sqrt{j_l(j_l+1)} & \quad\text{if}\ [\Gamma]= K,
  \end{array}
\right.
\end{equation}
are both diffeomorphism invariant. Thanks to diffeomorphism invariance, we can define the corresponding operators $\hat{V}^*(\{n_1,\dots\}_K)$ and $\hat{A}^*(\partial\{n_1,\dots\}_K)$ acting on $\Cyl_\Diff^*$ via \eref{O star}. The resulting operators are well defined on s-knot states.\footnote{Here, we have disregard the technicalities due to the discrete group $\mathrm{GS}_\Gamma$ of graph symmetries.} Furthermore, in principle, the eigenvalues of the area and the volume operators represent possible outcomes of the corresponding physical measurements\cite{Rovelli:1992vv}.

The situation is precisely the same as that in the classical theory. In classical GR, we distinguish between a 3-metric $q_{ab}$ and a 3-geometry $[q]$; the latter is an equivalence class of the former modulo diffeomorphisms. The notion of geometry is diffeomorphism invariant while the notion of metric is not. Given a metric $q_{ab}$ with coordinates $x^a$, we can compute the area of $\mathcal{S}$ or the volume of $\mathcal{R}$, if we define $\mathcal{S}$ by a map $(\sigma^1,\sigma^2)\rightarrow x^a(\sigma)$ and $\mathcal{R}$ by a map $(\sigma^1,\sigma^2,\sigma^3)\rightarrow x^a(\sigma)$. However, it makes no sense to ask what the area of $\mathcal{S}$ or the volume of $\mathcal{R}$ is in a given geometry $[q]$, because the coordinates have no significance in the geometry. Instead, given a geometry $[q]$, we should specify surfaces or regions \emph{intrinsically} on the geometry itself. For example, given the 3-geometry of the solar system, the region and the surface \emph{of the earth} are well defined, without any reference to coordinate localization, and it is meaningful to ask what the volume and area of the earth are. In this sense, a spin network can be regarded as a quantum (discretized) 3d metric, and an s-knot as a quantum 3d geometry.

The physical interpretation of the s-knots is extremely appealing: They represent different quantized 3d geometries, each of which is an abstract aggregate of chunks of discrete space separated by discrete adjacent surfaces. S-knots are not quantum excitations \emph{in space}; rather, they are excitations \emph{on top of one another}, as any reference to localization of the chunks and surfaces is dismissed. Also note that the ``empty'' s-knot $\obra{\emptyset}$ associated with the trivial spin network $\ket{\emptyset}$ describes a space with no volume and no area at all. We therefore have brought off the paradigm of background independence as advocated in \sref{sec:background-independent approach}. The manifestation of background independence will become even more prominent if we also include nongravitational matter fields as will be seen in \sref{sec:inclusion of matter fields}.

\section{Scalar constraint and quantum dynamics}\label{sec:scalar constraint and quantum dynamics}
We have constructed the $SU(2)$ and diffeomorphism invariant kinematical Hilbert space $\hil_\inv$ by imposing the Gauss and diffeomorphism constraints. While the quantum kinematics is well understood, the crux of the problem in LQG lies in the scalar constraint, implementation of which is supposed to reveal the quantum dynamics. One might attempt to repeat the group averaging procedure for the scalar constraint as we did for the diffeomorphism constraint, but this strategy turns out very difficult because the finite transformations generated by the scalar constraint are poorly understood even at the classical level. Instead, we adopt the strategy: First, we regularize the classical expression of the scalar constraint; and second, we promote the regulated classical constraint to a quantum operator and then remove the regulator. (We will follow closely the lines of Sec.~6.3 in \cite{Ashtekar:2004eh}.) Since the scalar constraint is very intricate, its implementation is far less clean and complete than that of the other two constraints. Consequently, the quantum dynamics remains a challenging open problem in LQG. Readers are referred to Refs.~\cite{Thiemann:1996ay,Thiemann:1996aw,Thiemann:1996av,Thiemann:1997rv} for more details.

\subsection{Regulated classical scalar constraint}\label{sec:classical scalar constraint}
The classical scalar constraint is given by \eref{C[N]}. Had we considered the Euclidean GR and chosen $\gamma=1$, we would have $\gamma^2=\sigma=1$ and the second term in \eref{C[N]} would vanish. Therefore, the first term has the interpretation of the scalar constraint for the Euclidean GR with $\gamma=1$. Accordingly, we rewrite the full Lorentzian constraint, i.e.\ \eref{C[N]} with $\sigma=-1$, as
\begin{equation}
\calC[N] = \sqrt{\gamma}\,\calC^\Eucl[N] -2(1+\gamma^2)\mathcal{T}[N],
\end{equation}
where
\begin{equation}
\calC^\Eucl[N]:=\frac{\sqrt{\gamma}}{16\pi G\gamma} \int_\Sigma d^3x
\frac{\tE^a_i\tE^b_j}{\sqrt{\det\tE}}
\,{\epsilon^{ij}}_k F_{ab}^k,
\end{equation}
and
\begin{equation}
\mathcal{T}[N]:=\frac{1}{16\pi G} \int_\Sigma d^3x
\frac{\tE^a_i\tE^b_j}{\sqrt{\det\tE}}\,
K^i_{[a}K^j_{b]}.
\end{equation}

One of the difficulties to deal with $\calC[N]$ is that it involves non-polynomial functions of $\tE^a_i$ through $\sqrt{\det\tE}$ and $K_a^i$. Fortunately, the non-polynomiality can be circumvented by Thiemann's trick\cite{Thiemann:1996ay,Thiemann:1996aw}.
First, by the canonical relation \eref{PB of A and E}, we can rewrite the combination for the cotriad $e_a^i$
\begin{equation}
e_a^i=\frac{1}{2}\,\epsilon_{abc}\, \epsilon^{ijk} \frac{\tE^b_j\tE^c_k}{\sqrt{\det\tE}}
\end{equation}
as a manageable Poisson bracket:
\begin{equation}
e_a^i(x) = \frac{1}{4\pi G\gamma}
\left\{A_a^i(x),V\right\},
\end{equation}
where $V=V(\Sigma)$ is the total volume of $\Sigma$ given by \eref{V(R)} with $\mathcal{R}=\Sigma$. This allows us to express $\calC^\Eucl[N]$ as
\begin{equation}\label{CEucl[N]}
\calC^\Eucl[N] = -\frac{1}{32\pi^2 G^2\gamma^{3/2}}
\int_\Sigma d^3x\, N(x)\, \epsilon^{abc}\,
\Tr \big(F_{ab}(x)\left\{A_c(x),V\right\}\big)
\end{equation}
in a form more suitable for loop quantization.

Second, we have
\begin{equation}
K_a^i=\frac{1}{8\pi\gamma}\left\{A_c(x),\bar{K}\right\}
\end{equation}
where $\bar{K}$ is defined as
\begin{equation}
\bar{K}:=\int_\Sigma d^3x\, K_a^i \tE^a_i,
\end{equation}
which can also be expressed as a Poisson bracket:
\begin{equation}
\bar{K}=\frac{1}{\gamma^{3/2}}
\left\{\calC^\Eucl[1],V\right\}.
\end{equation}
Therefore, $\mathcal{T}[N]$ can be recast as
\begin{eqnarray}\label{T[N]}
&&\mathcal{T}[N]\\
&=&
-\frac{2}{(8\pi G)^4\gamma^3}
\int_\Sigma d^3x\, N(x)\, \epsilon^{abc}\,
\Tr \left(\left\{A_a(x),\bar{K}\right\} \left\{A_b(x),\bar{K}\right\} \left\{A_c(x),V\right\}
\right).\nonumber
\end{eqnarray}

We have expressed $\calC^\Eucl[N]$ and $\mathcal{T}[N]$ in terms of $A$, $F$, and $V$. The next step to replace $A$ and $F$ with holonomies, which are the ``right'' variables to be used for loop quantization. For a small path $\gamma_{x,u}$ of coordinate length $\epsilon$ starting at $x$ and tangent to $u$, the holonomy $h_{\gamma_{x,u}}$ takes the expansion
\begin{subequations}\label{approx A}
\begin{eqnarray}
h_{\gamma_{x,u}} &=& 1 +\epsilon\, u^a A_a(x) + O(\epsilon^2),\\
h_{\gamma_{x,u}}^{-1}\equiv h_{\gamma_{x,u}^{-1}} &=& 1 -\epsilon\, u^a A_a(x) + O(\epsilon^2).
\end{eqnarray}
\end{subequations}
Similarly, for a rectangular loop $\alpha_{x,uv}$ with one vertex at $x$ and two sides tangent to $u$ and $v$, each of coordinate length $\epsilon$, the holonomy $h_{\alpha_{x,uv}}$ takes the expansion
\begin{subequations}\label{approx F}
\begin{eqnarray}
h_{\alpha_{x,uv}} &=& 1 + \epsilon^2 u^av^b F_{ab}(x) + O(\epsilon^3),\\
h_{\alpha_{x,uv}}^{-1}\equiv h_{\alpha_{x,uv}^{-1}} &=& 1 - \epsilon^2 u^av^b F_{ab}(x) + O(\epsilon^3).
\end{eqnarray}
\end{subequations}
Now, partition $\Sigma$ into small cubic cells, edges of which are of coordinate length $\epsilon$. Denote by $s_1$, $s_2$, and $s_3$ the edges of an elementary cell $\oblong$ based at a vertex $v_\oblong$, and by $\beta_1$, $\beta_2$, and $\beta_3$ the three oriented loops that are are the boundaries of the three rectangular faces based at $v_\oblong$ and orthogonal to $s_1$, $s_2$, and $s_3$, respectively (see Fig.~2 in \cite{Ashtekar:2004eh}). By Eqs.~(\ref{approx A}) and (\ref{approx F}), we can then regulate \eref{CEucl[N]} as $\sum_\oblong \calC^\Eucl_\oblong(N)$, where the contribution form each cell is
\begin{equation}
\calC^\Eucl_\oblong(N) =
\frac{N(v_\oblong)}{32\pi^2G^2\gamma^{3/2}}\sum_{I}\Tr
\left(
\left(h_{\beta_I}-h_{\beta_I^{-1}}\right)
h_{s_I^{-1}}\left\{h_{s_I},V\right\}
\right),
\end{equation}
and similarly regulate \eref{T[N]} as $\sum_\oblong \mathcal{T}_\oblong(N)$, where
\begin{eqnarray}
&&\mathcal{T}_\oblong(N)\\
&=&
\frac{2N(v_\oblong)}{(8\pi G)^4\gamma^3}\sum_{IJK}\epsilon^{IJK}\Tr
\left(
h_{s_I^{-1}}\left\{h_{s_I},\bar{K}\right\}
h_{s_J^{-1}}\left\{h_{s_J},\bar{K}\right\}
h_{s_K^{-1}}\left\{h_{s_K},V\right\}
\right).\nonumber
\end{eqnarray}
Note that we regulate $\left\{A,V\right\}$ by $h_\gamma^{-1}\left\{h_\gamma,V\right\}\equiv V-h_\gamma^{-1}Vh_\gamma$ instead of simply $\left\{h_\gamma,V\right\}$ and $\left\{A,\bar{K}\right\}$ by $h_\gamma^{-1}\left\{h_\gamma,\bar{K}\right\}\equiv \bar{K}-h_\gamma^{-1}\bar{K}h_\gamma$ instead of simply $\left\{h_\gamma,\bar{K}\right\}$, because we want to keep the scalar constraint to be $SU(2)$ invariant after the regularization.

The regularization can be more generic than the above prescription. Instead of the simple cubic partition, we can partition $\Sigma$ into cells $\oblong$ of arbitrary shape (particularly, the partition can be chosen to be a triangulation of $\Sigma$), and in every cell $\oblong$ we define edges $s_J$, $J=1,\dots,n_s$ and loops $\beta_i$, $i=1,\dots,n_\beta$ all based at a vertex $v_\oblong$ inside $\oblong$, where $n_s,n_\beta$ may be different for different cells (see Fig.~3 in \cite{Ashtekar:2004eh}); furthermore, we can choose an arbitrary representation $R^{(j)}$ of the $SU(2)$ group other than $R^{(1/2)}$. The entire prescription is denoted by $R_\epsilon$ and called a \emph{permissible classical regulator} if both the following conditions hold:
\begin{subequations}\label{C Repsilon}
\begin{eqnarray}
\lim_{\epsilon\rightarrow0} \calC^\Eucl_{R_\epsilon} &=& \calC^\Eucl[N],\\
\lim_{\epsilon\rightarrow0} \mathcal{T}_{R_\epsilon} &=& \mathcal{T}[N],
\end{eqnarray}
\end{subequations}
where
\begin{subequations}\label{CEucl Box}
\begin{eqnarray}
\calC^\Eucl_{R_\epsilon}[N] &=& \sum_\oblong \calC^\Eucl_{R_\epsilon\oblong}(N),\\
\calC^\Eucl_{R_\epsilon\oblong}(N) &=&
\frac{N(v_\oblong)}{32\pi^2G^2\gamma^{3/2}}\sum_{i,J}
C^{iJ}\Tr
\left(
\left(R^{(j)}(h_{\beta_i})-R^{(j)}(h_{\beta_i^{-1}})\right)\right.\nonumber\\
&&\qquad\qquad\qquad\quad\left.\mbox{}\times
R^{(j)}(h_{s_J^{-1}})\left\{R^{(j)}(h_{s_J}),V\right\}
\right),
\end{eqnarray}
\end{subequations}
\begin{subequations}\label{T Box}
\begin{eqnarray}
\mathcal{T}_{R_\epsilon}[N] &=& \sum_\oblong \mathcal{T}_{R_\epsilon\oblong}(N),\\
\mathcal{T}_{R_\epsilon\oblong}(N) &=&
\frac{2N(v_\oblong)}{(8\pi G)^4\gamma^3}\sum_{IJK}T^{IJK}\Tr
\left(
R^{(j)}(h_{s_I^{-1}})\left\{R^{(j)}(h_{s_I}),\bar{K}\right\}\right.\\
&&
\qquad\left.\mbox{}\times
R^{(j)}(h_{s_J^{-1}})\left\{R^{(j)}(h_{s_J}),\bar{K}\right\}
R^{(j)}(h_{s_K^{-1}})\left\{R^{(j)}(h_{s_K}),V\right\}
\right),\nonumber
\end{eqnarray}
\end{subequations}
and $C^{iJ}$ and $T^{IJK}$ are fixed constants, independent of the scale parameter $\epsilon$. There exists a great variety of permissible classical regulators, and this non-uniqueness is the source of quantization ambiguities.

\subsection{Quantum scalar constraint}\label{sec:quantum scalar constraint}
It is fairly straightforward to promote $\calC^\Eucl_{R_\epsilon}$ into a quantum operator acting on the space $\hil$ of cylindrical functions. Recall that the total volume operator $\hat{V}\equiv\hat{V}(\Sigma)$ and the holonomy operator $\hat{h}_\gamma$ are both well defined in $\hil$. Therefore, simply by replacing $V$ with $\hat{V}$, $h_\gamma$ with $\hat{h}_\gamma$, and Poisson brackets with commutators, we obtain the operator $\hat{\calC}^\Eucl_{R_\epsilon}$ on $\hil$:
\begin{subequations}\label{hat CEucl}
\begin{eqnarray}
\label{hat CEucl a}
\hat{\calC}^\Eucl_{R_\epsilon}[N] &=& \sum_\oblong \hat{\calC}^\Eucl_{R_\epsilon\oblong}(N),\\
\label{hat CEucl b}
\hat{\calC}^\Eucl_{R_\epsilon\oblong}(N) &=&
-\frac{iN(v_\oblong)}{32\pi^2G^2\gamma^{3/2}\hbar}\sum_{i,J}
C^{iJ}\Tr
\left(
\left(R^{(j)}(\hat{h}_{\beta_i})-R^{(j)}(\hat{h}_{\beta_i^{-1}})\right)\right.\nonumber\\
&&\qquad\qquad\qquad\qquad\left.\mbox{}\times
R^{(j)}(\hat{h}_{s_J^{-1}})\left[R^{(j)}(\hat{h}_{s_J}),\hat{V}\right]
\right).
\end{eqnarray}
\end{subequations}
Furthermore, because $\calC^\Eucl_{R_\epsilon}$ is regularized in a way preserving $SU(2)$ invariance, $\hat{\calC}^\Eucl_{R_\epsilon}$ is $SU(2)$ invariant and leaves the space $\hil_\mathrm{G}$ of spin networks invariant.
When acting on a given spin network state $\ket{\Gamma,\dots}$, the part $\big[R^{(j)}(\hat{h}_{s_J}),\hat{V}\big]$ will yield zero unless the segment $s_J$ intersects a node of $\Gamma$, because the volume operator has contributions only from nodes of the spin network. Also note that while nodes must be at least 4-valent to have nonvanishing volume, the operator $\hat{\calC}^\Eucl_{R_\epsilon}$ can have contributions from trivalent nodes, as $\hat{h}_{s_J}$ adds one more edge to the intersection node.

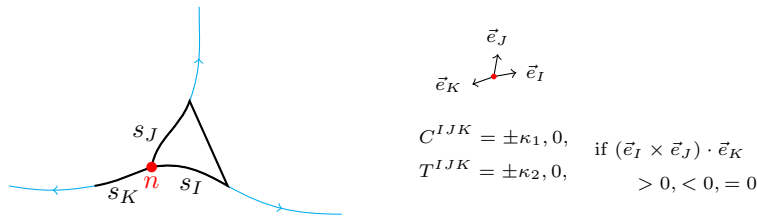
\begin{figure}

\center

\begin{tikzpicture}

%-------------------------------------------
%\draw[help lines] (-3,-2) grid (6,2);

\begin{scope}[rotate=0,scale=1.25]
%-------------------------------------------
%--- 1st edge s_J:
\draw [cyan,arrow data={0.7}{>}] (0,0) to [out=80,in=250] (0.4,0.7) to [out=70,in=270] (0.5,1.7);
\draw [mylabel=at 0.5 left with {$s_J$}] (0,0) to [out=80,in=250] (0.4,0.7);

%--- 2nd edge s_I:
\draw [cyan,arrow data={0.7}{>}] (0,0) to [out=10,in=150] (0.8,-0.2) to [out=-30,in=180] (2,-0.5);
\draw [mylabel=at 0.5 below with {$s_I$}] (0,0) to [out=10,in=150] (0.8,-0.2);

%--- 3rd edge s_K:
\draw [cyan,arrow data={0.7}{>}] (0,0) to [out=200,in=10] (-0.6,-0.2) to [out=190,in=-10] (-1.5,-0.2);
\draw [thick,mylabel=at 0.5 below with {$s_K$}] (0,0) to [out=200,in=10] (-0.6,-0.2);

%--- triangular loop:
\draw [thick] (0,0) to [out=80,in=250] (0.4,0.7) -- (0.8,-0.2) to [out=150,in=10] (0,0);

%--- node:
\draw [fill,red] (0,0) node [below] {$n$} circle [radius=0.05];

\end{scope}

\begin{scope}[shift={(4.5,1.2)},rotate=0,scale=0.3]
\draw [->] (0,0) -- (0.173648,0.984808) node [above] {\scriptsize $\vec{e}_J$};
\draw [->] (0,0) -- (0.984808,0.173648) node [right] {\scriptsize $\vec{e}_I$};
\draw [->] (0,0) -- (-0.939693,-0.34202) node [left] {\scriptsize $\vec{e}_K$};
\draw [fill,red] (0,0)  circle [radius=0.1];
\end{scope}

\begin{scope}[shift={(4.5,-0.3)}]
\node[align=left,above] at (0,0)
{\scriptsize $C^{IJK}=\pm\kappa_1,0$,\\
 \scriptsize $T^{IJK}=\pm\kappa_2,0$,};

\node[align=left,right] at (1.2,0.3)
{\scriptsize if $(\vec{e}_I\times\vec{e}_J)\cdot\vec{e}_K$\\
 \scriptsize \qquad $>0,<0,=0$};
\end{scope}

\end{tikzpicture}

\caption{The edges $s_I,s_J,s_K$ and the triangular loop $\beta_{IJ}$ at a trivalent node $n$. The pattern is similar at other multivalent nodes. $C^{IJK}$ and $T^{IJK} $ are given by constants in accordance with the orientation of $\vec{e}_I,\vec{e}_J,\vec{e}_K$ tangent to $s_I,s_J,s_K$.}
\label{fig:edges at node}

\end{figure}

As $\hat{\calC}^\Eucl_{R_\epsilon}$ is well defined in $\hil$, here comes the critical issue: We have to ensure that the resulting quantum operator is covariant under diffeomorphisms. For this purpose, we have to restrict our regularization scheme to a \emph{diffeomorphism covariant quantum regulator}, which is a family of permissible classical regulators that are not fixed but transform covariantly as a graph $\Gamma$ is moved under diffeomorphisms. The precise definition of a diffeomorphism covariant quantum regulator is not important here, but the essential point is that such regulators exist and lead to the regulated operator $\hat{\calC}^\Eucl_{R_\epsilon}$ that is densely defined in the full Hilbert space $\hil$ with domain $\Cyl$ and diffeomorphism covariant\cite{Thiemann:1996aw,Thiemann:1997rv}.
The simplest and most convenient case of such regulators can be summarized as follows. For a given graph $\Gamma$, make the partition refined enough such that each cell $\oblong$ contains at most one node of $\Gamma$. For a cell $\oblong$ containing a node $n$, the edges $s_I$ are assigned to the proper segments of the links of $\Gamma$ incident at $n$; orientations of $s_I$ are all assigned to be outgoing at $n$. The loops $\beta_i\equiv\beta_{[IJ]}\equiv\beta_{IJ}$ are chosen to be the triangular loops spanned by $s_I$ and $s_J$; the loop $\beta_{IJ}$ contains no other points of $\Gamma$ except for the edges $s_I,s_J$, and its orientation is defined as the same as the plane by the ordered pair of $s_I$ and $s_J$. See \fref{fig:edges at node} for the illustration. Finally, the constants $C^{iK}\equiv C^{IJK}$ are given by $\pm\kappa_1,0$, and the constants $T^{IJK}$ given by $\pm\kappa_2,0$, depending on the orientation of a triple of vectors tangent to $s_I,s_J,s_K$ relative to the background orientation of $\Sigma$, where $\kappa_1,\kappa_2$ are fixed constants. With this regulator, the sum $\sum_\oblong$ in \eref{hat CEucl a} effectively becomes the sum $\sum_{n\in\Gamma}$ over the nodes of $\Gamma$, and the lapse function $N(v_\oblong)$ and the volume operator $\hat{V}$ in \eref{hat CEucl b} become $N_n$ and $\hat{V}(\{n\})$, respectively, for each node.

As $\hat{\calC}^\Eucl_{R_\epsilon}$ is well defined in $\hil$, we can define the corresponding operator acting on $\Cyl^*$ via its action on the states of $\Cyl$ by
\begin{equation}
\oinner{\hat{\calC}^\Eucl_{R_\epsilon}\Psi}{\Phi}:=
\oinner{\Psi}{\hat{\calC}^\Eucl_{R_\epsilon}\Phi}
\end{equation}
for every $\obra{\Psi}\in\Cyl^*$ and $\ket{\Phi}\in\Cyl$.\footnote{To be precise, we should call the operator on the left-hand side $\hat{\calC}^{\Eucl*\dag}_{R_\epsilon}$ in the same spirit of \eref{O star}, but we neglect $*$ and $\dag$ to keep the notation simple.} The final step is to remove the regulator and obtain the operator $\hat{\calC}^\Eucl$ acting on $\Cyl^*$. It does not work if one attempts to take the limit directly
\begin{equation}
\oinner{\hat{\calC}^\Eucl\Psi}{\Phi}\overset{?}{=}
\oinner{\Psi}{\lim_{\epsilon\rightarrow0}\hat{\calC}^\Eucl_{R_\epsilon}\Phi},
\end{equation}
since $\hat{\calC}^\Eucl_{R_\epsilon}$ becomes ill-defined in $\Cyl$ in the limit $\epsilon\rightarrow0$. Instead, we should \emph{define} the operator $\hat{\calC}^\Eucl$ acting on $\Cyl^*$ by
\begin{equation}\label{CEucl diff}
\oinner{\hat{\calC}^\Eucl\Psi}{\Phi}:=
\lim_{\epsilon\rightarrow0}\oinner{\Psi}{\,\hat{\calC}^\Eucl_{R_\epsilon}\Phi},
\end{equation}
where the limit is now a limit of a sequence of numbers, instead a sequence of operators. The limit exists if $\obra{\Psi}$ is a diffeomorphism invariant state, namely $\obra{\Psi}\in\Cyl_\Diff^*$, as we can see in the following crucial observation. Given $\ket{\Phi}$ a spin network $\ket{\Gamma,\dots}$, the operator $\hat{\calC}^\Eucl_{R_\epsilon}$ on the right-hand side of \eref{CEucl diff} modifies $\ket{\Gamma,\dots}$ in two ways by changing its graph $\Gamma$ and its coloring. The graph $\Gamma$ is changed by the two classes of operators: $\hat{h}_{s_K}$ and $\hat{h}_{\beta_{IJ}}$. The former superimposes an edge of coordinate length $\epsilon$ to a link of $\Gamma$, and the latter adds the triangular loop as depicted in \fref{fig:edges at node}. When $\epsilon$ is sufficiently small, changing $\epsilon$ in the operator changes the resulting state, but the resulting state remains in the same diffeomorphism equivalence class. Consequently, the value of $\oinner{\Psi}{\,\hat{\calC}^\Eucl_{R_\epsilon}\Phi}$ for $\obra{\Psi}\in\Cyl_\Diff^*$ becomes independent of $\epsilon$ once $\epsilon$ is sufficiently small, and thus the $\epsilon\rightarrow0$ limit in \eref{CEucl diff} is finite. Therefore, we can well define the quantum operator $\hat{\calC}^\Eucl$ acting on the domain $\Cyl_\Diff^*$, as the regulator is removed trivially.

The Hamiltonian constraint operator, albeit defined upon spin networks with reference to coordinates in the first place, turns out to be well defined and independent of the regularization scale $\epsilon$ on the domain of diffeomorphism invariant states, i.e.\ s-knots.  To sum up, the small $\epsilon$ of coordinate length loses its physical significance at the diffeomorphism invariant level and the $\epsilon\rightarrow0$ limit becomes finite because making the regulator smaller cannot modify anything below the Planck scale as there is really nothing below the the short-scale discreteness\cite{Rovelli:1993bm}. This striking feature as a consequence of the intimate interplay between diffeomorphism invariance and short-scale discreteness profoundly cures the ultraviolet pathology that has long plagued quantum field theory of gravity.

The term $\mathcal{T}[N]$ in the scalar constraint can be dealt with in a completely parallel manner. Specifically, we promote \eref{T Box} to the operator
\begin{subequations}\label{hat T}
\begin{eqnarray}
\hat{\mathcal{T}}_{R_\epsilon}[N] &=& \sum_\oblong \hat{\mathcal{T}}_{R_\epsilon\oblong}(N),\\
\hat{\mathcal{T}}_{R_\epsilon\oblong}(N) &=&
\frac{2iN(v_\oblong)}{(8\pi G)^4\gamma^3\hbar^3}\sum_{IJK}T^{IJK}\Tr
\left(
R^{(j)}(\hat{h}_{s_I^{-1}})\left[R^{(j)}(\hat{h}_{s_I}),\hat{\bar{K}}_{R_\epsilon}\right]\right.\\
&&
\qquad\left.\mbox{}\times
R^{(j)}(\hat{h}_{s_J^{-1}})\left[R^{(j)}(\hat{h}_{s_J}),\hat{\bar{K}}_{R_\epsilon}\right]
R^{(j)}(\hat{h}_{s_K^{-1}})\left[R^{(j)}(\hat{h}_{s_K}),\hat{V}\right]
\right),\nonumber
\end{eqnarray}
\end{subequations}
where $\hat{\bar{K}}_{R_\epsilon}$ is defined as
\begin{equation}
\hat{\bar{K}}_{R_\epsilon}:=
\frac{i}{\hbar\,\gamma^{3/2}}\left[\hat{V},\hat{\calC}^\Eucl_{R_\epsilon}\right],
\end{equation}
and in the end define the operator $\hat{\mathcal{T}}$ on the domain $\Cyl_\Diff^*$ by
\begin{equation}
\oinner{\hat{\mathcal{T}}\Psi}{\Phi}:=
\lim_{\epsilon\rightarrow0}\oinner{\Psi}{\hat{\mathcal{T}}_{R_\epsilon}\Phi}.
\end{equation}
The total quantum scalar operator
\begin{equation}
\hat{\calC}[N]=\sqrt{\gamma}\,\hat{\calC}^\Eucl[N] -2(1+\gamma^2)\hat{\mathcal{T}}[N]
\end{equation}
is well defined on the domain $\Cyl_\Diff^*$, namely
\begin{equation}
\hat{\calC}[N]:\Cyl_\Diff^*\rightarrow\Cyl^*.
\end{equation}
Up to diffeomorphisms, the operator $\hat{\calC}_{R_\epsilon}[N]$ with the diffeomorphism covariant quantum regulator is independent of $\epsilon$; that is, for any sufficiently small $\epsilon$ and $\epsilon'$, given any $\ket{\Phi}\in\Cyl$, there is a diffeomorphism $\varphi$ such that
\begin{equation}\label{Re Re'}
\hat{\calC}_{R_{\epsilon'}}[\varphi^*N]\ket{\Phi} = \hat{U}_\varphi\,\hat{\calC}_{R_\epsilon}[N]\,\hat{U}_\varphi^\dag\ket{\Phi}.
\end{equation}

It should be noted, however, that the operator $\hat{\calC}[N]$ does not leave $\Cyl_\Diff^*$ invariant, because $\calC[N]$ does not commute with $\calC_\Diff[\vec{N}]$ even at the classical level. (Also note that $\calC[N]$ depends on coordinates via $N(x)$.) More precisely, let $\obra{\eta(\Psi_\Gamma)}\in\Cyl_\Diff^*$ be a group-averaged state as defined in \eref{eta Psi}. As defined in \eref{CEucl diff}, the action of $\hat{\calC}[N]$ on the diffeomorphism invariant state $\obra{\eta(\Psi_\Gamma)}$ is given by
\begin{eqnarray}\label{action of C}
\oinner{\hat{\calC}[N]\eta(\Psi_\Gamma)}{\Phi}
&:=& \lim_{\epsilon\rightarrow0}\oinner{\eta(\Psi_\Gamma)}{\hat{\calC}_{R_\epsilon}[N]\Phi}
\equiv \lim_{\epsilon\rightarrow0}\sum_{\varphi\in\Diff/\Diff_\Gamma} \inner{\hat{U}_\varphi\hat{P}_{\Diff,\Gamma}\Psi_\Gamma}{\,\hat{\calC}_{R_\epsilon}[N]\Phi} \nonumber\\
&=& \lim_{\epsilon\rightarrow0}\sum_{\varphi\in\Diff/\Diff_\Gamma} \inner{\hat{P}_{\Diff,\Gamma}\Psi_\Gamma}{\,\hat{U}_\varphi^\dag\, \hat{\calC}_{R_\epsilon}[N]\, \Phi} \nonumber\\
&=& \lim_{\epsilon\rightarrow0}\sum_{\varphi\in\Diff/\Diff_\Gamma} \inner{\hat{P}_{\Diff,\Gamma}\Psi_\Gamma}{\,\hat{\calC}_{R_{\epsilon'}}[{\varphi^{-1*}}N]\, \hat{U}_\varphi^\dag\, \Phi} \nonumber\\
&=& \lim_{\epsilon\rightarrow0}\sum_{\varphi\in\Diff/\Diff_\Gamma} \inner{\hat{U}_\varphi\, \hat{\calC}_{R_{\epsilon'}}[\varphi^{-1*}N]^\dag \hat{P}_{\Diff,\Gamma}\Psi_\Gamma}{\Phi},
\end{eqnarray}
where we have used Eqs.~(\ref{eta Psi}) and (\ref{Re Re'}), and the regulator $R_{\epsilon'}=R_{\epsilon'}(\epsilon,\varphi)$ nontrivially depends on $\epsilon$ and $\varphi$. It follows from \eref{action of C} that
\begin{equation}\label{idea of action of C}
\obra{\hat{\calC}[N]\eta(\Psi_\Gamma)} \sim
\lim_{\epsilon\rightarrow0}\sum_{\varphi\in\Diff/\Diff_\Gamma} \bra{\hat{U}_\varphi\, \hat{\calC}_{R_{\epsilon'(\epsilon,\varphi)}}[\varphi^{-1*}N]^\dag \hat{P}_{\Diff,\Gamma}\Psi_\Gamma}.
\end{equation}
The resulting state on the right-hand side is a distribution belonging to $\Cyl^*$ as the sum is over the huge group $\Diff/\Diff_\Gamma$, but it is not in $\Cyl_\Diff^*$ as the summands are not arranged in the style of group averaging. The quantum theory defined by the quantum scalar constraint $\obra{\hat{\calC}[N]\Psi}=0$ nevertheless is diffeomorphism invariant, because what matters is the kernel of $\hat{\calC}[N]$, which is a proper subspace of $\Cyl_\Diff^*$ and independent of $N(x)$.

On the other hand, $\hat{\calC}[N]$ is manifestly $SU(2)$ invariant, as the regularization in \sref{sec:classical scalar constraint} has been adopted to be $SU(2)$ invariant. Taking into account both the Gauss constraint and the diffeomorphism constraint, we have the quantum operator $\hat{\calC}[N]$ well defined on the domain $\Cyl_\inv^*$, namely
\begin{equation}
\hat{\calC}[N]:\Cyl_\inv^*\rightarrow\Cyl_\mathrm{G}^*.
\end{equation}
The Cauchy completion of the kernel of $\hat{\calC}[N]$, which is a proper subspace of $\Cyl_\inv^*$, is the sought-after physical Hilbert space $\hil_\phys$ in \eref{quantization scheme}.

\subsection{Solutions to the scalar constraint}\label{sec:solutions to the scalar constraint}
The scalar constraint operator $\hat{\calC}[N]$ has been rigorously defined on the domain $\Cyl_\inv^*$. The next step is to construct the physical Hilbert space $\hil_\phys$ by solving the solutions to the scalar constraint $\obra{\hat{\calC}[N]\Psi}=0$.

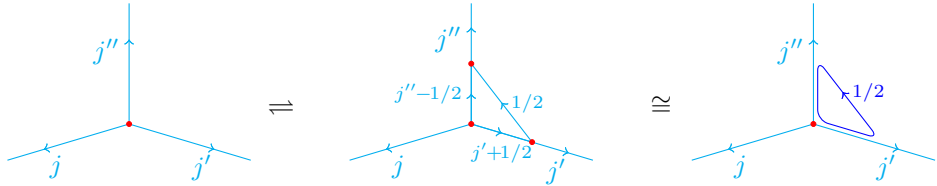
\begin{figure}

\center

\begin{tikzpicture}

%-------------------------------------------
%\draw[help lines] (-3,-2) grid (10,2);

\begin{scope}[shift={(0,0)},scale=0.8]
%-------------------------------------------
%--- 3 links:
\draw [cyan,arrow data={0.7}{>},mylabel=at 0.6 left with {$j''$}] (0,0) to (0,2);
\draw [cyan,arrow data={0.7}{>},mylabel=at 0.6 below with {$j'$}] (0,0) to (2,-0.6);
\draw [cyan,arrow data={0.7}{>},mylabel=at 0.6 below with {$j$}] (0,0) to (-2,-0.6);
%--- node:
\draw [fill,red] (0,0) circle [radius=0.04];
\end{scope}

\begin{scope}[shift={(4.5,0)},scale=0.8]
%-------------------------------------------
%--- 3 links:
\draw [cyan,arrow data={0.8}{>},mylabel=at 0.7 left with {$j''$}] (0,0) to (0,2);
\draw [cyan,arrow data={0.8}{>},mylabel=at 0.7 below with {$j'$}] (0,0) to (2,-0.6);
\draw [cyan,arrow data={0.7}{>},mylabel=at 0.6 below with {$j$}] (0,0) to (-2,-0.6);
%--- loop:
\draw [cyan,arrow data={0.5}{>},mylabel=at 0.5 below with {\scriptsize $j'\!\!+\!\!1/2$}] (0,0) to (1,-0.3);
\draw [cyan,arrow data={0.5}{>},mylabel=at 0.5 right with {\scriptsize $1/2$}] (1,-0.3) to (0,1);
\draw [cyan,arrow data={0.5}{>},mylabel=at 0.5 left with {\scriptsize $j''\!\!-\!\!1/2$}] (0,0) to (0,1);
%--- nodes:
\draw [fill,red] (0,0) circle [radius=0.04] (1,-0.3) circle [radius=0.04] (0,1) circle [radius=0.04];
\end{scope}

\begin{scope}[shift={(9,0)},scale=0.8]
%-------------------------------------------
%--- 3 links:
\draw [cyan,arrow data={0.7}{>},mylabel=at 0.6 left with {$j''$}] (0,0) to (0,2);
\draw [cyan,arrow data={0.7}{>},mylabel=at 0.6 below with {$j'$}] (0,0) to (2,-0.6);
\draw [cyan,arrow data={0.7}{>},mylabel=at 0.6 below with {$j$}] (0,0) to (-2,-0.6);

 \begin{scope}[shift={(0.07,0.06)},scale=1]
 %--- loop:
 \draw [blue,rounded corners,arrow data={0.4}{>},mylabel=at 0.4 right with {\scriptsize $1/2$}]
 (0.5,-0.15) to (1,-0.3) to (0,1) to (0,0) to (0.5,-0.15);
 \end{scope}

%--- nodes:
\draw [fill,red] (0,0) circle [radius=0.04];
\end{scope}

%---:
\node [above] at (2,0) {$\rightleftharpoons$};
\node [above] at (7,0) {$\cong$};

\end{tikzpicture}

\caption{When an extraordinary loop is added or removed at a node, the change is between the left diagram and the middle one, which can be understood as the right diagram. Here, we show the case that the extraordinary loop is labelled by $R^{(j=1/2)}$.}
\label{fig:action of hat C}

\end{figure}

The actions of the resulting operators $\hat{\calC}^\Eucl$ and $\hat{\mathcal{T}}$ on a given $SU(2)$ and diffeomorphism invariant state $\obra{\eta(\Psi_{\Gamma,j,i})}$ are rather simple. As indicated in \eref{idea of action of C} with Eqs.~(\ref{hat CEucl}) and (\ref{hat T}), it turns out that, if $\ket{\Psi_{\Gamma,j,i}}\equiv\ket{\Gamma,j_l,i_n}$ contains any \emph{extraordinary} loops (labelled by $R^{(j)}$), $\hat{\calC}^\Eucl$ will remove one such loop and $\hat{\mathcal{T}}$ will remove two in $\obra{\eta(\Psi_{\Gamma,j,i})}$, in addition to possible changes in the intertwiner at the node\cite{Thiemann:1996aw}. The extraordinary loop is of the type introduced by the regulator at any of the nodes as depicted in \fref{fig:edges at node}. The details of the actions of $\hat{\calC}^\Eucl$ and $\hat{\mathcal{T}}$ are discussed in \cite{Borissov:1997ji}. (Also see Ref.~\cite{Rovelli:1995vu} for the general structure of the scalar constraint.) The action of adding or removing an extraordinary loop at a node is schematically illustrated in \fref{fig:action of hat C}.

An s-knot state $\obra{\eta(\Psi_{\Gamma_0,j_0})}$ is said to be \emph{simple} if any spin networks belonging to the same diffeomorphism class of $\obra{\eta(\Psi_{\Gamma_0,j_0})}$ do not contain an extraordinary loop labelled by $R^{(j)}$. Obviously, simple s-knots are annihilated by both $\hat{\calC}^\Eucl$ and $\hat{\mathcal{T}}$ and thus are solutions to both the Euclidean constraint and the full (Lorentzian) scalar constraint. In a sense, these simple solutions are analogues of time-symmetric solutions to the classical Hamiltonian constraint. Particularly, s-knot states of loops or multiloops are trivial examples of simple solutions.

More solutions can be obtained by starting from the simple solutions. Let $\obra{\eta(\Psi_{\Gamma_0,j_0}^{(n)})}$ be an s-knot obtained from $\obra{\eta(\Psi_{\Gamma_0,j_0}^{(0)})}\equiv\obra{\eta(\Psi_{\Gamma_0,j_0})}$ by attachment of $n$ extraordinary loops labelled by $R^{(j)}$ (note that there are many ways of the attachment), and denote by $\mathcal{D}_{\Gamma_0,j_0}^{(n)}$ the subspace of $\Cyl_\Diff^*$ spanned by those s-knot states with $n$ extraordinary loops. The resulting spaces $\mathcal{D}_{\Gamma_0,j_0}^{(n)}$ are finite-dimensional and have trivial intersection with one another, i.e.,
\begin{equation}
\mathcal{D}_{\Gamma_0,j_0}^{(n)} \cap \mathcal{D}_{\Gamma'_0,j'_0}^{(n')} = \emptyset,
\quad \text{if}\ (\Gamma_0,j_0),n\neq(\Gamma'_0,j'_0),n'.
\end{equation}
Every $\obra{\Psi}\in\Cyl_\Diff^*$ can be uniquely decomposed as
\begin{equation}\label{Dn decomposition}
\obra{\Psi} = \sum_{(\Gamma,j),n}\obra{\eta(\Psi_{\Gamma,j}^{(n)})},
\quad \text{where}\ \obra{\eta(\Psi_{\Gamma,j}^{(n)})}\in\mathcal{D}_{\Gamma,j}^{(n)}.
\end{equation}
This unique decomposition enables us to find the solutions to the scalar constraint in a systematic way.

First, for the solutions to the Euclidean constraint, we have
\begin{equation}
\obra{\Psi}\hat{\calC}^\Eucl[N]=0
\quad\Leftrightarrow\quad
\obra{\eta(\Psi_{\Gamma,j}^{(n)})}\hat{\calC}^\Eucl[N]=0,
\quad\text{for every}\ (\Gamma,j)\ \text{and}\ n.
\end{equation}
That is, $\obra{\Psi}$ is a solution to the Euclidean constraint if and only if its components with respect to the decomposition \eref{Dn decomposition} are all solutions. This property reduces the problem of finding the solutions to the Euclidean scalar constraint to that of finding solutions in finite-dimensional subspaces, which is equivalent to the tractable task (by a computer) to find the kernel of finite matrices.

Now, consider the full (Lorentzian) scalar constraint. In the scheme with respect to the decomposition \eref{Dn decomposition}, the problem of finding solutions to the full scalar constraint is reduced to a hierarchy of steps. That is, the scalar constraint equation
\begin{equation}
\obra{\Psi}\hat{\calC}[N] = \sqrt{\gamma}\,\obra{\Psi}\hat{\calC}^\Eucl[N] -2(1+\gamma^2)\obra{\Psi}\hat{\mathcal{T}}[N] = 0
\end{equation}
is equivalent to the hierarchy of equations
\begin{eqnarray}
\obra{\eta(\Psi_{\Gamma,j}^{(1)})}\hat{\mathcal{T}}[N] &=& 0,\nonumber\\
2(1+\gamma^2)\obra{\eta(\Psi_{\Gamma,j}^{(2)})}\hat{\mathcal{T}}[N] &=&
\sqrt{\gamma}\,\obra{\eta(\Psi_{\Gamma,j}^{(1)})}\hat{\calC}^\Eucl[N],\nonumber\\
&\vdots&\nonumber\\
2(1+\gamma^2)\obra{\eta(\Psi_{\Gamma,j}^{(n+1)})}\hat{\mathcal{T}}[N] &=&
\sqrt{\gamma}\,\obra{\eta(\Psi_{\Gamma,j}^{(n)})}\hat{\calC}^\Eucl[N],\nonumber\\
&\vdots&
\end{eqnarray}
since $\hat{\calC}^\Eucl$ removes one extraordinary loop and $\hat{\mathcal{T}}$ removes two.
This hierarchical procedure gives a good control on the solutions and suggests a sound postulation that the series should terminate at a finite number of steps.

Even though we have a good understanding of solutions to the scalar constraint, our knowledge about them is far less clear and complete than that about s-knot states. Additionally, we encounter various quantization ambiguities, notably the choice of $R^{(j)}$ for the $SU(2)$ representation, which further complicate the solutions. It should also be remarked that the scalar constraint operator $\hat{\calC}[N]$ is evidently not hermitian in the space $\hil_\Diff$, as it removes extraordinary loops but does not add them. One can simply obtain the hermitian scalar constraint operator by replacing it with $(\hat{\calC}+\hat{\calC}^\dag)/2$, and the hermitian operator is expected to be better behaved for some technical issues as well as for some aspects of the classical limit. However, as a nonhermitian scalar constraint operator does not lead to any inconsistency, there is no logical necessity of being hermitian. Meanwhile, a great number of variant approaches of the quantum scalar constraint have been considered in the literature (see Sec.~7.4.1 of Ref.~\cite{Rovelli:2004tv}).

\subsection{Quantum dynamics}\label{sec:quantum dynamics}
The space of solutions to the quantum scalar constraint is the kernel of $\hat{\calC}[N]$, which is a proper subspace of $\Cyl_\inv^*$. Consequently, the physical inner product between any two solutions are naturally inherited from the inner product of the space $\Cyl_\inv^*$; i.e., for $\obra{\Psi},\obra{\Phi}\in\text{Kernel of}\ \hat{\calC}$, the physical inner product is given by
\begin{equation}
\ooinner{\Psi}{\Phi}_\phys := \ooinner{\Psi}{\Phi},
\end{equation}
where the right-hand side is defined in \eref{ooinner}.
With respect to the physical inner product $\ooinner{\cdot}{\cdot}_\phys$, the Cauchy completion of the kernel of $\hat{\calC}$ is the physical Hilbert space $\hil_\phys$.

Even though $\hil_\phys$ can been rigorously constructed (at least formally), the quantum dynamics in terms of evolution remains very obscure. Given physical state $\obra{\Psi}\in\hil_\phys$, to read out the ``evolution'' (with respect to the arbitrary coordinate time $t$ used in the ADM  foliation), one might attempt to designate the quantum counterpart of the classical Hamilton's equation \eref{Adot Edot} as
\begin{equation}
\frac{d}{dt}\obra{\Psi}\,\hat{\mathcal{O}}(t)\oket{\Psi}_\phys = \frac{\hbar}{i}\obra{\Psi}\big[\hat{\mathcal{O}}(t),\hat{H}]\big]\oket{\Psi}_\phys,
\end{equation}
where the quantum Hamiltonian $\hat{H}:= (8\pi G\gamma)^{-1} \big(\hat{\calC}_\mathrm{G}[(\omega\cdot t)]+\hat{\calC}_\Diff[\vec{N}]+\hat{\calC}[N]\big)$, and $\hat{\mathcal{O}}(t)$ is the operator of an observable $\mathcal{O}$ corresponding to a measurement performed at coordinate time $t$. However, this strategy does not work, because the right-hand side simply vanishes as physical states are annihilated by all three constraints. In a sense, the evolution, in the conventional notion, is completely ``frozen''. This difficulty is closely related to the renowned ``problem of time'' in QG\cite{Isham:1992ms,Kuchar:1991qf} and has entailed the \emph{timeless} description of quantum mechanics (see Chap.~5 of Ref.~\cite{Rovelli:2004tv} for an in-depth elaboration). In the literature, there are mainly two strategies to unveil the quantum dynamics: the ``complete observable approach'' vs.\ the ``partial observable approach''.\footnote{It should be remarked that there are many various approaches in the literature that are more or less in one form or the other of the two main strategies but different in finer detail. The aim here is neither to give decisive definitions of these two strategies nor to exhaust all possible approaches but to give general ideas about how the problem of quantum dynamics could be tackled.} \emph{Complete observables}, also referred to as ``Dirac observables'' or ``physical observables'', are those that commute with all three constraints; by contrast, \emph{partial observables} are those that do not commute with all three constraints (the scalar constraint in particular). In other words, complete observables are free of any gauge ambiguities, while partial observables still have a remnant of gauge dependence in the strict sense. (For more on the distinction, see Refs.~\cite{Dittrich:2004cb,Dittrich:2005kc}.)

In the complete observable approach, instead of $\hat{\mathcal{O}}(t)$, we construct a family of Dirac operators $\widehat{\mathcal{O}|_\Phi}$ parameterized by $\Phi$. The operator $\widehat{\mathcal{O}|_\Phi}$ is nontrivially constructed so that it is well defined in $\hil_\phys$ and commutes with all constraint operators for any value of $\Phi$. We interpret the operator $\widehat{\mathcal{O}|_{\Phi_0}}$ as representing a measurement $\mathcal{O}$ performed at the instance when the observable $\Phi$ takes the value $\Phi_0$.\footnote{In the timeless language, a measurement is said to be conducted at some ``instance'', rather than at some ``instant''.} For a given physical state $\obra{\Psi}\in\hil_\phys$, the quantum dynamics is portrayed in terms of the expectation values
\begin{equation}\label{Psi Ophi Psi}
\obra{\Psi}\,\widehat{\mathcal{O}|_\Phi}\oket{\Psi}_\phys,
\end{equation}
which do not describe the evolution of observables with respect to a preferred time variable but, instead, the correlation between observables ($\mathcal{O}$ and $\Phi$). In this regard, the variable $\Phi$ is said to serve as the ``internal time'' (also known as ``internal clock'').
Unfortunately, it is extremely difficult to construct Dirac operators that are nontrivial and physically sound. Usually, one also has to include nongravitational matter into the system in order to provide a privileged dynamical reference as the internal time.\footnote{The strategy of the complete observable approach is closely related to the idea of the \emph{reduced phase space quantization}, in which one finds a one-parameter family of complete observables $\mathcal{O}|_\Phi$ at the classical level \emph{before} quantization. See \sref{sec:reduced phase space}.} In some symmetry-reduced theories, such as LQC, the Dirac operators can be explicitly formulated and consequently the quantum dynamics can be clearly deciphered, as will be shown in \sref{sec:LQC}.

The partial observable approach, on the other hand, adopts a very different philosophy. While it is asserted in the complete observable approach that only complete observables are physical, the partial observable approach contends that it would be too restrictive to dismiss all partial observables, because one can never grasp ``totality'' of the whole system and therefore some gauge degrees of freedom, especially those for the reparametrization invariance in time, nevertheless bear physical significance as far as one concerns real measurements, which are conducted without the knowledge of totality. Unattainability of totality is even more crucial in quantum mechanics, as one can never disregard the observer, who is not viewed as a part of the system to be observed. By the standard Copenhagen interpretation of quantum mechanics, it is impossible to know and even invalid to ask what the physical state $\obra{\Psi}\in\hil_\phys$ is for the whole world. In this sense, the quantum dynamics described in the style of \eref{Psi Ophi Psi} seems inadequate and heuristic at best.

In accord with the Copenhagen interpretation, the partial observable approach describes the quantum dynamics in terms of the ``if-then-what'' prediction. Provided that a measurement $\mathcal{O}\!_A$ performed at some instance yields the outcome $a$, the quantum dynamics is to predict what the probability is for the measurement $\mathcal{O}\!_B$ performed at another instance to yield the outcome $b$. The only difference from the conventional quantum mechanics is that we do not need to specify the time separation between the two measurements, because any notion of time separation is in principle encoded in the measurement outcomes $a$ and $b$ themselves in the timeless description. (In fact, we do not even need to specify the time-ordering of the two measurements. See Sec.~3.5 of Ref.~\cite{Chiou:2010mw} for more discussions.)

For LQG, to make sense of the quantum dynamics of spacetime, we naturally choose the relevant measurements to be those for the intrinsic areas and volumes as studied in \sref{sec:quantum geometry}. Note that the corresponding operators defined in Eqs.~(\ref{intrinsic V op}) and (\ref{intrinsic A op}) correspond to partial observables as opposed to complete observables, as they do not commute with the scalar constraint $\hat{\calC}[N]$. Consequently, after a complete measurement of the quantum geometry, the kinematical state is collapsed into an eigenstate of the quantum geometry, which is an s-knot state $\obra{K,c}\in\hil_\inv$. The quantum dynamics of spacetime is then posed as a predictive question: What then is the probability for another complete measurement of the quantum geometry to yield the outcome associated with the s-knot $\obra{K',c'}$? It is natural to postulate that the probability is given by the $\abs{W(K,c;K',c')}^2$, where $W(K,c;K',c')$ is the \emph{transition amplitude}
\begin{equation}\label{W(KK')}
W(K,c;K',c'):=\obra{K,c}\hat{P}_\calC\oket{K',c'}
\end{equation}
and $\hat{P}_\calC$ is the ``projector'' that projects an s-knot state in $\hil_\inv$ into the subspace $\hil_\phys$. More precisely, the projector $\hat{P}_\calC:\hil_\inv\rightarrow\hil_\phys\subset\hil_\inv$ is formally given by
\begin{equation}
\hat{P}_\calC := \sum_{\obra{\Psi}\in\hil_\phys} \oket{\Psi}\obra{\Psi},
\end{equation}
where $\{\obra{\Psi}\in\hil_\phys\}$ forms an orthonormal basis of $\hil_\phys$. Therefore, the quantum dynamics in principle can be inferred from the complete knowledge about the physical Hilbert space $\hil_\phys$. The transition amplitude given by \eref{W(KK')} bears close similarity to that in the spin foam formalism as will be discussed in \sref{sec:spin foam theory}.

Both the complete observable approach and the partial observable approach are far from fully developed and remain tentative, and our understanding about the quantum dynamics of spacetime is still very limited. The obstacles not only lie in the mathematical difficulties but are also deeply rooted from the conceptual (and even philosophical) riddles of interpreting the fundamental notions of space, time, quantum measurements, etc.

\section{Inclusion of matter fields}\label{sec:inclusion of matter fields}
So far, to bring out the main ideas of LQG, we have ignored nongravitational matter fields. Inclusion of matter fields does not require a major revamp of the underlying framework. The quantum states of space plus matter naturally extend the notion of s-knots with additional degrees of freedom. We will follow the lines of Sec.~7.2 of Ref.~\cite{Rovelli:2004tv}. For more details, see Chap.~12 of Ref.~\cite{Thiemann:2007zz} and references therein (and also see Chap.~9 of Ref.~\cite{Ashtekar:1991hf} for the corresponding classical theories).

\subsection{Yang-Mills fields}
The easiest extension is to incorporate Yang-Mills fields. Let $G_\mathrm{YM}$ be the Yang-Mills group. The two 3d connections, Ashtekar connection $A$ and Yang-Mills connection $A_\mathrm{YM}$, can be considered together as a single connection $\mathcal{A}=(A,A_\mathrm{YM})$ associated with the group $SU(2)\times G_\mathrm{YM}$. As the holonomies of $A_\mathrm{YM}$ and surface integrals of the Yang-Mills electric field can be defined exactly in the same ways as those of $A$ and $\tE$, the construction of $\hil$, $\hil_\mathrm{G}$, and $\hil_\inv$ extends straightforwardly to the total connection $\mathcal{A}$ without much difficulty.

The gauge ($G$) and diffeomorphism ($\Diff$) invariant quantum states of space plus Yang-Mills fields are given by the s-knot states that are classified by knotted graphs and carry irreducible representations of the group $G$ on the links and the corresponding intertwiners at the nodes. Because $G$ is the direct product of $SU(2)$ and $G_\mathrm{YM}$, its irreducible representations are simply the products of those of $SU(2)$ and $G_\mathrm{YM}$. Consequently, the coloring for each link is extended as $(j_l,k_l)$ and the coloring for each nodes is extended as $(i_n,w_n)$, where $k_l$ is the irreducible representation of $G_\mathrm{YM}$ representing the Yang-Mills electric flux passing through the surface $l$, and $w_n$ is the intertwiner of $G_\mathrm{YM}$ representing the Yang-Mills field strength at the node $n$.

\subsection{Fermions}
The next is to include fermions. Let $\eta(x)$ be a Grassmann-valued fermionic field. It transforms as an irreducible representation $k$ under the gauge transformation of the Yang-Mills group $G_\mathrm{YM}$ and as a fundamental ($j=1/2$) representation under the $SU(2)$ transformation. It is more convenient in LQG to take the densitized field $\psi(x):=\sqrt{\det\tE}\,\eta(x)$ as the fundamental field variable\cite{Thiemann:1997rq}.

As an extension of \eref{cylindrical fun}, the cylindrical function of the connection $\mathcal{A}$ and the fermion field $\psi$ is a smooth (Grassmann-valued) function of $L$ group elements of $G=SU(2)\times G_\mathrm{YM}$ and $2N\abs{k}$ Grassmann variables ($\abs{k}$ is the dimension of $k$) defined as
\begin{eqnarray}
\Psi_{\Gamma,f}[\mathcal{A},\psi,\bar{\psi}]
&=&f\big(h_{\gamma_1}(\mathcal{A}),\dots,h_{\gamma_L}(\mathcal{A}),\nonumber\\
&&\quad\;
\dots,\psi^{m_i}(x_n),\dots,\bar{\psi}^{\bar{m}_i}(x_n),\dots\big),
\end{eqnarray}
where $\Gamma$ is the graph comprised of $L$ edges $\gamma_l$ and $N$ nodes located at $x_{n=1,\dots,N}$, $m_{i=1,\dots.\abs{k}}\in w(k)$ are indices (i.e.\ weights) of the representation $k$, and $\bar{m}_{i=1,\dots,\abs{k}}\in w(\bar{k})$ are indices of the conjugate representation $\bar{k}$. Since Grassmann variables anticommute with one another, cylindrical functions are at most linear in $\psi^m_n\equiv\psi^m(x_n)$ and $\bar{\psi}^{\bar{m}}_n\equiv\bar{\psi}^{\bar{m}}(x_n)$ for each pair $(n,m)$.
If two cylindrical functions are defined for the same graph $\Gamma$, as an extension of \eref{Gamma f Gamma g}, the inner product between them is defined as
\begin{eqnarray}
&&\inner{\Psi_{\Gamma,f}}{\Psi_{\Gamma,g}}\nonumber\\
&:=&
\int_{G^{L}}\prod_{l=1}^L d\mu_l
\prod_{m\in w(k)} \prod_{n=1}^N
\left(\mathcal{I}_{\psi_n^m,\psi_n^{m*}}+\int d\psi_n^m d\psi_n^{m*}\right)
\left(\mathcal{I}_{\bar{\psi}_n^{\bar{m}},\bar{\psi}_n^{\bar{m}*}}+\int d\bar{\psi}_n^{\bar{m}} d\bar{\psi}_n^{\bar{m}*}\right) \nonumber\\
&&\quad\mbox{}\times
{f(h_{\gamma_1},\dots,h_{\gamma_L},
\psi^{m_1}_1,\dots,\psi^{m_\abs{k}}_N,
\bar{\psi}^{\bar{m}_1}_1,\dots,\bar{\psi}^{\bar{m}_\abs{k}}_N)}^*
\nonumber\\
&&\quad\mbox{}\times
g(h_{\gamma_1},\dots,h_{\gamma_L},\psi^{m_1}_1,\dots,\psi^{m_\abs{k}}_N,
\bar{\psi}^{\bar{m}_1}_1,\dots,\bar{\psi}^{\bar{m}_\abs{k}}_N),
\end{eqnarray}
where $d\mu$ is the Haar measure on $G$, $\mathcal{I}_{\psi,\psi^*}$ is a linear transformation that maps the integrand into a new (Grassmann-valued) function by assigning $\psi=0$ and $\psi^*=0$, $\int d\psi d\psi^*$ is the Grassmann integral over the superspace spanned by $\psi$ and $\psi^*$, and $\psi_n^m,\bar{\psi}_n^{\bar{m}}$ are treated as independent Grassmann variables. The extension of this inner product to any two cylindrical functions is completely analogous to that in the case of pure gravity. The construction of $\hil$, $\hil_\mathrm{G}$, and $\hil_\inv$ can then be readily repeated.

The gauge and diffeomorphism invariant quantum states are again given by the s-knots with extended colorings. For each node $n$, in addition to the intertwiner $(i_n,w_n)$, we also have to specify an integer $F_n$ as the degree of the monomial in $\psi^m(x_n)$, which determines the fermion number in the region of the node, and an integer $\bar{F}_n$ as the degree of the monomial in $\bar{\psi}^{\bar{m}}(x_n)$, which determines the antifermion number in the region of the node.

In the presence of fermions, at each node $n$, the intertwiner $(i_n,w_n)$ is not only intertwined with the coloring $(j_l,k_l)$ of the links attached to $n$ but also with the (anti)fermion numbers $F_n,\bar{F}_n$. This is because fermions carry the $j=1/2$ representation indices of $SU(2)$ and the $k$ representation indices of $G_\mathrm{YM}$, which have to be properly contracted with the matrix elements of $R^{(j_l)}\left(h_{\gamma_l}(A)\right)$ and $R^{(k_l)}\left(h_{\gamma_l}(A_\mathrm{YM})\right)$ in a way such that the generalized Clebsh-Gordon conditions are satisfied for both $SU(2)$ and $G_\mathrm{YM}$. In other words, fermions are quanta of ``charged'' particles, with both $SU(2)$ and $G_\mathrm{YM}$ charges, and the quanta of electric flux labelled by $j_l$ for $SU(2)$ and $k_l$ for $G_\mathrm{YM}$ can emerge from and end at the charged particles. (The fact that fermions are $SU(2)$ charged implies that they contribute to the spectrum of the area operator. See Ref.~\cite{Montesinos:1998dw} for more discussions.) In the same regard, both the gravitational and Yang-Mills fields are self-sourcing, as they yield colorings for both links and nodes.

\subsection{Scalar fields}
Scalar fields can also be included in LQG, but in a less natural manner than those for Yang-Mills fields and fermions.

Let $\phi(x)$ be a scalar field, which transforms as a trivial ($j=0$) representation under the $SU(2)$ transformation and as an irreducible representation $k'$ under the gauge transformation of the Yang-Mills group $G_\mathrm{YM}$. Thus, $\phi$ takes the value in the vector space of the $k'$ representation. The problem is that the $k'$ representation vector space is noncompact with respect to natural $G_\mathrm{YM}$-invariant measures, thus rendering the definition of the inner product between cylindrical functions difficult. (Fermions do not suffer from this problem, since the Grassmann integral naturally gives the appropriate measure.) One way to get around this problem is to assume $k'$ to be the \emph{adjoint} representation of $G_\mathrm{YM}$ and replace $\phi(x)$ with the associated ``point holonomy'' $U(x):=\exp\left(\phi(x)\right)$, which takes the value of $G_\mathrm{YM}$ and admits the natural Haar measure.

The cylindrical function of $\mathcal{A}$, $\psi$, and $\phi$ is a smooth (Grassmann-valued) function of $L$ group elements of $G=SU(2)\times G_\mathrm{YM}$, $2N\abs{k}$ Grassmann variables, and $N$ group elements of $G_\mathrm{YM}$ defined as
\begin{eqnarray}
\Psi_{\Gamma,f}[\mathcal{A},\psi,\bar{\psi},\phi]
&=&f\big(h_{\gamma_1}(\mathcal{A}),\dots,h_{\gamma_L}(\mathcal{A}),
\psi^{m_1}(x_1),\dots,\psi^{m_\abs{k}}(x_N),\nonumber\\
&&\quad\; \bar{\psi}^{\bar{m}_1}(x_1),\dots\bar{\psi}^{\bar{m}_\abs{k}}(x_N),
U(x_1),\dots,U(x_{N})\big).
\end{eqnarray}
The inner product between two cylindrical functions of the same graph is then naturally defined as
\begin{eqnarray}
\inner{\Psi_{\Gamma,f}}{\Psi_{\Gamma,g}}&:=&
\int_{G^L}\prod_{l=1}^L d\mu_l
\int_{G_\mathrm{YM}^N}\prod_{n=1}^N d\mu_n \\
&&\;\prod_{m\in w(k)} \prod_{n=1}^N
\left(\mathcal{I}_{\psi_n^m,\psi_n^{m*}}+\int d\psi_n^m d\psi_n^{m*}\right)
\left(\mathcal{I}_{\bar{\psi}_n^{\bar{m}},\bar{\psi}_n^{\bar{m}*}}+\int d\bar{\psi}_n^{\bar{m}}d\bar{\psi}_n^{\bar{m}*}\right) \nonumber\\
&&\quad\mbox{}\times
{f(h_{\gamma_1},\dots,h_{\gamma_L},
\psi^{m_1}_1,\dots,\psi^{m_\abs{k}}_N, \bar{\psi}^{\bar{m}_1}_1,\dots\bar{\psi}^{\bar{m}_\abs{k}}_N,
U_1,\dots,U_N)}^*\nonumber\\
&&\quad\mbox{}\times
g(h_{\gamma_1},\dots,h_{\gamma_L},
\psi^{m_1}_1,\dots,\psi^{m_\abs{k}}_N, \bar{\psi}^{\bar{m}_1}_1,\dots\bar{\psi}^{\bar{m}_\abs{k}}_N,
U_1,\dots,U_N),\nonumber
\end{eqnarray}
where $d\mu_l$ is the Haar measure on $G$, and $d\mu_n$ the Haar measure on $G_\mathrm{YM}$. Extension to any two cylindrical functions is straightforward as before.

In the presence of scalar fields, of the resulting s-knot state, the coloring for each node $n$ is augmented with an additional integer number $S_n$, which represents the total adjoint charge, or equivalently the total number, of the scalar particles in the region of the node.
Like fermions, scalar fields carry $G_\mathrm{YM}$ (adjoint) charge and accordingly $S_n$ is coupled with $w_n,k_l,F_n$ in the generalized Clebsh-Gordon condition for $G_\mathrm{YM}$; unlike fermions, on the other hand, scalar fields are trivial for $SU(2)$ and $S_n$ is decoupled from $i_n,j_l$ in the generalized Clebsh-Gordon condition for $SU(2)$.

\subsection{S-knots of geometry and matter}
In summary, with inclusion of all kinds of matter (Yang-Mills fields, fermions, scalar fields), the gauge and diffeomorphism quantum states are s-knots $\obra{K,c}\equiv\obra{\Gamma,j_l,i_n,k_l,w_n,F_n,\bar{F}_n,S_n}$ labelled by the following quantum numbers:
\begin{itemize}

\item $\Gamma$: an abstract knotted graph with oriented links $l$ and nodes $n$.

\item $j_l$: an irreducible $j$ representation of $SU(2)$ associated with each link $l$.

\item $i_n$: an $SU(2)$ intertwiner associated with each node $n$.

\item $k_l$: an irreducible $k$ representation of $G_\mathrm{YM}$ associated with each link $l$.

\item $w_n$: a $G_\mathrm{YM}$ intertwiner associated with each node $n$.

\item $F_n,\bar{F}_n$: two integers associated with each node $n$.

\item $S_n$: an integer associated with each node $n$.

\end{itemize}
These quantum numbers correspond to physical quantities as listed in \tref{tab:quantum numbers}.
\begin{table}[ht]
\tbl{Physical quantities for the quantum numbers of s-knots.\label{tab:quantum numbers}}
{\begin{tabular}{cl}
\toprule
%\Hline
%\Hline
Quantum number & Physical quantity \\
\colrule
$\Gamma$ & adjacency relation of regions $n$ and surfaces $l$ \\
$j_l$ & area of the surface $l$ \\
$i_n$ & volume of the region $n$ \\
$k_l$ & Yang-Mills electric flux through the surface $l$ \\
$w_n$ & Yang-Mills field strength at the region $n$ \\
$F_n,\bar{F}_n$ & numbers of fermions and antifermions at the region $n$ \\
$S_n$ & number (total adjoint charge) of scalars at the region $n$ \\
\botrule
\end{tabular}
}
%\begin{tabnote}
%$^{\text a}$Sample table footnote.
%\end{tabnote}\label{ra_tbl1}
\end{table}

The physical interpretation of s-knots as discussed in \sref{sec:quantum geometry} can be directly generalized to the s-knots of both geometry and matter in the obvious way. The paradigm of background independence is even more remarkable with inclusion of matter fields, as geometry and matter fields are truly on the equal footing and live on top of one another via their contiguous relations without any reference to a given background.

Furthermore, in the presence of nongravitational matter fields, there is no difficulty to perform the ADM foliation and obtain the three constraints. Particularly, the scalar constraint of the classical theory can be regulated in the same fashion as that in \sref{sec:classical scalar constraint} and then promoted to the quantum operator as in \sref{sec:quantum scalar constraint}. Therefore, the quantum theory of dynamics of spacetime plus matter in principle can be constructed in the same manner as in Secs.~\ref{sec:solutions to the scalar constraint} and \ref{sec:quantum dynamics}.

\section{Low-energy physics}\label{sec:low-energy physics}
The essential premise of LQG is that quantum states are background-independent excitations out of nothingness---a concept in direct conflict with the foundation of conventional QFTs. This poses the big challenge to LQG to show how the low-energy description of conventional QFTs arises from the underlying Planckian world in an appropriate sense (of coarse graining). The final resolution is still far out of reach, but the research is being undertaken step by step. The low-energy physics remains one of the most wanted missing pieces in LQG---Not until we figure out the low-energy physics of LQG and show its agreement with Einstein's theory of classical GR and compatibility with the QFT of the Standard Model can we declare LQG to be an adequate quantum theory of GR, because both classical GR and the Standard Model have been intensively tested in the low-energy regime. In this section, we present the basic ideas of recovering low-energy physics and refer readers to Chap.~11 of Ref.~\cite{Thiemann:2007zz} for more systematic treatments and more details.

\subsection{Weave states}
It is impressive that LQG yields discrete spectra of geometry as a natural consequence of quantization. The eigenstates of geometry, i.e.\ s-knots, however do not look akin at all to the smooth macroscopic spaces (curved or flat) we are familiar with. What then is the connection between the discrete quantum geometry and the smooth classical geometry?

Consider a very huge spin network with a very large number of nodes and links, each of which is of area or volume in the Planck scale. This is a big lattice of Planck-scale lattice size, but it appears as a smooth 3d geometry when probed at a macroscopic scale much larger than the Planck scale. This is analogous to the fabric of a piece of cloth, which is composed of thousands of granular threads but appear smooth at a distance.

More precisely, given a fixed classical 3d metric $q_{ab}(x)=\delta_{ij}e_a^i(x)e_b^j(x)$, it is possible to construct a spin network state $\ket{S}$ that approximates the metric at scale $l\gg\Pl,$, much larger than the Planck length $\Pl$. That is, for any region $\mathcal{R}$ and surface $\mathcal{S}$, we have
\begin{subequations}\label{weave state}
\begin{eqnarray}
\hat{A}(\mathcal{S})\ket{S} &=& \left(\left.A(\mathcal{S})\right|_e+O(\Pl/l)\right)\ket{S}, \\
\hat{V}(\mathcal{R})\ket{S} &=& \left(\left.V(\mathcal{R})\right|_e+O(\Pl/l)\right)\ket{S},
\end{eqnarray}
\end{subequations}
where $\left.A(\mathcal{S})\right|_e$ is the classical area of $\mathcal{S}$ defined in \eref{A(S)} and $\left.V(\mathcal{R})\right|_e$ is the classical volume of $\mathcal{R}$ defined in \eref{V(R)} with the given cotriad $e_a^i$, and $O(\Pl/l)$ denotes small corrections in $\Pl/l$. Such a spin network $\ket{S}$ is called a \emph{weave state} of the metric $q_{ab}$. This definition is given for spin networks but can be easily carried over to the diffeomorphism invariant level for s-knots.

Several weave states have been constructed for flat space, Schwarzschild space, space with gravitational waves, and more\cite{Ashtekar:1992tm}. It should be noted that \eref{weave state} does not determine a unique weave state; when all macroscopic physics is taken into account, it seems more likely that the proper quantum state for a macroscopic geometry is a coherent superposition of the weave state solutions.

\subsection{Loop states vs.\ Fock states}
The basic variables of LQG are holonomies (Wilson loops) of the connection $A$ along 1-dimensional curves and the fluxs of the conjugate momentum $\tE$ across 2-dimensional surfaces. These variables however fail to be well defined in the Fock space of excitations of $A$ in the perturbative QFTs. A further inquiry into the low-energy physics of LQG then is to understand how the perturbative description of quantum excitations in terms of Fock states in conventional QFTs arise as a low-energy limit of the nonperturbative theory in terms of loop states.

These issues have been studied for simple examples. The main effort so far is to construct mathematical and conceptual tools that will facilitate a systematic analysis of quantum fields on semiclassical states of quantum geometry\cite{Sahlmann:2002qj,Sahlmann:2002qk}. Particularly, the connection between loop states and Fock states has been studied in detail for the Maxwell field\cite{Varadarajan:1999it,Varadarajan:2001nm,Ashtekar:2001xp,Velhinho:2001qj} and linearized gravity\cite{Varadarajan:2002ht} in Minkowski spacetime.

\subsection{Holomorphic coherent states}\label{sec:coherent states}
The relation between quantum states and the classical theory is most likely to be understood by the techniques of coherent states. Various constructions of coherent states have been suggested. Here we briefly describe the formulation of \emph{holomorphic coherent states}\cite{Ashtekar:1994nx,Thiemann:2002vj,Bahr:2007xa, Bahr:2007xn,Flori:2009rw,Bianchi:2009ky} by following the lines of Ref.~\cite{Bianchi:2009ky} for the \emph{Euclidean} theory.

Consider the heat kernel $K_t(h,h_0)$ on $SU(2)$, which is given by the Peter-Weyl expansion:
\begin{equation}
K_t(h,h_0) = \sum_j (2j+1)\, e^{-j(j+1)t}\,\Tr\, R^{(j)}(hh_0^{-1}),
\end{equation}
where the positive real number $t$ is the ``heat kernel time''.
For a given graph $\Gamma$ of $N$ nodes, a holomorphic coherent state associated with $\Gamma$ is defined as the $SU(2)$-invariant projection of a product over links of heat kernels:
\begin{equation}\label{holomorphic coherent state}
\Psi_{\Gamma,H_{ab},t_{ab}}(h_{ab}) = \int_{SU(2)^N} \prod dg_a \prod_{ab} K_{t_{ab}}(h_{ab},g_aH_{ab}g_b^{-1}),
\end{equation}
where, to simplify the notation, we use $a,b,\dots=1,\dots,N$ to denote the nodes of $\Gamma$ and the (ordered) couples $ab$ to denote the (oriented) links.\footnote{This notation scheme is well-suited for complete graphs, but generalization to arbitrary (non-complete) graphs is obvious.}
The label $H_{ab}$ for each link $ab$ is an element of $SL(2,\mathbb{C})$, which is diffeomorphic to the $SU(2)$ cotangent bundle $T^*SU(2)\cong SU(2)\times su(2)^*\cong SU(2)\times su(2)$.

The state given in \eref{holomorphic coherent state} is analogous to the standard wave packet of nonrelativistic quantum theory
\begin{equation}
\psi_{x_0,p_0,\sigma} = \frac{1}{(2\pi\sigma^2)^{1/4}}\, e^{-\frac{(x-x_0)^2}{4\sigma^2}}e^{\frac{i}{\hbar}p_0(x-x_0)},
\end{equation}
which is peaked in the position $x_0$ as well as in the momentum $p_0$ and can be written (up to a constant phase) as a Gaussian function
\begin{equation}
\psi_{X_0,\sigma} = \frac{1}{(2\pi\sigma^2)^{1/4}}\,
e^{-\frac{(x-X_0)^2}{4\sigma^2}}
\end{equation}
peaked on a \emph{complex} position $X_0:=x_0+2i\sigma^2\hbar^{-1}p_0$. As $K_t(h,H_0)$ is the analytic continuation to $SL(2,\mathbb{C})$ of the heat kernel $K_t(h,h_0)$ on $SU(2)$ and $t=0$ reduces $K_t(h,h_0)$ to a delta function, $H_{ab}$ is analogous to $X_0$ (while $h_{ab}$ analogous to $x_0$) and $t_{ab}$ is analogous to $\sigma$.

As $SL(2,\mathbb{C})$ is diffeomorphic to $SU(2)\times su(2)$, we can decompose each $SL(2,\mathbb{C})$ element as
\begin{equation}
H_{ab} =  h_{ab}\, e^{2it L_{ab}},
\end{equation}
where $h_{ab}\in SU(2)$ and $L_{ab}\in su(2)$.
Alternatively, $H_{ab}$ can be written in the form
\begin{equation}
H_{ab} = n_{ab}\, e^{-i(\xi_{ab}+i\eta_{ab})\frac{\sigma_3}{2}}n_{ba}^{-1},
\end{equation}
where $n_{ab},n_{ba}\in SU(2)$ are two \emph{unrelated} group elements of $SU(2)$, $\xi\in[0,2\pi)$ is an angle, and $\eta_{ab}\in\mathbb{R}^+$ is a positive real numbers.
Any element $n$ of $SU(2)$ can be associated with a unit vector $\vec{n}\in\mathbb{R}^3$ via
\begin{equation}
\vec{n}:=R^{(j=1)}(n)\cdot \vec{z},
\end{equation}
where $\vec{z}=(0,0,1)\in\mathbb{R}^3$ and $R^{(j=1)}(n)$ acts as a rotation matrix. Consequently, for each link $l\equiv ab$, we have four labels of two unit vectors, one angle, and one real number: $(\vec{n}_{i(l)},\vec{n}_{f(l)},\xi_l,\eta_l)\equiv(\vec{n}_{ab},\vec{n}_{ba},\xi_{ab},\eta_{ab})$, where $i(l)$ and $f(l)$ denote the initial and final endpoints (nodes) of the link $l$.

Therefore, apart from the labels $t_{ab}$ (its meaning will be clarified shortly), the state given in \eref{holomorphic coherent state} associated with a spin network graph $\Gamma$ is labelled by a number $\eta_l$ and an angle $\xi_l$ for each link $l$, and for each node a set of unit vectors $\vec{n}$, one for each link at that node.
These variables together admit a geometric interpretation of a simplicial 3-complex as follows. The graph $\Gamma$ is assumed to be dual to a simplicial decomposition of the spatial manifold, as each node is dual to a 3-simplex (chunk of space) and each link is dual to a face of two adjacent simplices. The vectors $\vec{n}$'s at a node are outgoing unit vectors normal to the faces of the simplex dual to the node,\footnote{Note that in general $\vec{n}_{ab}\neq-\vec{n}_{ba}$. The difference between $\vec{n}_{ab}$ and $-\vec{n}_{ba}$ encodes connections of the intrinsic geometry.} and the positive parameter $\eta_l$ for a link $l$ is related to a spin $j_l^0$, which is the average of the area of the face dual to the link $l$. Furthermore, the simplicial extrinsic curvature is specified by an angle for each face of two adjacent simplices and is identified by $\xi_l$. To sum up, the state in \eref{holomorphic coherent state} is specified by both an intrinsic geometry (labelled by $\vec{n}$ and $j_l^0$) and an extrinsic geometry (labelled by $\xi_l$) for a simplicial 3-complex due to $\Gamma$.

To see that the holomorphic coherent states represent semiclassical states at some appropriate large-scale limit, we study their asymptotics for large $\eta_l$. Using the asymptotic formula
\begin{equation}
R^{(j_{ab})}(e^{-i(\xi_{ab}+i\eta_{ab}){\sigma_3}/{2}})
\approx e^{-i\xi_{ab}j_{ab}} e^{\eta_{ab}j_{ab}} \ket{j_{ab},+j_{ab}}\bra{j_{ab},+j_{ab}}
\end{equation}
for $\eta_{ab}\rightarrow\infty$, it was shown in Ref.~\cite{Bianchi:2009ky} that the holomorphic coherent state in \eref{holomorphic coherent state} is given by the superposition:
\begin{eqnarray}\label{asymptotic coherent state}
\Psi_{\Gamma,H_{ab},t_{ab}}(h_{ab})
&\approx& \sum_{j_{ab}}
\left( \prod_{ab}\, (2j_{ab}+1)\, c_{j_{ab}^0,\xi_{ab},\sigma_{ab}^0}(j_{ab})
\right) \nonumber\\
&&\qquad \mbox{}\times
\Psi_{\Gamma,j_{ab},\Phi_a(\vec{n}_{ab})}(h_{ab}),
\end{eqnarray}
where $c_{j^0,\xi,\sigma^0}(j)$ is a Gaussian function multiplied by a phase:
\begin{equation}\label{c(j)}
c_{j^0,\xi,\sigma^0}(j) := \exp\left(-\frac{(j-j^0)^2}{2\sigma^0}\right) e^{-i\xi j}
\end{equation}
with
\begin{equation}
(2j_{ab}^0+1):=\frac{\eta_{ab}}{t_{ab}},
\qquad \text{and} \qquad
\sigma_{ab}^0 :=\frac{1}{2t_{ab}}.
\end{equation}
Here, $\Phi_a(\vec{n}_{ab})$ is the \emph{coherent intertwiner} introduced in Ref.~\cite{Livine:2007vk}, which is a linear superposition of an orthonormal intertwiner basis $\{i_a\}$, given by
\begin{equation}
{\Phi_a(\vec{n}_{ab})^{m'_1\cdots}}_{m_1\cdots} = \sum_{i_a}\Phi_{i_a}(\vec{n}_{ab}) {v_{i_a}^{m'_1\cdots}}_{m_1\cdots}
\end{equation}
with
\begin{equation}
\Phi_{i_a}(\vec{n}_{ab}) = v_{i_a}\cdot \left(\bigotimes_b \ket{j_{ab},\vec{n}_{ab}}\right),
\end{equation}
where $\ket{j_{ab},\vec{n}_{ab}} :=R^{(j_{ab})}(n_{ab})\ket{j_{ab},+j_{ab}}$.
The states $\Psi_{\Gamma,j_{ab},\Phi_a(\vec{n}_{ab})}$ are spin networks with coherent intertwiners:
\begin{equation}
\Psi_{\Gamma,j_{ab},\Phi_a(\vec{n}_{ab})}(h_{ab})
:=
\sum_{i_a}\left(\prod_a \Phi_{i_a}(\vec{n}_{ab}) \right) \Psi_{\Gamma,j_{ab},i_a}(h_{ab}).
\end{equation}

The asymptotic states given by \eref{asymptotic coherent state} are exactly the boundary semiclassical states used in Ref.~\cite{Bianchi:2009ri} and the coefficients $c_{j^0,\xi,\sigma^0}(j)$ in \eref{c(j)} are the same as proposed in Ref.~\cite{Rovelli:2005yj}, which demand that the states are peaked both on the area ($j^0$) and on the extrinsic angle ($\xi$) with appropriate spread widths (specified by $\sigma^0\equiv1/(2t)\approx(j^0)^k$ with $k<2$). These states represent the boundaries of Euclidean 4-geometries and are used to compute transition amplitudes for given boundary states in the Euclidean theory (see \sref{sec:covariant LQG}). The analysis of graviton propagators provides a methodology for studying the low-energy physics, and particularly the comparison to the classical theory of Euclidean GR confirms that the value $\xi$ is fixed by $\xi=\gamma\cos^{-1}(-1/4)$. The construction of holomorphic coherent states for the Lorentzian theory however is much less established and demands more works of further research.

\section{Spin foam theory}\label{sec:spin foam theory}
Conventional QFTs admit two different formulations: the canonical (Hamiltonian) formalism and the sum-over-histories (path integral) formalism. So far we have concentrated on the canonical formulation of LQG. It is time to discuss the ``spin foam'' theory, which can be viewed as the covariant approach of LQG, alternative to the canonical approach. While the general structure of spin foam models matches nicely with the canonical theory of LQG, the precise relation between these two formalisms however is still not entirely clear and remains a key topic in current research (see \sref{sec:LQG vs spin foam}). Up to now, the canonical theory of LQG and the theory of spin foams should be regarded as two closely related but independent approaches of QG.
In this section, we present the basic ideas of the spin foam theory and refer readers to Refs.~\cite{Rovelli:2014book, Perez:2003vx,Oriti:2003wf,Perez:2004hj,Perez:2012wv} for detailed construction of spin foam models. Also see Chap.~9 of Ref.~\cite{Rovelli:2004tv} and Chap.~14 of Ref.~\cite{Thiemann:2007zz} for shorter accounts.

\subsection{From s-knots to spin foams}
Spin foams can be regarded as the ``worldsurfaces'' swept out by s-knots traveling and transmuting in time. A spin foam represents a quantized spacetime, in the same sense that an s-knot represents a quantized space.

Consider the transition amplitude from one s-knot $\obra{k'}$ to another $\obra{k}$
\begin{equation}
W(k,k'):=\obra{k}\hat{P}_\calC\oket{k'}
\end{equation}
as given in \eref{W(KK')}. The operator $\hat{P}_\calC$ projects s-knots into the kernel of $\hat{\calC}[N]$. Heuristically, we can write the projector as
\begin{equation}\label{Pc}
\hat{P}_\calC = \lim_{t\rightarrow\infty}e^{-\hat{\calC}[N]t}
= \lim_{t\rightarrow\infty}\sum_{E\geq0}\oket{E}\,e^{-Et}\obra{E}
= \sum_{E=0}\oket{E}\obra{E}
\end{equation}
assuming the operator $\hat{\calC}[N]$ has a nonnegative spectrum of $E$.\footnote{As commented in \sref{sec:quantum scalar constraint}, the projector $\hat{P}_\calC$ is independent of the choice of $N(x)$, but we have to assume hermiticity and positive definiteness of $\hat{\calC}[N]$ in \eref{Pc}. Hermiticity can be easily prescribed as discussed in the last paragraph in \sref{sec:solutions to the scalar constraint}. Positive definiteness, however, is not guaranteed, but heuristically we can replace the local scalar constraint $C(x)$ with $\sqrt{C(x)^2}$ before regularization and quantization to yield positive definiteness. This heuristic tweak in fact makes good sense from the standpoint of the \emph{Master constraint program}, which will be outlined in \sref{sec:Master constraint program}.} Consequently,
\begin{equation}
W(k,k')=\lim_{t\rightarrow\infty}\obra{k}e^{-\hat{\calC}[N]t}\oket{k'}.
\end{equation}
In the same fashion of the path integral in quantum mechanics, inserting resolutions of the identity $1=\sum_k\oket{k}\obra{k}$, we can expand the above expression as a product of small time-step evolutions:
\begin{eqnarray}\label{W(k,k') product of steps 1}
W(k,k') &=&  \lim_{t\rightarrow\infty}\lim_{M\rightarrow\infty}\sum_{k_1,\dots,k_M}
\obra{k}e^{-\hat{\calC}[N]\Delta t}\oket{k_M} \obra{k_M}e^{-\hat{\calC}[N]\Delta t}\oket{k_{M-1}}\nonumber\\
&&\qquad\qquad\qquad\qquad
\cdots
\obra{k_2}e^{-\hat{\calC}[N]\Delta t}\oket{k_1}
\obra{k_1}e^{-\hat{\calC}[N]\Delta t}\oket{k'},
\end{eqnarray}
where $\Delta t\equiv t/M$.
For a fixed $t$, we can always make $M$ big enough such that $\Delta t$ smaller than any given number. When $\Delta t$ is sufficiently small, we have
\begin{equation}\label{exp expansion}
\obra{k_{n+1}}e^{-\hat{\calC}[N]\Delta t}\oket{k_n}
\approx\delta_{k_{n+1},k_n} -\Delta t\, N(v) \obra{k_{n+1}}\,\hat{\calC}[1]\oket{k_n},
\end{equation}
where $\obra{k_{n+1}}\,\hat{\calC}[N]\oket{k_n}\equiv N(v)\obra{k_{n+1}}\,\hat{\calC}[1]\oket{k_n}$ is nonvanishing only between two s-knots $k_n,k_{n+1}$ that differ by the action of $\hat{\calC}$ at a node $v$. As discussed in \sref{sec:solutions to the scalar constraint}, the action of $\hat{\calC}$ is to add or remove one or two extraordinary loops and also modify the colorings. Therefore, while $\delta_{k_{n+1},k_n}$ keeps $k_{n+1}$ to be the same as $k_n$, the action of $\obra{k_{n+1}}\,\hat{\calC}[N]\oket{k_n}$ is on the nodes of s-knots as schematically illustrated in \fref{fig:action of hat C}.
Consequently, for $k\neq k'$, \eref{W(k,k') product of steps 1} leads to
\begin{eqnarray}\label{W(k,k') product of steps 2}
W(k,k') &=&  \lim_{\Delta t\rightarrow0} \sum_{N=0}^\infty (-\Delta t)^{N+1}
\sum_{k_1,\dots,k_N \atop k_{n+1}\neq k_n}
\obra{k}{\hat{\calC}[1]}\oket{k_N} \obra{k_N}{\hat{\calC}[1]}\oket{k_{N-1}}\nonumber\\
&&\qquad\qquad\qquad\qquad\qquad\qquad
\cdots
\obra{k_2}{\hat{\calC}[1]}\oket{k_1}
\obra{k_1}{\hat{\calC}[1]}\oket{k'},
\end{eqnarray}
where we have absorbed the factor $N(v)$ into $\Delta t$ and also identify $k'\equiv k_0$ and $k\equiv k_{N+1}$ for shorthand.

Now, recall that in \sref{sec:quantum scalar constraint}, as a consequence of the intimate interplay between 3d diffeomorphism invariance and 3d short-scale discreteness, the value of $\oinner{\Psi}{\,\hat{\calC}^\Eucl_{R_\epsilon}\Phi}$ in \eref{CEucl diff} turns out to be independent of $\epsilon$ once $\epsilon$ is sufficiently small and consequently the regulator for $\epsilon\rightarrow0$ can be removed. The regulator $\Delta t$ on the right-hand side of \eref{W(k,k') product of steps 2} is analogous to $\epsilon$ in the sense that reparametrization invariance in time now plays the role of 3d diffeomorphism invariance. That is, as long as $\Delta t$ is small enough, making it smaller will not change anything further. In addition to discreteness of space at the Planck length scale, let us postulate also discreteness of time at the Planck time scale. It is then suggested, somehow by the interplay between reparametrization invariance and short-scale discreteness in time, that the dependence on $\Delta t$ in \eref{W(k,k') product of steps 2} should go away as $\Delta t$ becomes sufficiently small and consequently the regulator for $\Delta t\rightarrow0$ should be removed.\footnote{The short-scale discreteness in time is only postulated. Whether the canonical theory of LQG gives rise to temporal discreteness is unknown. The suggestion is only heuristic. Rigourously, to make sense of it, we might have to reformulate the regulator $R_\epsilon$ in \sref{sec:quantum scalar constraint} such that it is intricately matched with the lapse function $N$ in a 4d diffeomorphism covariant manner.}

By removing the regulator $\Delta t$, \eref{W(k,k') product of steps 2} then leads to
\begin{equation}\label{W(k,k')}
W(k,k') = \sum_{N=0}^\infty w(N) \sum_{\sigma^{(N)}} A(\sigma^{(N)})
\equiv \sum_{\sigma=(k,\dots,k') \atop k_{n+1}\neq k_n} w\left(N(\sigma)\right)\, A(\sigma),
\end{equation}
where $w(N)$ is a weight factor for different $N$ arising from $(-\Delta t)^{N+1}$, and $\sigma^{(N)}:=(k\equiv k_{N+1},k_N,k_{N-1},\dots,k_1,k'\equiv k_0)$ with $k_{n+1}\neq k_n$ is a discrete sequence of s-knots, which represents a ``history'' from $k'$ to $k$ with $N+1$ times of intermediate state change. The amplitude associated with a particular history $\sigma=(k,\cdots,k_{v+1},k_v,\cdots,k')$ is given by
\begin{equation}\label{A(sigma)}
A(\sigma) = \prod_v A_v(\sigma)
\end{equation}
where $v$ labels the steps of the history, and $A_v(\sigma)$ is determined by
\begin{equation}\label{Av(sigma)}
A_v(\sigma) \sim \obra{k_{v+1}}\,\hat{\calC}[1]\oket{k_v}
\end{equation}
up to a constant proportional factor that can be absorbed into rescaling of $\Delta t$.
In \eref{W(k,k')}, the transition amplitude is cast in a sum-over-histories formulation, which is not a functional integral over continuous histories of fields but a sum over discrete histories of s-knots.

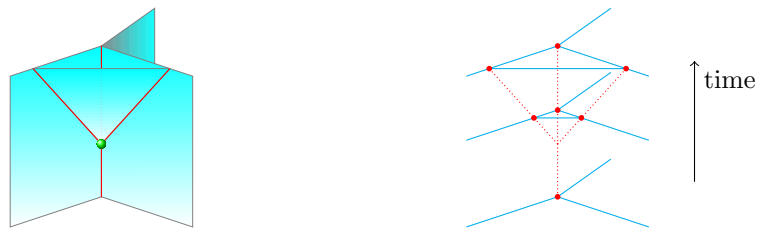
\begin{figure}

\center

\begin{tikzpicture}

%-------------------------------------------
%\draw[help lines] (-3,-2) grid (10,2);

%==============================================================================
%=== PART I:

\begin{scope}[shift={(0,0)},scale=1]
%--- 1st face:
\path[draw,help lines,right color=cyan] (0.7,-0.5) -- (0.7,1.5) -- (0,1) -- (0,-1) -- cycle;
%--- 2nd face:
\path[draw,help lines,top color=cyan] (0,-1) -- (0,1) -- (-1.2,0.6) -- (-1.2,-1.4) -- cycle;
%--- 3rd face:
\path[draw,help lines,top color=cyan] (0,-1) -- (0,1) -- (1.2,0.6) -- (1.2,-1.4) -- cycle;
%--- edge:
\draw [red] (0,-1) -- (0,1);
%--- 4th face:
\path[draw,help lines,top color=cyan] (0,-0.3) -- (0.9,0.7) -- (-0.9,0.7) -- cycle;
%--- edges:
\draw [red] (0,-0.3) -- (0.9,0.7);
\draw [red] (0,-0.3) -- (-0.9,0.7);

%\draw [blue] (1.2,0) -- (0.484615, 0.238462);
%\draw [blue] (1.2,-0.3) -- (0.276923, 0.00769231);
%\draw [blue] (1.2,-0.2) -- (0.346154, 0.0846154);
%\draw [blue] (1.2,-0.25) -- (0.311538, 0.0461538);

%--- vertex:
\shade[shading=ball, ball color=green] (0,-0.3) circle (0.06);
\end{scope}

%==============================================================================
%=== PART II:
\begin{scope}[shift={(6,0)},scale=1]

 %--- edges (worldlines of nodes):
 \draw [densely dotted,red] (0,-1) -- (0,1);
 \draw [densely dotted,red] (0,-0.3) -- (0.9,0.7);
 \draw [densely dotted,red] (0,-0.3) -- (-0.9,0.7);

 %--- top s-knot:
 \begin{scope}[shift={(0,0)}]
 \draw [cyan] (0,1) -- (1.2,0.6);
 \draw [cyan] (0,1) -- (-1.2,0.6);
 \draw [cyan] (0,1) -- (0.7,1.5);
 \draw [fill,red] (0,1) circle [radius=0.03];
 \end{scope}
 \draw [cyan] (0.9,0.7) -- (-0.9,0.7);
 \draw [fill,red] (0.9,0.7) circle [radius=0.03] (-0.9,0.7) circle [radius=0.03];

 %--- middle s-knot:
 \begin{scope}[shift={(0,-0.85)}]
 \draw [cyan] (0,1) -- (1.2,0.6);
 \draw [cyan] (0,1) -- (-1.2,0.6);
 \draw [cyan] (0,1) -- (0.7,1.5);
 \draw [fill,red] (0,1) circle [radius=0.03];
 \end{scope}
 \draw [cyan] (0.311538, 0.0461538) -- (-0.311538, 0.0461538);
 \draw [fill,red] (0.311538, 0.0461538) circle [radius=0.03] (-0.311538, 0.0461538) circle [radius=0.03];

 %--- bottom s-knot:
 \begin{scope}[shift={(0,-2)}]
 \draw [cyan] (0,1) -- (1.2,0.6);
 \draw [cyan] (0,1) -- (-1.2,0.6);
 \draw [cyan] (0,1) -- (0.7,1.5);
 \draw [fill,red] (0,1) circle [radius=0.03];
 \end{scope}

%--- time arrow:
\draw [->] (1.8,-0.8) -- (1.8,0.8) node [below right] {time};

\end{scope}

\end{tikzpicture}

\caption{\emph{Left}: A typical spin foam vertex (colorings of adjacent edges and faces are not indicated). \emph{Right}: Time-slicing (foliation) of the left diagram gives the evolution of s-knots (time is chosen to to be vertically upward). The trivalent node on the bottom is branched into three trivalent nodes on the top.}
\label{fig:spin foam vertex}

\end{figure}

\begin{figure}

\center

\begin{tikzpicture}

%-------------------------------------------
%\draw[help lines] (-3,-3) grid (10,3);

%==============================================================================
%=== PART I:

%-------------------------------------------
%--- bottom s-knot:
\begin{scope}[shift={(0.5,-2)},scale=1]
\draw [cyan] (0, 0) ellipse (1.2 and 0.4);
\draw [cyan] (-0.5,-0.363624) to [out=20,in=200] (0.5,0.363624);
\end{scope}

%--- 1st face:
\path[top color=cyan] (0,0.4+2) to [out=200,in=20] (-0.8,-0.298142+2) to [out=280,in=80] (-0.3,-1) to [out=260,in=110] (-0.5+0.5,-0.363624-2) to [out=20,in=200] (0.5+0.5,0.363624-2) to [out=120,in=275] (0.5,0.5) to [out=95,in=290] (0,0.4+2);

%--- edges:
\draw [red] (-0.8,-0.298142+2) to [out=280,in=80] (-0.3,-1) to [out=260,in=110] (-0.5+0.5,-0.363624-2);
\draw [red] (0.5+0.5,0.363624-2) to [out=120,in=275] (0.5,0.5) to [out=95,in=290] (0,0.4+2);

%--- 2nd face:
\path [draw,red,top color=cyan] plot [smooth,tension=1] coordinates {(-0.3,-1) (1.37,-0.22) (0.5,0.5)} to [out=220,in=90] (0.2,-0.1) to [out=270,in=50] (-0.3,-1);
%\draw [fill,red] (1.4,-0.1) circle (0.01);

%--- vertices:
\shade[shading=ball, ball color=green] (-0.3,-1) circle (0.06);
\shade[shading=ball, ball color=green] (0.5,0.5) circle (0.06);

%--- test:
%\shade[shading=ball, ball color=blue] (1.2,0.1) circle (0.06);
%\shade[shading=ball, ball color=blue] (0.2,-0.1) circle (0.06);

%-------------------------------------------
%--- top s-knot:
\begin{scope}[shift={(0,2)},scale=1]
\draw [cyan] (0, 0) ellipse (1.2 and 0.4);
\draw [cyan] (-0.8,-0.298142) to [out=20,in=200] (0,0.4);
\end{scope}

%-------------------------------------------
%--- re-draw the middle link of the bottom s-knot:
\begin{scope}[shift={(0.5,-2)},scale=1]
\draw [cyan] (-0.5,-0.363624) to [out=20,in=200] (0.5,0.363624);
\end{scope}

%---cylinder profile lines:
\draw [cyan] (-1.2,2) to [out=280,in=90] (-0.8,0) to [out=270,in=100] (-1.2+0.5,0-2);
\draw [cyan] (1.2,2) to [out=280,in=90] (1.4,-0.1) to [out=270,in=100] (1.2+0.5,0-2);

%==============================================================================
%======= PART II:

\begin{scope}[shift={(4.2,0)},scale=1]

%--- edges:
\draw [densely dotted,red] (-0.8,-0.298142+2) to [out=280,in=80] (-0.3,-1) to [out=260,in=110] (-0.5+0.5,-0.363624-2);
\draw [densely dotted,red] (0.5+0.5,0.363624-2) to [out=120,in=275] (0.5,0.5) to [out=95,in=290] (0,0.4+2);

\draw [densely dotted,red] plot [smooth,tension=1] coordinates {(-0.3,-1) (1.37,-0.22) (0.5,0.5)} to [out=220,in=90] (0.2,-0.1) to [out=270,in=50] (-0.3,-1);

%-------------------------------------------
%--- top s-knot:
\begin{scope}[shift={(0,2)},scale=1]
\draw [cyan] (0,0) ellipse (1.2 and 0.4);
\draw [cyan] (-0.8,-0.298142) to [out=20,in=200] (0,0.4);
\draw [fill,red] (-0.8,-0.298142) circle [radius=0.03];
\draw [fill,red] (0,0.4) circle [radius=0.03];
\node [right] at (1.4,0) {$\obra{k}$};
\end{scope}

%-------------------------------------------
%--- bottom s-knot:
\begin{scope}[shift={(0.5,-2)},scale=1]
\draw [cyan] (0, 0) ellipse (1.2 and 0.4);
\draw [cyan] (-0.5,-0.363624) to [out=20,in=200] (0.5,0.363624);
\draw [fill,red] (-0.5,-0.363624) circle [radius=0.03];
\draw [fill,red] (0.5,0.363624) circle [radius=0.03];
\node [right] at (1.4,0) {$\obra{k'}$};
\end{scope}

%-------------------------------------------
%--- middle s-knot:
\begin{scope}[shift={(0.25+1.4-0.25-1.1,-0.1)},scale=1]
\draw [cyan] (0,0) ellipse (1.1 and 0.4);
\draw [cyan] (-0.55,-0.34641) to [out=30,in=270-35] (-0.1,0.05) to [out=90-35,in=180+30] (0.22,0.391158);
\draw [cyan] (-0.1,0.05) to [out=-20,in=160] (1.1,0);
\draw [fill,red] (-0.55,-0.34641) circle [radius=0.03];
\draw [fill,red] (-0.1,0.05) circle [radius=0.03];
\draw [fill,red] (0.22,0.391158) circle [radius=0.03];
\draw [fill,red] (1.1,0) circle [radius=0.03];
\end{scope}

%--- time arrow:
%\draw [->] (2,-0.8) -- (2,0.8) node [below right] {time};

\end{scope}

%==============================================================================
%======= PART III: I and II combined:

%-------------------------------------------
%--- move PART I:
\begin{scope}[shift={(8.4,0)},scale=1]

%--- 1st face:
\path[top color=cyan] (0,0.4+2) to [out=200,in=20] (-0.8,-0.298142+2) to [out=280,in=80] (-0.3,-1) to [out=260,in=110] (-0.5+0.5,-0.363624-2) to [out=20,in=200] (0.5+0.5,0.363624-2) to [out=120,in=275] (0.5,0.5) to [out=95,in=290] (0,0.4+2);

%--- 2nd face:
\path [draw,red,top color=cyan] plot [smooth,tension=1] coordinates {(-0.3,-1) (1.37,-0.22) (0.5,0.5)} to [out=220,in=90] (0.2,-0.1) to [out=270,in=50] (-0.3,-1);
%\draw [fill,red] (1.4,-0.1) circle (0.01);

%--- vertices:
%\shade[shading=ball, ball color=green] (-0.3,-1) circle (0.06);
%\shade[shading=ball, ball color=green] (0.5,0.5) circle (0.06);

%--- edges:
\draw [red] (-0.8,-0.298142+2) to [out=280,in=80] (-0.3,-1) to [out=260,in=110] (-0.5+0.5,-0.363624-2);
\draw [red] (0.5+0.5,0.363624-2) to [out=120,in=275] (0.5,0.5) to [out=95,in=290] (0,0.4+2);

%---cylinder profile lines:
\draw [cyan] (-1.2,2) to [out=280,in=90] (-0.8,0) to [out=270,in=100] (-1.2+0.5,0-2);
\draw [cyan] (1.2,2) to [out=280,in=90] (1.4,-0.1) to [out=270,in=100] (1.2+0.5,0-2);

\end{scope}

%-------------------------------------------
%--- move PART II:
\begin{scope}[shift={(8.4,0)},scale=1]

%-------------------------------------------
%--- top s-knot:
\begin{scope}[shift={(0,2)},scale=1]
\draw [cyan] (0, 0) ellipse (1.2 and 0.4);
\draw [cyan] (-0.8,-0.298142) to [out=20,in=200] (0,0.4);
%\draw [fill,red] (-0.8,-0.298142) circle [radius=0.03];
%\draw [fill,red] (0,0.4) circle [radius=0.03];
\end{scope}

%-------------------------------------------
%--- bottom s-knot:
\begin{scope}[shift={(0.5,-2)},scale=1]
\draw [cyan] (0, 0) ellipse (1.2 and 0.4);
\draw [cyan] (-0.5,-0.363624) to [out=20,in=200] (0.5,0.363624);
%\draw [fill,red] (-0.5,-0.363624) circle [radius=0.03];
%\draw [fill,red] (0.5,0.363624) circle [radius=0.03];
\end{scope}

%-------------------------------------------
%--- middle s-knot:
\begin{scope}[shift={(0.25+1.4-0.25-1.1,-0.1)},scale=1]
\draw [cyan] (0,0) ellipse (1.1 and 0.4);
\draw [cyan] (-0.55,-0.34641) to [out=30,in=270-35] (-0.1,0.05) to [out=90-35,in=180+30] (0.22,0.391158);
\draw [cyan] (-0.1,0.05) to [out=-20,in=160] (1.1,0);
%\draw [fill,red] (-0.55,-0.34641) circle [radius=0.03];
%\draw [fill,red] (-0.1,0.05) circle [radius=0.03];
%\draw [fill,red] (0.22,0.391158) circle [radius=0.03];
%\draw [fill,red] (1.1,0) circle [radius=0.03];
\end{scope}

\end{scope}

\end{tikzpicture}

\caption{\emph{Left}: A simple spin foam composed of 2 vertices, 6 edges, and 6 faces (3 faces inside the cylinder are shaded; the other 3 on the surface of the cylinder are not). Its top and bottom boundaries, $\obra{k}$ and $\obra{k'}$, are given by two $\theta$-shaped s-knots. \emph{Middle}: Time-slicing of the left spin foam shows the evolution from $\obra{k'}$ to $\obra{k}$ through an intermediate s-knot state. \emph{Right}: The left and right diagrams overlapped for better visualization.}
\label{fig:simple spin foam}

\end{figure}
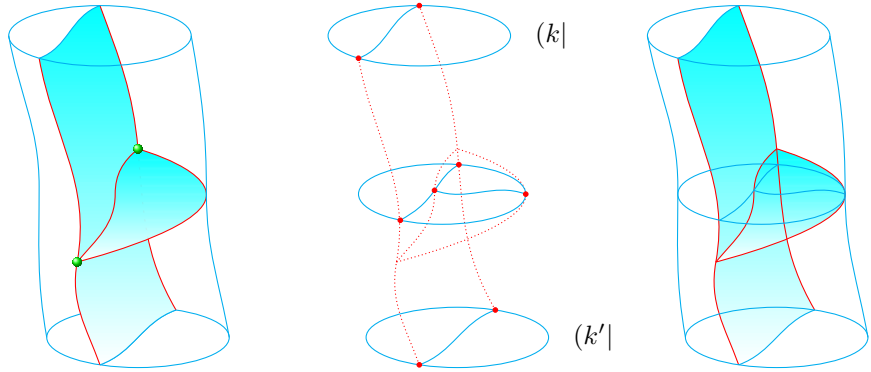

A history $\sigma=(k,\cdots,k_{v+1},k_v,\cdots,k')$, $k_{n+1}\neq k_n$, is called a \emph{spin foam}. More precisely, imagine that a graph of an s-knot in an abstract 4d space moves upward along the ``time'' direction and the graph is changed by branching of its edges at each step under the action of $\hat{\calC}$; the worldsheet swept out by the moving and changing graph is the spin foam. Call ``faces'' and denote by $f$ the worldsurfaces of the links of the graph; call ``edges'' and denote by $e$ the worldlines of the nodes of the graph; and call ``vertices'' and denote by $v$ the points at which the edges branch. \Fref{fig:spin foam vertex} depicts a vertex of a spin foam corresponding to the action in \fref{fig:action of hat C}, and \fref{fig:simple spin foam} shows a simple examples of spin foams.

%------------------------------------------------------------------------------
%------------------------------------------------------------------------------

The combinatorial object defined by the collection and the adjacency relation of faces $f$, edges $e$, and vertices $v$ is called a ``two-complex'' and denoted by $\Gamma$. As an s-knot is identified not only by its graph but also the colorings of its links (irreducible representations) and nodes (intertwiners), accordingly a spin form denoted by
\begin{equation}
\sigma=(\Gamma,j_f,i_e)
\end{equation}
is determined by a two-complex $\Gamma$, the coloring of irreducible representations $j_f$ associated with faces $f$, and the coloring of intertwiners $i_e$ associated with edges $e$.

\subsection{Spin foam formalism}
The discussion in the previous subsection motivates a sum-over-histories formulation of QG. For a given spin foam $\sigma=(\Gamma,j_f,i_e)$, it is natural to implement $A_v(\sigma)$ in \eref{Av(sigma)} as a function $A_v(j_f,i_e)$, called the vertex amplitude associated with the vertex $v$, where $j_f$ and $i_e$ are the colorings for the faces and edges adjacent to $v$. Consequently, Eqs.~(\ref{W(k,k')}) and (\ref{A(sigma)}) lead to the formal spin foam expression of the amplitude transition:
\begin{equation}\label{spin foam 1}
W(k,k') = \sum_{\partial\sigma=k\cup k'}
w\left(\Gamma(\sigma)\right)
\sum_{j_f,i_e}\prod_v
A_v(j_f,i_e),
\end{equation}
where $\partial\sigma=k\cup k'$ indicates that the boundary (i.e., initial and final s-knots) of $\sigma$ is given by the union of $k'$ and $k$, and $w(\Gamma)$ is a weight factor that depends only on the two-complex. The weight factor $w(\Gamma)$ is a generalized form of $w(N)$, where $N=N(\Gamma)$ is the number of vertices of $\Gamma$.

It is often more convenient to recast \eref{spin foam 1} in the extended form
\begin{equation}\label{spin foam 2}
W(k,k') = \sum_{\partial\sigma=k\cup k'}
w\left(\Gamma(\sigma)\right)
\sum_{j_f,i_e}
\prod_f A_f(j_f)
\prod_e A_e(j_f,i_e)
\prod_v A_v(j_f,i_e),
\end{equation}
where $A_f$ and $A_e$ are the amplitudes associated with faces and edges, which in principle can be absorbed into a redefinition of $A_v$.
For most models in the literature, $A_f(j_f)$ is simply given by the dimension $\dim(j_f)$ of the irreducible representation $j_f$. Consequently the transition amplitude is given in the form
\begin{equation}\label{spin foam 3}
W(k,k') = \sum_{\partial\sigma=k\cup k'}
w\left(\Gamma(\sigma)\right)
\sum_{j_f,i_e}
\prod_f \dim(j_f)
\prod_e A_e(j_f,i_e)
\prod_v A_v(j_f,i_e),
\end{equation}
and accordingly the partition function is given by
\begin{equation}\label{spin foam 4}
Z = \sum_{\sigma}
w\left(\Gamma(\sigma)\right)
\sum_{j_f,i_e}
\prod_f \dim(j_f)
\prod_e A_e(j_f,i_e)
\prod_v A_v(j_f,i_e).
\end{equation}
A spin foam model is then formally defined by the specification of
\begin{enumerate}

\item a category of two-complexes $\Gamma$ and the weight factor $w(\Gamma)$.

\item a Lie (or deformed Lie) group and associated irreducible representations $j_f$ and intertwiners $i_e$.

\item functions of the vertex amplitude $A_v(j_f,i_e)$ and the edge amplitude $A_e(j_e,i_e)$.

\end{enumerate}

There are various spin foam models that have been elaborated\cite{Perez:2003vx,Oriti:2003wf,Perez:2004hj,Perez:2012wv}. They provide tentative quantum theories of 3d GR, 4d BF theories, 4d Euclidean GR, etc. A remarkable feature is that many very different models all coincide with the form of \eref{spin foam 4}, suggesting that this expression can be viewed as a general background-independent formalism of covariant QFTs (also see \sref{sec:covariant LQG}).

\section{Black hole thermodynamics}\label{sec:black hole thermodynamics}
From theorems proved by Hawking \textit{et al.}\cite{Hawking:1971vc,Bardeen:1973gs}, which reveal a remarkable resemblance between a set of laws obeyed by black holes and the principles of thermodynamics, Bekenstein suggested that a Schwarzschild black hole should carries an entropy proportional to the area of its event horizon divided by the Planck area\cite{Bekenstein:1972tm,Bekenstein:1973ur,Bekenstein:1974ax}:
\begin{equation}\label{S-BH a}
S_\mathrm{BH} = a\, k_\mathrm{B}\frac{A}{G\hbar}
\end{equation}
where $k_\mathrm{B}$ is Boltzmann's constant, $a$ is a proportional constant, and $A=4\pi(2GM)^2$ is the area of the event horizon with $M$ being the mass of the black hole.
By the standard thermodynamical relation $T^{-1}=dS/dE\equiv dS/dM$, this implies that the black hole has a finite temperature
\begin{equation}
T = \frac{\hbar}{a32\pi k_\mathrm{B} GM}.
\end{equation}
Shortly after Bekenstein's conjecture, by studying QFT in a gravitational background, Hawking showed that black holes emit thermal Hawking radiation\cite{Hawking:1974rv,Hawking:1974sw} corresponding to the Hawking temperature
\begin{equation}
T=\frac{\hbar}{8\pi k_\mathrm{B}GM},
\end{equation}
which tightly confirms Bekenstein's conjecture and fixes the constant $a=1/4$. \Eref{S-BH a} with $a=1/4$ is often referred to as the Bekenstein-Hawking formula:
\begin{equation}\label{S-BH}
S_\mathrm{BH} = k_\mathrm{B}\frac{A}{4G\hbar},
\end{equation}
(where the subscript ``BH'' coincidentally stands for both ``black hole'' and ``Bekenstein-Hawking'').

In the modern interpretation of entropy in statistical mechanics, entropy (divided by $k_\mathrm{B}$) is defined as the logarithm of the number of admissible microstates for a given macroscopic state. What then are the microscopic degrees of freedom responsible for the black hole entropy? Can we derive \eref{S-BH} from first principles? As $\hbar$ enters in \eref{S-BH}, answers to these questions require a quantum theory of gravity. In LQG, a detailed description of black hole thermodynamics has become an active direction of research. One of the major achievements of LQG is to derive the Bekenstein-Hawking formula from the first principles for the Schwarzschild and other black holes\cite{Ashtekar:1997yu,Ashtekar:1999wa,Ashtekar:2000eq,Ashtekar:2000hw, Ashtekar:2001is,Ashtekar:2001jb,Ashtekar:2003jh,Ashtekar:2003zx,Ashtekar:2004nd}.

In Secs.~\ref{sec:statistical ensemble} and \ref{sec:Bekenstein-Hawking entropy}, we present the basic ideas, following Refs.~\cite{Rovelli:1996dv,Rovelli:1996ti} (also see Sec.~8.2 of Ref.~\cite{Rovelli:2004tv}). Readers are referred to Chap.~15 of Ref.~\cite{Thiemann:2007zz} for a rigourous treatment and Sec.~8 of Ref.~\cite{Ashtekar:2004eh} for a shorter account. More recent advances in the black hole entropy can be found in the excellent review paper Ref.~\cite{BarberoG.:2012ae}; we mention some of the important results in \sref{sec:more on black hole entropy}. Also see \cref{ch:Carlip} for other aspects of black hole thermodynamics.

\subsection{Statistical ensemble}\label{sec:statistical ensemble}
The microscopic degrees of freedom responsible for the black hole entropy $S_\mathrm{BH}$ are those that can participate in the energy exchanges with the exterior. Because, by definition of black holes, the states of gravity and matter inside a black hole have no effect on the exterior, the microstates of the interior of the black hole are irrelevant for $S_\mathrm{BH}$ measured from the exterior.\footnote{For instance, in the Kruskal extension of a Schwarzschild black hole, the region III may contains billions of galaxies, but these do not bear any detectable consequences for us in the region I.} Observed from the exterior, therefore, the entropy of the black hole (for its thermal interaction with the surrounding exterior) is completely determined by the geometry of the event horizon.

More precisely, the microcanonical ensemble is composed of the microstates of the horizon geometry that give rise to an exactly specified total area. Denote by $N(A)$ the number of the microcanonical ensemble of area $A$, the quantity $S(A)=k_\mathrm{B}\ln N(A)$ is then the entropy for the black hole as far as its thermal interaction with the exterior is concerned.

Given any arbitrary surface (say, not a horizon of any kind), we can of course ask how many microstates of the surface correspond to a specific area, but the number of the microstates usually has no thermodynamical significance. As heat or information can flow across a surface without changing its geometry, the geometry of a surface and thus the number of its microstates in general have nothing to do with the heat exchange or the notion of entropy. As originally argued by Jacobson\cite{Jacobson:1995ab}, it is for any \emph{causal horizon} that the heat flow always accompanies a stress-energy tensor that distorts the horizon geometry and consequently the entropy measured by the observer who is separated from the system by the causal barrier is given by the geometry of the causal horizon (see also Refs.~\cite{Wall:2010cj,Bianchi:2013rya} and see Ref.~\cite{Bianchi:2012ui} in the context of LQG). The black hole entropy given by the geometry of the event horizon is a special case of Jacobson's argument.

\subsection{Bekenstein-Hawking entropy}\label{sec:Bekenstein-Hawking entropy}
Let us now compute the number $N(A)$ from the quantum geometry of LQG. First, consider that the quantum state of the geometry of a 3d ADM leaf $\Sigma_t$ is given by an spin network state $\ket{s}$. The horizon is a 2d surface $\mathcal{S}$ imbedded in $\Sigma_t$ and its geometry is determined by its intersections with the spin network $\ket{s}$. As we are interested in the geometry as probed from the exterior, the surface $\mathcal{S}$ we consider is, more rigorously, the one that is outside but infinitesimally close to the horizon. Therefore, we exclude the possibility that the spin network nodes lie on $\mathcal{S}$ and, instead of \eref{area spectrum generic}, the area of $\mathcal{S}$ is simply given by \eref{area spectrum}
\begin{equation}\label{A for horizon}
{A}(\mathcal{S}) = 8\pi G\hbar\gamma \sum_{i} \sqrt{j_i(j_i+1)}\,,
\end{equation}
where $j_1,\dots,j_n$ are the colorings of the links intersecting the surface $\mathcal{S}$.

For a given $A$, how many admissible microstates of the horizon are there then? From the perspective of an external observer, an admissible microstate corresponds to a possible choice of the sequence $j_1,\dots,j_n$ that gives rise to $A$ via \eref{A for horizon} and a possible way of ``ending'' the links to the exterior. A possible ``end'' of a link with coloring $j$ is simply a vector in the vector space of the $j$ representation, which has a ($2j+1$)-fold multiplicity. To sum up, the admissible microstates are obtained by considering all sequences $j_1,\dots,j_n$ that give the area $A$, and for each sequence there is a multiplicity of $\prod_{i}(2j_i+1)$ possible states.

For a large $A$ much greater than the Planck area $\Pl^2\equiv G\hbar$, we assume that the number of admissible microstates is dominated by the case of $j_i=1/2$ for all intersecting links (we will come back to this assumption shortly). In this case, the area of a single link is given by
\begin{equation}
A_{j=1/2} = 4\pi\sqrt{3}\, G\hbar\gamma
\end{equation}
and thus the number of intersections is
\begin{equation}
n = \frac{A}{A_{j=1/2}} = \frac{A}{4\pi\sqrt{3}\, G\hbar\gamma}.
\end{equation}
Each $j=1/2$ link has a multiplicity of $(2j+1)=2$, so the total number of admissible microstates for a given large $A$ is given by
\begin{equation}
N(A) = 2^n = 2^{\frac{A}{4\pi\sqrt{3}\, G\hbar\gamma}}.
\end{equation}
Consequently, we obtain the entropy of the black hole
\begin{equation}
S_\mathrm{BH} = k_\mathrm{B}\ln N(A)
=\frac{1}{\gamma}\frac{\ln 2}{4\pi\sqrt{3}}\, k_\mathrm{B} \frac{A}{G\hbar},
\end{equation}
which agrees perfectly with the Bekenstein-Hawking entropy \eref{S-BH} provided that the Barbero-Immirzi parameter is fixed at the value
\begin{equation}\label{gamma value naive}
\gamma=\frac{\ln 2}{\pi\sqrt{3}} = 0.127384\dots
\end{equation}
The Bekenstein-Hawking entropy can be calculated for different kinds of black holes by LQG. The striking feature is that they all lead to the same value of $\gamma$ and thus ensure the consistency of the black hole thermodynamics under the framework of LQG.

We have made the assumption that the leading contribution to the entropy comes entirely from the lowest states ($j_i=1/2$), as had been suggested in Ref.~\cite{Ashtekar:2000eq} and considered true for several years. The detailed computation of Refs.~\cite{Domagala:2004jt,Meissner:2004ju} based on a practical rephrasing of the combinatorial problem pointed out that in fact all quantum states (not only the lowest ones) must be taken into account. This leads to a revision from \eref{gamma value naive} for the value of $\gamma$, which can be numerically calculated at arbitrary accuracy\cite{Meissner:2004ju} as
\begin{equation}\label{gamma value}
\gamma = \gamma_\mathrm{M} \equiv 0.23753295796592\dots
\end{equation}
The simple calculation presented above on the flawed assumption nevertheless grasps the conceptual essence of relating the number of microstates to the black hole entropy.

\subsection{More on black hole entropy}\label{sec:more on black hole entropy}
Section~\ref{sec:Bekenstein-Hawking entropy} gives a simple account for the black hole entropy from first principles of LQG. A more rigorous approach is the modeling of black holes in LQG by using \emph{isolated horizons} as inner boundaries\cite{Ashtekar:1997yu,Ashtekar:1999wa,Ashtekar:2000eq}. Quantization of the degrees of freedom for the isolated horizon can be understood as a Chern-Simons theory. Obtained by tensoring the Chern-Simons boundary Hilbert space and the LQG bulk Hilbert space, the Hilbert space of the resulting model provides the groundwork to address the statistical problems of black holes.

The combinatorial problem associated with the computation of the black hole entropy can be rephrased in a more manageable way\cite{Domagala:2004jt,Meissner:2004ju}, allowing one to compute the asymptotic behavior of the black hole entropy in the limit of large horizon area. For $A\gg\Pl^2$, the computation gives an exact formula to the subleading order as
\begin{equation}
k_\mathrm{B}^{-1}S = \ln N(A)
= \frac{\gamma_\mathrm{M}}{4\gamma}\frac{A}{G\hbar}
-\frac{1}{2}\ln\frac{A}{G\hbar}
+O\left((A/\Pl^2)^0\right),
\end{equation}
where $\gamma_\mathrm{M}$ is a constant, whose numerical value is given by \eref{gamma value}.
This formula agrees with \eref{S-BH a} on the proportionality to $A$  in the leading order and also obtains the precise subleading order correction as a logarithmic function of $A$. Comparison to the Bekenstein-Hawking formula in \eref{S-BH} fixes the Barbero-Immirzi parameter $\gamma$ to be $\gamma=\gamma_\mathrm{M}$.

For small areas, the precise number counting was first suggested in Refs.~\cite{Corichi:2006bs,Corichi:2006wn} and later has been thoroughly investigated by employing number-theoretical and combinatorial methods (see Ref.~\cite{Agullo:2010zz} for a detailed account and Ref.~\cite{BarberoG.:2012ae} for a review). The key discovery is that, for microscopic black holes, the so-called black hole degeneracy spectrum when plotted as a function of the area exhibits a persistent ``periodicity'' (more precisely, a modulation with a regular period of growing magnitude). This produces an effectively evenly-spaced area spectrum, despite the fact that the area spectrum in LQG is not evenly spaced, and makes contact in a nontrivial way with the evenly-spaced black hole horizon area spectrum predicted by Bekenstein and Mukhanov under general conditions\cite{Bekenstein:1995ju}. The periodicity in the degeneracy spectrum leads to a striking ``staircase'' structure with regular steps when the black hole entropy is plotted as a function of the area. The staircase behavior eventually disappears in the large area limit.

The framework of precise number counting also predicts the subleading correction to the Bekenstein-Hawking law. For various model settings, the subleading correction generically takes the form $a_1k_\mathrm{B}\ln A/\Pl^2$, where the coefficient $a_1$ is independent of the value of $\gamma$ but \emph{differs} for different models. The logarithmic correction is qualitatively in agreement with those obtained by different approaches (including those based on asymptotic symmetries, horizon symmetries, and certain string theories), despite very different physical assumptions; there are some indications that even the coefficient $a_1$ might be universal, up to differences on the treatment of angular momentum and conserved charges\cite{Carlip:2000nv}.

Recently, an explicit $SU(2)$ formulation for the black hole entropy has been developed based on covariant Hamiltonian methods\cite{Engle:2009vc,Engle:2010kt,Engle:2011vf}. This formulation avoids the partial gauge fixing of the standard approach and gives rigorous support for the earlier proposal that the quantum black hole degrees of freedom could be described by an $SU(2)$ Chern-Simons theory.

\section{Loop quantum cosmology}\label{sec:LQC}
Loop quantum cosmology (LQC) is a finite, symmetry-reduced model of LQG that applies principles of the full theory to cosmological settings. Thanks to the mathematical simplifications, many obscure aspects (e.g.\ quantum dynamics and semiclassical physics) in the full theory become transparent in LQC. The framework of LQC provides a ``bottom-up'' approach to the full theory, in which many ideas of LQG can be explicitly implemented and tested.

The distinguishing feature of LQC is that the quantum geometry of LQG gives rise to a brand new quantum force that is inappreciable at low spacetime curvature but rises very rapidly and opposes the classical gravitational force in the Planck regime. As a consequence, for a variety of models of LQC, the cosmological singularity (big bang, big crunch, big rip, etc.) is avoided by the opposing force, therefore affirming the long-held conviction that singularities in GR signal a breakdown of the classical theory and should be resolved by the quantum effects of gravity. Particularly, the big bang singularity is replaced by a \emph{quantum bounce}, which bridges the present expanding universe with a preexistent contracting universe. The new cosmological scenario suggests a change of the paradigm in the standard big-bang cosmology.

In what follows, to illustrate the basic ideas of LQC, we consider the simplest setting---LQC in the $k=0$ Friedmann-Lema\^{\i}tre-Robertson-Walker (FLRW) model and particularly follow the lines of Ref.~\cite{Ashtekar:2006wn}. See Refs.~\cite{Bojowald:2008zzb,Ashtekar:2011ni,Bojowald:2011zzb,Bojowald:2011book} for more detailed construction and more models of LQC.

\subsection{Symmetry reduction}
The first step is to impose symmetries of cosmology---spatial homogeneity and oftentimes also isotropy---on the phase space variables \emph{before} quantization. Symmetry reduction deliberately ignores infinitely many degrees of freedom and thus brings up the question whether the results from the full theory of LQG, if can be obtained after all, should resemble any predictions by by LQC. The arguments in Sec.~1.2 of Ref.~\cite{Ashtekar:2011ni} suggest that the answer is likely to be affirmative, provided that the framework of LQC captures the essential features of the full theory.

In the $k=0$ FLRW model, both homogeneity and isotropy are assumed and the spacetime metric in the comoving coordinates is given by
\begin{equation}\label{FLRW metric}
d\tau^2 \equiv g_{\mu\nu} x^\mu dx^\nu = -N(t)dt^2 +a(t)^2dx^2.
\end{equation}
With the imposition of homogeneity and isotropy, the fundamental fields $A_a^i(x)$ and $\tE^a_i(x)$ are reduced to two variables $c$ and $p$:
\begin{eqnarray}\label{c and p}
A_a^i(x) \rightarrow c, \qquad
\tE^a_i(x) \rightarrow p,
\end{eqnarray}
which are independent of $x$ because of homogeneity and satisfy the canonical relation
\begin{equation}\label{PB of c and p}
\{c,p\}
=\frac{8}{3}\pi G\gamma,
\end{equation}
which is the symmetry-reduced counterpart of \eref{PB of A and E}. As $\tE^a_i$ is related to ``area'' in the sense of \eref{A(S)}, the symmetry-reduced variable $p$ represents the area of the surfaces of a finite-sized cubic cell $\mathcal{V}$ (which is chosen to make sense of the spatial integral $\int_\Sigma d^3x\rightarrow \int_\mathcal{V}d^3x$); more precisely,
\begin{equation}
p=a^2L^2,
\end{equation}
where $a$ is the scalar factor in \eref{FLRW metric} and $L$ is the coordinate length of the edges of $\mathcal{V}$.

Under the symmetry reduction, the Gauss constraint is trivially satisfied, and the diffeomorphism constraint is no longer an issue as the diffeomorphism invariance is gauge fixed in accord with the homogeneity. Only the scalar constraint remains significant. In the $k=0$ FLRW setting, we have $F=dA+A\wedge A=A\wedge A$ and $K=(A-\Gamma)/\gamma=A/\gamma$, which lead to $F=A\wedge A=\gamma^2 K\wedge K$, or equivalently ${\epsilon^{ij}}_k F_{ab}^k=2A^i_{[a}A^j_{b]}=2\gamma^2K^i_{[a}K^j_{b]}$. Correspondingly, \eref{C[N]} becomes
\begin{equation}\label{C(N) FLRW}
\calC[N]=-\frac{N}{8\pi G\gamma^2} \int_\mathcal{V} d^3x
\frac{\tE^a_i\tE^b_j}{\sqrt{\abs{q}}}
\,{\epsilon^{ij}}_k F_{ab}^k,
\equiv
-\frac{N}{8\pi G\gamma^2} \int_\mathcal{V} d^3x
\frac{\tE^a_i\tE^b_j}{\sqrt{\abs{q}}}
\,2A^i_{[a}A^j_{b]},
\end{equation}
where $N=N(t)$ is independent of $x$, and the spatial integral $\int_\Sigma d^3x$ is restricted to $\int_\mathcal{V} d^3x \equiv \int_0^Ldx_1\int_0^Ldx_2\int_0^Ldx_3$ to make $\calC[N]$ finite. By \eref{c and p}, it follows from \eref{C(N) FLRW} that the gravitational part of the scalar constraint in terms of $c$ and $p$ is given by
\begin{equation}\label{Cgrav cl}
C_\grav(N) = -\frac{6N}{8\pi G\gamma^2} \frac{p^2}{\sqrt{p^3}}\,c^2,
\end{equation}
where the lapse function $N(x,t)=N(t)$ is chosen to be homogeneous.

In the full theory, instead of $A(x)$ and $\tE(x)$, the holonomy $h_\gamma$ and the electric flux over a surface $E_{\mathcal{S},f}$ are used as the fundamental variables for loop quantization. In the symmetry-reduced theory, analogous to \eref{h gamma}, the variable $c$ is replaced by the ``holonomy'' of $c$, defined as
\begin{equation}
\mathcal{N}_\mu:=e^{i\mu c/2}
\end{equation}
for an arbitrary $\mu\in\mathbb{R}$. On the other hand, corresponding to \eref{E[S,f]}, $p$ remains the good variable, as it is redundant to specify $\mathcal{S}$ and $f(x)$ in the homogeneous and isotropic setting. The variables $\mathcal{N}_\mu$ and $p$ form the closed ``loop algebra'':
\begin{equation}
\{\mathcal{N}_\mu,p\}=i\frac{4\pi G\gamma}{3}\mu\,\mathcal{N}_\mu
\qquad\text{and}\qquad
\{\mathcal{N}_\mu,\mathcal{N}_\nu\}=0.
\end{equation}

The symmetry reduction into the $k=0$ FLRW model is summarised in \tref{tab:symmetry reduction}.

\begin{table}[ht]
\tbl{Symmetry reduction in the $k=0$ FLRW model.\label{tab:symmetry reduction}}
{\begin{tabular}{cc}
\toprule
%\Hline
%\Hline
in the full theory & in the $k=0$ FLRW model  \\
\colrule
$A_a^i(x)$ & $c$ \\
$\tE^a_i(x)$ & $p$ \\
$\{A_a^i(x),\tE^b_j(y)\}=8\pi G\gamma\,\delta^i_j\delta_a^b\delta^3(x-y)$ & $\{c,p\}=\frac{8}{3}\pi G\gamma$ \\
\colrule
$\calC_\mathrm{G}[\lambda]$ & --- \\
$\calC_\Diff[\vec{N}]$ & --- \\
$\calC[N]$ in \eref{C[N]} & $C_\grav(N)$ in \eref{Cgrav cl} \\
\colrule
holonomy: $h_\gamma$ in \eref{h gamma}  &  $\mathcal{N}_\mu:=e^{i\mu c/2},\ \mu\in\mathbb{R}$ \\
electric flux: $E_{\mathcal{S},f}$ in \eref{E[S,f]} & $p$ \\
loop algebra: $\{h_\gamma,E_{\mathcal{S},f}\}=\dots$ &
$\{\mathcal{N}_\mu,p\}=i\frac{4\pi G\gamma}{3}\mu\mathcal{N}_\mu$ \\
\botrule
\end{tabular}
}
%\begin{tabnote}
%$^{\text a}$Sample table footnote.
%\end{tabnote}\label{ra_tbl1}
\end{table}

\subsection{Quantum kinematics}
Analogous to the linear space $\Cyl$ of cylindrical functions of $A$, i.e.\ $\Psi_{\Gamma,f}[A]$, defined in \eref{cylindrical fun} in the full theory, we begin with the linear space $\Cyl^\mathrm{S}$ of \emph{almost periodic functions} of $c$ defined as
\begin{equation}
\Psi(c) = \sum_k \xi_k \mathcal{N}_{\mu_k}(c)
\equiv \sum_k \xi_k e^{i\mu_kc/2},
\end{equation}
where $k$ runs over a finite number of integers, $\mu_k\in\mathbb{R}$, and $\xi_k\in\mathbb{C}$.
Analogous to the inner product of $\Cyl$ defined in Eqs.~(\ref{Gamma f Gamma g}) and (\ref{Gamma' f' Gamma'' g''}), the inner product of $\Cyl^\mathrm{S}$ is defined via
\begin{equation}\label{inner product sym}
\inner{\mathcal{N}_{\mu_1}}{\mathcal{N}_{\mu_2}}
\equiv \inner{\mu_1}{\mu_2} =\delta_{\mu_1\mu_2},
\end{equation}
where the right-hand side is the Kronecker delta, but $\mu_1,\mu_2\in\mathbb{R}$ take continuous values. This inner product is very different from that defined via $\inner{\mu_1}{\mu_2} =\delta(\mu_1-\mu_2)$ with the Dirac delta function. To highlight this nuance, $\Cyl^\mathrm{S}$ is said to be in the ``polymer representation'' of $\mu$.

The (gravitational part) of the kinematical Hilbert space $\hil_\kin^\mathrm{S}$ is the Cauchy completion of $\Cyl^\mathrm{S}$ with respect to the inner product \eref{inner product sym}. That is, $\hil^\mathrm{S}_\kin=L^2(\mathbb{R}_\mathrm{Bohr},d\mu_\mathrm{Bohr})$, where $\mathbb{R}_\mathrm{Bohr}$ is the Bohr compactification of $\mathbb{R}$ and $d\mu_\mathrm{Bohr}$ is the corresponding measure. An orthonormal basis of $\hil^\mathrm{S}_\kin$ is given by $\{\mathcal{N}_\mu|\,\mu\in\mathbb{R}\}$. By Dirac's bra-ket notation, the map $c\mapsto e^{i\mu c/2}$ is denoted by $\ket{\mu}$ with
\begin{equation}
e^{i\frac{\mu c}{2}} = \inner{c}{\mu},
\end{equation}
and consequently we have
\begin{equation}
\ket{\Psi} = \sum_\mu \ket{\mu}\inner{\mu}{\Psi} \equiv \sum_\mu \Psi(\mu)\ket{\mu}.
\end{equation}

Analogous to the holonomy, area, and volume operators defined in \sref{sec:operators and quantum geometry}, we have corresponding operators acting on $\hil^\mathrm{S}_\kin$:
\begin{eqnarray}
\hat{\mathcal{N}}_\mu \Psi(c) &\equiv&  \widehat{e^{i\frac{\mu c}{2}}}\Psi(c) := e^{i\frac{\mu c}{2}}\Psi(c),\\
\hat{p}\, \Psi(c) &:=&  -i\frac{8\pi\gamma}{3}\,\Pl^2\frac{\partial\Psi(c)}{\partial c},\\
\hat{V} \Psi(c) &:=& \abs{\hat{p}}^{3/2} \Psi(c),
\end{eqnarray}
where $\Pl\equiv\sqrt{G\hbar}$ is the Planck length. When acting on the basis state $\ket{\mu}$, these operator behave as
\begin{eqnarray}
\label{op Nmu}
\hat{\mathcal{N}}_{\mu'}\ket{\mu} &\equiv& \widehat{e^{i\frac{\mu c}{2}}}\ket{\mu}  = \ket{\mu+\mu'},\\
\label{op p}
\hat{p}\, \ket{\mu} &=&  \frac{8\pi\gamma}{6}\,\Pl^2\,\mu\ket{\mu},\\
\label{op V}
\hat{V} \ket{\mu} &=& V_\mu\ket{\mu} \equiv \left(\frac{8\pi\gamma}{6}\abs{\mu}\right)^{3/2}\Pl^3\ket{\mu}.
\end{eqnarray}

The correspondence for quantum kinematics between the full theory and the symmetry-reduced theory is listed in \tref{tab:LQC kinematics}.

\begin{table}[ht]
\tbl{Quantum kinematics of LQG and LQC in the $k=0$ FLRW model.\label{tab:LQC kinematics}}
{\begin{tabular}{ll}
\toprule
%\Hline
%\Hline
\multicolumn{1}{c}{LQG} & \multicolumn{1}{c}{LQC in the $k=0$ FLRW model}\\
%LQG & $k=0$ FLRW LQC  \\
\colrule
$\Cyl$ & $\Cyl^\mathrm{S}$ \\
cylindrical functions: & almost periodic functions: \\
\quad $\Psi_{\Gamma,f}[A]$ in \eref{cylindrical fun} &
\quad$\Psi(c) = \sum_k \xi_k\mathcal{N}_{\mu_k} \equiv \sum_k \xi_k e^{i\mu_kc/2}$ \\
inner product: & inner product: \\
\quad$\inner{\Psi_{\Gamma,f}}{\Psi_{\Gamma',g'}}$ in Eqs.~(\ref{Gamma f Gamma g}), (\ref{Gamma' f' Gamma'' g''})& \quad$\inner{\mathcal{N}_{\mu_1}}{\mathcal{N}_{\mu_2}}
\equiv \inner{\mu_1}{\mu_2} =\delta_{\mu_1\mu_2}$ \\
$\hil=L^2[\mathcal{A},d\mu_\mathrm{AL}]$ & $\hil^\mathrm{S}_\kin=L^2(\mathbb{R}_\mathrm{Bohr},d\mu_\mathrm{Bohr})$\\
\colrule
$\ket{\Gamma,j_l,\alpha_l,\beta_l}$ in Eqs.~(\ref{basis states}), (\ref{basis states 2}) & \quad$\ket{\mu}$ \\
spin networks $\ket{\Gamma,j_l,i_n}$ & \quad $\ket{\mu}$\quad(Gauss constraint trivial)\\
s-knots $\obra{K,c}$ & \quad $\ket{\mu}$\quad(no diffeomorphism constraint) \\
\colrule
holonomy operator: $\hat{h}_\gamma$ in Eqs.~(\ref{hat h}), (\ref{hat S})  &  \quad$\hat{\mathcal{N}}_{\mu'}\ket{\mu} \equiv \widehat{e^{i\mu'c/2}}\ket{\mu} = \ket{\mu+\mu'}$ \\
area operator: $\hat{E}(\mathcal{S})$ in \eref{area spectrum generic} & \quad$\hat{p}\, \ket{\mu} =  \frac{8\pi\gamma}{6}\,\Pl^2\,\mu\ket{\mu}$ \\
volume operator: $\hat{V}(\mathcal{R})$ in \eref{V on spin network} &
\quad$\hat{V} \ket{\mu} =\left(\frac{8\pi\gamma}{6}\abs{\mu}\right)^{3/2}\!\Pl^3\ket{\mu} \equiv V_\mu\ket{\mu}$ \\
\botrule
\end{tabular}
}
%\begin{tabnote}
%$^{\text a}$Sample table footnote.
%\end{tabnote}\label{ra_tbl1}
\end{table}

\subsection{Quantum constraint operator}
Just as in the full theory, we have to regularize the scalar constraint before quantization. In \eref{C(N) FLRW}, the nonpolynomial factor $1/\sqrt{\abs{q}}$ is dealt with Thiemann's trick and $F_{ab}$ is replaced by a holonomy long a square loop $\oblong$. Corresponding to Eqs.~(\ref{C Repsilon}) and (\ref{CEucl Box}) (with $R^{(j)}=R^{(1/2)}$, $C^{iJ}=1$), it turns out $C_\grav$ with $N=1$ can be can be written as
\begin{subequations}
\begin{eqnarray}
C_\grav &=& \lim_{\lambda\rightarrow0}C^{(\lambda)}_\grav, \\
\label{Cgrav}
C^{(\lambda)}_\grav &=& -\frac{4\,\sgn(p)}{8\pi\gamma^3\lambda^3G} \sum_{ijk} \epsilon^{ijk}
\Tr \left( h_i^{(\lambda)} h_j^{(\lambda)} \big(h_i^{(\lambda)}\big)^{-1} \big(h_j^{(\lambda)}\big)^{-1}
h_k^{(\lambda)} \left\{\big(h_k^{(\lambda)}\big)^{-1}, V\right\}
\right) \nonumber\\
&=& \sin\lambda c\left[
-\frac{4\,\sgn(p)}{8\pi\gamma^3\lambda^3G} \sum_k
\Tr\, \tau_k\, h_k^{(\lambda)} \left\{\big(h_k^{(\lambda)}\big)^{-1}, V\right\}
\right] \sin\lambda c,
\end{eqnarray}
\end{subequations}
where
\begin{equation}
h_i^{(\lambda)}=\cos\frac{\lambda c}{2} + 2\tau_i\sin\frac{\lambda c}{2}
\end{equation}
is the holonomy along the edge of $\oblong$, which is in the $x^i$ direction and of coordinate length $\lambda L$, and $\lambda$ plays the role of $\epsilon$ in \eref{C Repsilon} (more precisely, $\lambda L=\epsilon$).

It is straightforward to obtain the quantum operator $\hat{C}^{(\lambda)}_\grav$ associated with $C^{(\lambda)}_\grav$, but the limit $\lambda\rightarrow0$ is ill-defined. Unlike the full theory of LQG, the dependence of $\hat{C}^{(\lambda)}_\grav$ on the regulator $\lambda$ does not go away because the diffeomorphism has been gauge fixed. In order to inherit the spatial discreteness from LQG appropriately, we take the prescription \emph{by hand} to shrink the square loop $\oblong$ to a finite area equal to the \emph{area gap} $\Delta$, i.e. the smallest nonzero eigenvalue of area, given by \eref{Delta}.\footnote{As this is only a phenomenological prescription, different numerical values for $\Delta$ of the same order of magnitude are also used in the literature.} Consequently, we are led to choose for $\lambda$ a specific function $\lambda=\mubar(p)$ given by
\begin{equation}
\text{Area of}\ \oblong=  a^2 (\lambda L)^2 = \mubar^2 \abs{p} =\Delta,
\end{equation}
or equivalently
\begin{equation}
\mubar = \sqrt{\frac{\Delta}{\abs{p}}} = \frac{\sqrt{3\sqrt{3}}}{\sqrt{2}} \frac{1}{\sqrt{\abs{\mu}}}.
\end{equation}

\Eref{op Nmu} suggests that, in a heuristical sense,
\begin{equation}
\widehat{e^{i\frac{\mubar c}{2}}}\Psi(\mu) = e^{\mubar\frac{d}{d\mu}}\Psi(\mu),
\end{equation}
even though the differential operator $\frac{d}{d\mu}$ is ill-defined in $L^2(\mathbb{R}_\mathrm{Bohr},d\mu_\mathrm{Bohr})$.
Let us define the affine parameter
\begin{equation}
v := K\,\sgn(\mu) \abs{\mu}^{3/2}, \quad \text{with}\ K=\frac{2\sqrt{2}}{3\sqrt{3\sqrt{3}}},
\end{equation}
such that formally
\begin{equation}
\mubar\frac{d}{d\mu} = \frac{d}{dv}.
\end{equation}
In the $v$-representation, the action of $\widehat{e^{i\frac{\mubar c}{2}}}$ can then be defined as
\begin{equation}
\widehat{e^{i\frac{\mubar c}{2}}}\Psi(v) = \Psi(v+1),
\end{equation}
and correspondingly \eref{op V} leads to
\begin{equation}
\hat{V}\ket{v} = \left(\frac{4\pi\gamma}{3}\right)^{3/2}\frac{\abs{v}}{K}\,\Pl^3\, \ket{v}.
\end{equation}
%===========================================
%Changing variables from $\mu$ to $v$ is trivial in the polymer representation in the sense that we simply have $L^2(\mathbb{R}_\mathrm{Bohr},d\mu_\mathrm{Bohr})=L^2(\mathbb{R}_\mathrm{Bohr},dv_\mathrm{Bohr})$ as opposed to $L^2(\mathbb{R},d\mu)=L^2(\mathbb{R},(\frac{d\mu}{dv})dv)$, which involves the factor $\frac{d\mu}{dv}$. Accordingly, \eref{inner product sym} implies
%\begin{equation}
%\inner{v_1}{v_2}=\delta_{v_1v_2}.
%\end{equation}

As $\widehat{e^{i\frac{\mubar c}{2}}}$ and $\hat{V}$ are now well defined, $C^{(\lambda)}_\grav$ in \eref{Cgrav} can be promoted to the quantum operator
\begin{equation}
\hat{C}_\grav = \sin(\mubar c)
\left[
\frac{3i\,\sgn(\mu)}{\pi\gamma^3\mubar^3\Pl^2}
\left(
\sin\frac{\mubar c}{2}\,\hat{V}\cos\frac{\mubar c}{2}
-\cos\frac{\mubar c}{2}\,\hat{V}\sin\frac{\mubar c}{2}
\right)
\right]
\sin(\mubar c),
\end{equation}
where we have chosen the symmetric ordering to make $\hat{C}_\grav$ hermitian. Its action on $\Psi(v)$ is given by
\begin{equation}
\hat{C}_\grav\Psi(v) = f_+(v)\Psi(v+4) + f_0(v)\Psi(v) + f_-(v)\Psi(v-4),
\end{equation}
where
\begin{subequations}
\begin{eqnarray}
f_+(v) &=& \frac{27}{16}\sqrt{\frac{4\pi}{3}}\frac{K\Pl}{\gamma^{3/2}} \abs{v+2}\Abs{\abs{v+1}-\abs{v+3}},\\
f_-(v) &=& f_+(v-4),\\
f_0(v) &=& -f_+(v)-f_-(v).
\end{eqnarray}
\end{subequations}

Meanwhile, to make the quantum dynamics more transparent, we add a free massless scalar $\phi(x,t)=\phi(t)$ to serve as the internal time. At the classical level, the matter part of the scalar constraint is given by
\begin{equation}
C_\matt = 8\pi G \abs{p}^{-3/2} p_\phi^2,
\end{equation}
where
\begin{equation}\label{p phi}
p_\phi=a^3L^3\dot{\phi}
\end{equation}
is the momentum of $\phi$ and $\dot{\phi}$ is the derivative with respect to the proper time. The matter part of the kinematical Hilbert space is given by the ordinary construction $L^2(\mathbb{R},d\phi)$ without any ``holonomization''.
%In other words, the total (gravity+matter) kinematical Hilbert space is given by $\hil_\kin=L^2(\mathbb{R}_\mathrm{Bohr},dv_\mathrm{Bohr})\otimes L^2(\mathbb{R},d\phi)$.
However, the inverse triad factor $\abs{p}^{-3/2}$ still has to be regularized by Thiemann's trick as for $1/\sqrt{\abs{q}}$ in $C_\grav$. Up to quantization ambiguities, the quantum operator for $\abs{p}^{-3/2}$ turns out to be
\begin{equation}\label{inverse p}
\widehat{\abs{p}^{-\frac{3}{2}}}\Psi(v) = \left(\frac{3}{4\pi\gamma\Pl^2}\right)^{3/2} B(v) \Psi(v),
\end{equation}
where
\begin{equation}
B(v) = \left(\frac{3}{2}\right)^3 K \abs{v}\, \Abs{\abs{v+1}^{1/3}-\abs{v-1}^{1/3}}^3.
\end{equation}

Consequently, the total quantum constraint equation is given by
\begin{equation}
\hat{C}\Psi(v,\phi) =\left(\hat{C}_\grav+\hat{C}_\matt\right)\Psi(v,\phi)=0,
\end{equation}
which is explicitly expressed as
\begin{eqnarray}\label{constraint eq}
\partial_\phi^2 \Psi(v,\phi) &=& B(v)^{-1}
\left(C_+(v)\Psi(v+4,\phi)+C_0(v)\Psi(v,\phi)+C_-(v)\Psi(v-4,\phi)\right)\nonumber\\
&=:& -\Theta \Psi(v,\phi),
\end{eqnarray}
where
\begin{subequations}
\begin{eqnarray}
C_+(v) &=& \frac{3\pi KG}{8}\, \abs{v+2} \Abs{\abs{v+1}-\abs{v+3}},\\
C_-(v) &=& C_+(v-4),\\
C_0(v) &=& -C_+(v)-C_-(v).
\end{eqnarray}
\end{subequations}

\subsection{Physical Hilbert space}
The form of the constraint equation \eref{constraint eq} suggests that it is natural to regard $\phi$ as the internal time, with respect to which $\Psi(v)$ evolves. To implement this idea, an appropriate kinematical Hilbert space for both geometry and the scalar field is chosen to be $\hil_\kin=L^2(\mathbb{R}_\mathrm{Bohr},B(v)dv_\mathrm{Bohr}) \otimes L^2(\mathbb{R},d\phi)$. It is straightforward to show that both the operators $\hat{p}_\phi=-i\hbar\partial_\phi$ and $\Theta$ are hermitian in $\hil_\kin$, and furthermore $\Theta$ is positive definite.

The operator $\Theta$ is a second-order \emph{difference} (cf.\ differential) operator. Consequently, the space of physical states (i.e., the space of solutions to the constraint equation) is naturally divided into sectors each of which is preserved by the ``evolution''. Denote by $\mathcal{L}_\epsilon$ the ``lattice'' of points $\{\abs{\epsilon}+4n;n\in\mathbb{Z}\}\cup\{-\abs{\epsilon}+4n;n\in\mathbb{Z}\}$ in the $v$-axis and denote by $\hil^\grav_\epsilon$ the subspace of $L^2(\mathbb{R}_\mathrm{Bohr},B(v)dv_\mathrm{Bohr})$ with states whose support is restricted to lattice $\mathcal{L}_\epsilon$. Each of $\hil^\grav_\epsilon$ is mapped to itself by $\Theta$.

Let $e_k(v)$ in $\hil^\grav_\epsilon$ be the eigenstate of $\Theta$:
\begin{equation}\label{ek(v)}
\Theta e_k(v) = \omega(k)^2 e_k(v),
\end{equation}
where $\omega(k)^2$ is the corresponding eigenvalue.
The general solution to the quantum constraint \eref{constraint eq} with initial data in $\hil^\grav_\epsilon$ is then given by
\begin{equation}\label{general sol}
\Psi(v,\phi) = \int_{-\infty}^\infty dk
\left(\tilde{\Psi}_+(k)e_k(v)e^{i\omega\phi}
+\tilde{\Psi}_-(k)e_k(v)^*e^{-i\omega\phi}\right),
\end{equation}
where $\tilde{\Psi}_\pm(k)$ are in $L^2(\mathbb{R},dk)$.
The positive/negative-frequency solutions satisfy a Schr\"{o}dinger type first order differential equation in $\phi$:
\begin{equation}\label{Schrodinger type eq}
\mp i\frac{\partial\Psi_\pm}{\partial\phi} = \sqrt{\Theta}\, \Psi_\pm.
\end{equation}
Therefore, the positive/negative-frequency solutions with the initial state $\Psi(v,\phi_0)$ are given by
\begin{equation}
\Psi_\pm(v,\phi) = e^{\pm i\sqrt{\Theta}\,(\phi-\phi_0)} \Psi(v,\phi_0).
\end{equation}
The positive-frequency and negative-frequency solutions can be discussed separately, and we focus on the former ones in what follows.

The sector of the physical Hilbert space $\hil_\phys^\epsilon$ labelled by $\epsilon\in[0,2]$ consists of positive-frequency solutions $\Psi(v,\phi)$ to \eref{Schrodinger type eq} with initial data $\Psi(v,\phi_0)$ in the symmetric sector of $\hil^\grav_\epsilon$.\footnote{The sign flip $v\rightarrow-v$ corresponds to reverse of the triad orientation. Because of no parity violating processes, we are led to choose the symmetric sector in which $\Psi(-v)=\Psi(v)$.} The physical inner product is defined as
\begin{equation}
\inner{\Psi_1}{\Psi_2}_\phys^\epsilon
:= \sum_{v\in\{\pm\abs{\epsilon}+4n;n\in\mathbb{Z}\}} B(v)\Psi_1(v,\phi_0)^* \Psi_2(v,\phi_0).
\end{equation}
It is easy to show that this definition is independent of the choice of $\phi_0$ as long as $\Psi_{1,2}(v,\phi)$ are solutions to \eref{Schrodinger type eq}.

Dirac operators acting on $\hil_\phys$ are $\hat{p}_\phi$ and a family of operators $\widehat{\abs{v}_\phi}$ parameterized by $\phi$, defined as
\begin{eqnarray}
\hat{p}_\phi \Psi(v,\phi) &:=& -i\hbar \frac{\partial\Psi(v,\phi)}{\partial\phi},\\
\label{op v-phi}
\widehat{\abs{v}_\phi} \Psi(v,\phi) &:=& e^{i\sqrt{\Theta}\,(\phi-\phi_0)} \abs{v} \Psi(v,\phi_0).
\end{eqnarray}
Both $\hat{p}_\phi$ and $\widehat{\abs{v}_\phi}$ preserve the sector $\hil_\phys^\epsilon$ and are hermitian with respect to $\inner{\cdot}{\cdot}_\phys^\epsilon$. The expectation value $\bra{\Psi}\,\hat{p}_\phi\ket{\Psi}_\phys$ gives the momentum of $\phi$, which is a constant of motion, i.e., independent of $\phi$. On the other hand, the expectation value $\bra{\Psi}\widehat{\abs{v}_\phi}\ket{\Psi}_\phys$ tells how big the volume of $\mathcal{V}$ is when the internal time takes the value $\phi$, in the style of \eref{Psi Ophi Psi}. It should be noted that the definition of $\widehat{\abs{v}_\phi}$ in \eref{op v-phi} is independent of the choice of $\phi_0$.

\subsection{Quantum dynamics}
The eigenvalue problem \eref{ek(v)} can be numerically solved (by a computer), and the eigenfunctions $e_k(v)$ can be used to build coherent states. Particularly, in \eref{general sol}, a natural choice is to set $\tilde{\Psi}_-(k)=0$ and
\begin{equation}
\tilde{\Psi}_+(k) = e^{-\frac{k-k^\star}{2\sigma^2}} e^{-i\omega \phi^\star},
\end{equation}
with a suitable small spread $\sigma$, to obtain a coherent physical state $\Psi_{p_\phi^\star,v_{\phi_0}^\star}(v,\phi)$. The resulting state is peaked at $\bra{\Psi_{p_\phi^\star,v_{\phi_0}^\star}}\, \hat{p}_\phi \ket{\Psi_{p_\phi^\star,v_{\phi_0}^\star}}_\phys =p_\phi^\star\equiv-\sqrt{12\pi G}\,\hbar\, k^\star$ and at $\bra{\Psi_{p_\phi^\star,v_{\phi_0}^\star}}\, \widehat{\abs{v}_{\phi_0}} \ket{\Psi_{p_\phi^\star,v_{\phi_0}^\star}}_\phys = v_{\phi_0}^\star$, where $v_{\phi_0}^\star$ is determined by $\phi^\star$.

To give a coherent state that is semiclassical at late times, we have to choose large values for $p_\phi^\star$ (i.e., $k^\star\ll-1$) and for $v_{\phi_0}^\star$ (i.e., $v_{\phi_0}^\star\gg1$). Given such a coherent physical state, the evolution with respect to the internal time $\phi$ can be read out from the expectation value
\begin{equation}
\langle v(\phi)\rangle:=
\frac{\bra{\Psi_{p_\phi^\star,v_{\phi_0}^\star}}\, \widehat{\abs{v}_{\phi}} \ket{\Psi_{p_\phi^\star,v_{\phi_0}^\star}}_\phys}
{\inner{\Psi_{p_\phi^\star,v_{\phi_0}^\star}} {\Psi_{p_\phi^\star,v_{\phi_0}^\star}}_\phys}.
\end{equation}
The numerical results show that the trajectory of $\langle v(\phi)\rangle$ follows the classical solution at large scale but at the Planck regime undergoes the \emph{quantum bounce}, which bridges the expanding classical solution with the contracting one (see Fig.~4 in Ref.~\cite{Ashtekar:2006wn}). The quantum bounce takes place when the energy density  $\rho_\phi$ of the scalar field approaches the critical density $\rho_\mathrm{crit}\approx 3/(8\pi G\gamma^2\Delta) = \sqrt{3}/(16\pi^2\gamma^3G^2\hbar)$ of the Planck scale. More precisely, the energy density operator is defined as
\begin{equation}
\widehat{\rho_\phi|_\phi} := \widehat{\frac{p_\phi^2}{2\,\abs{p^3}_\phi}}
\equiv \frac{K^2}{2}\left(\frac{3}{4\pi G\hbar\gamma}\right)^3 \widehat{\frac{p_\phi^2}{\abs{v^3}_\phi}}
\end{equation}
and its expectation value
\begin{equation}
\langle \rho_\phi(\phi) \rangle := \frac{\bra{\Psi_{p_\phi^\star,v_{\phi_0}^\star}}\, \widehat{\rho_\phi|_\phi} \ket{\Psi_{p_\phi^\star,v_{\phi_0}^\star}}_\phys}
{\inner{\Psi_{p_\phi^\star,v_{\phi_0}^\star}} {\Psi_{p_\phi^\star,v_{\phi_0}^\star}}_\phys}
\approx \frac{K^2}{2}\left(\frac{3}{4\pi G\hbar\gamma}\right)^3 \frac{(p_\phi^\star)^2}{\langle v(\phi)\rangle^3}
\end{equation}
is bounded above by $\rho_\mathrm{crit}$ (see Fig.~6 in Ref.~\cite{Ashtekar:2006wn}).

If the internal time $\phi$ is expressed in terms of the proper time $t$ via $a^3L^3d\phi=\langle\hat{p}_\phi\rangle dt=p_\phi^\star dt$ in accord with \eref{p phi},\footnote{The change of variables from $\phi$ to $t$ makes sense only for sharply peaked coherent states. In the quantum theory, the internal time is more fundamental than the proper time, which is a derived notion only meaningful for sharply peaked coherent states.} the trajectory of $\langle v(\phi)\rangle$ in relation to $\langle \rho_\phi(\phi) \rangle$ can be described rather accurately by the effective equation for the scalar factor $a(t)$:
\begin{equation}\label{modified Friedmann eq}
\left(\frac{\dot{a}}{a}\right)^2 = \frac{8\pi G}{3}\rho_\phi \left(1-\frac{\rho_\phi}{\rho_\mathrm{crit}}\right),
\end{equation}
which is the Friedmann equation modified by the opposing quantum force.

\subsection{Other models}
A slight simplification in the model discussed above leads to the ``solvable LQC'' (sLQC)\cite{Ashtekar:2007em}. In sLQC, analytic solutions are available and it is shown that the quantum bounce is generic not only for coherent states but for any arbitrary physical states and the matter density has an absolute upper bound given by
\begin{equation}\label{rho crit}
\rho_\mathrm{crit}= \frac{3}{8\pi G\gamma^2\Delta}.
\end{equation}

The construction of LQC has been extended to a variety of models with various degrees of rigor, including $k=+1$\cite{Ashtekar:2006es} and $k=-1$\cite{Vandersloot:2006ws} FLRW models, possibly with a nonzero cosmological constant\cite{Bentivegna:2008bg,Kaminski:2009pc,Pawlowski:2011zf}, Bianchi I\cite{Chiou:2006qq,Chiou:2007sp,Chiou:2007mg,Ashtekar:2009vc,MartinBenito:2009qu}, Bianchi II\cite{Ashtekar:2009um} and Bianchi IX\cite{WilsonEwing:2010rh} models, Kantowski-Sachs model\cite{Modesto:2004wm,Chiou:2008eg}, unimodular model\cite{Chiou:2010ne}, higher-order holonomy extension\cite{Chiou:2009hk,Chiou:2009yx}, and much more. The bouncing scenario has been shown to be robust in different models.

Particularly, the formalism of LQC in the Kantowski-Sachs model can be used to study the loop quantum geometry of the interior of a Schwarzschild black hole\cite{Ashtekar:2005qt,Modesto:2005zm,Modesto:2006mx, Bohmer:2007wi,Chiou:2008nm}. The study suggests that, just like resolution of the cosmological singularity, the classical black hole singularity is also resolved by the loop quantum effects and replaced by the quantum bounce, which either bridges the black hole interior to a white hole interior or gives birth to a baby black hole\cite{Chiou:2008nm}.\footnote{Validity of this approach however might be questionable, since the Schwarzschild interior is extensible beyond its boundary (event horizon) and thus cannot be considered as a self-contained universe (i.e., it is not a genuine Kantowski-Sachs spacetime). The more rigourous approach is to consider the black hole interior and exterior as a whole in the framework of spherically symmetric midisuperspaces (see \sref{sec:spherically symmetric loop gravity}.)}

Furthermore, inhomogeneity has also been taken into consideration in the Gowdy model (the simplest of the
inhomogeneous models)\cite{MartinBenito:2008ej,Garay:2010sk,Brizuela:2009nk, MartinBenito:2010bh,MartinBenito:2010up,Brizuela:2011ps} as well as in the framework of ``lattice LQC''\cite{Bojowald:2006qu,Bojowald:2007ra}.

\section{Current directions and open issues}\label{sec:current directions}
In this review article, we gave a self-contained introduction to the canonical formulation of LQG and briefly covered the ideas of the spin foam theory, black hole thermodynamics, and LQC. The core formalism of LQG provides a coherent framework in which the fundamental principles of GR and QFT consort with each other in harmony without invoking additional hypotheses. The key result of LQG is a compelling physical picture of quantum space, which is made up of quantized areas and volumes of the Planck scale in a profoundly background-independent fashion. Despite many remarkable achievements, the theory of LQG is still far from complete, especially for the aspects of quantum dynamics and low-energy limit. A wide range of research has been developed for the open problems. We conclude this review by outlining a non-exhaustive list of recent advances that are not covered in the main text and also addressing some open issues along the list. One can easily appreciate that LQG has grown into a vast and active area of research along many various directions.

\subsection{The Master constraint program}\label{sec:Master constraint program}
The scalar constraint remains the major unsolved problem in LQG as noted in \sref{sec:scalar constraint and quantum dynamics}. The Master constraint program\cite{Thiemann:2003zv} (also see Sec.~10.6 of Ref.~\cite{Thiemann:2007zz} for a systematic account) proposed an elegant solution to the difficulties concerning the algebra of commutators among smeared scalar constraints in terms of the so-called \emph{Master constraint}, which combines the smeared scalar constraints $\calC[N]:=\int_\Sigma d^3x N(x)C(x)$ for all smearing functions $N(x)$ into a single constraint
\begin{equation}
\boldsymbol{\mathsf{M}} = \int_\Sigma d^3x\, \frac{C(x)^2}{\sqrt{\abs{q}}},
\end{equation}
where the factor $1\sqrt{\abs{q}}$ has been incorporated to make the integrand a scalar density of weight 1 so that $\boldsymbol{\mathsf{M}}$ is diffeomorphism invariant, i.e., $\{\calC_\Diff[\vec{N}],\boldsymbol{\mathsf{M}}\}=0$. It is clear that $\boldsymbol{\mathsf{M}}=0$ implies the infinitely many constraints $C(x)=0$ for $\forall x\in\Sigma$ and \textit{vice versa}.

The single constraint $\boldsymbol{\mathsf{M}}$ in place of $\calC[N]$ greatly reduces the number of constraints and drastically simplifies the Poisson bracket structures among constraints: \eref{PB CDiff C} is replaced by $\{\calC_\Diff[\vec{N}],\boldsymbol{\mathsf{M}}\}=0$, and more notably the complicated structure of \eref{PB C C} is now reduced to the trivial algebra $\{\boldsymbol{\mathsf{M}},\boldsymbol{\mathsf{M}}\}=0$. As the master constraint renders the constraint algebra as a genuine Lie algebra, the program opens a new avenue for understanding the quantum dynamics and establishing the semiclassical limit.

The Master constraint program has various advantages in an attempt to define the physical inner product and formulate rigorous path integrals\cite{Han:2009aw,Han:2009ay}, and it seems to be more directly related to the spin foam formalism than the standard formulation of LQG.

\subsection{Algebraic quantum gravity}\label{sec:AQG}
The Master constraint program has evolved into a fully combinatorial theory known as algebraic quantum gravity (AQG)\cite{Giesel:2006uj,Giesel:2006uk,Giesel:2006um,Giesel:2007wn} (also see Sec.~10.6.5 of Ref.~\cite{Thiemann:2007zz}). AQG is closely related to yet very different from LQG by the fact that no topology or differential structure of space is assumed \textit{a priori}. In this sense, background independence of AQG is even more compelling, and it is possible to talk about topology change. A viable semiclassical machinery that involves only non-graph-changing operators has been suggested, and considerable progress has been made in providing contact with the low-energy physics.

\subsection{Reduced phase space quantization}\label{sec:reduced phase space}
There are two major approaches to the canonical quantization of a field theory with gauge symmetries: the so-called ``Dirac approach'' and the ``reduced phase space approach''. In the Dirac approach, one first construct the Hilbert space that represents the partial (gauge-variant) observables and then impose the constraints at the quantum level to select the physical (gauge-invariant) states. In the reduced phase space approach, one first constructs the physical (gauge-invariant) observables at the classical level and then directly construct the Hilbert space that represents the physical observables. (Of course, it is possible to take a ``mixed'' approach by implementing different strategies for different gauge symmetries.)

The advantage of the Dirac approach is that the algebra of partial observables is usually simple enough and thus the (kinematical) Hilbert space is easy to construct, but one has to deal with spurious degrees of freedom, which are the possible source of quantization ambiguities and quantum anomalies. On the other hand, the advantage of the reduced phase space approach is that one never has to care about the kinematical Hilbert space at all, but the induced algebra of physical observables is typically so complicated that the corresponding (physical) Hilbert space is extremely difficult to find.

The standard formulation of LQG adopts the Dirac approach, as constructing the reduced phase space of GR with standard matter is uncontrollably nontrivial. However, there is a hope of obtaining the reduced phase space with a manageable induced algebra, if one adds pressure-free dust\cite{Brown:1994py} or a massless scalar field\cite{Kuchar:1995xn} to the theory. This is essentially because, at the price of introducing additional matter, the constraints can be \emph{deparameterized} with respect to the dust or the scalar field, which serves as a material reference system.

Schematically (consider only the scalar constraint), when the system is deparameterized, one can find a partial variable $T$ serving as an ``internal clock'' such that the local scalar constraint (at least locally) reads as $C[q^a,p_a]=P+\mathsf{h}[q^a,p_a]$, where $P$ is the momentum conjugate to $T$ and both $T$ and $P$ are independent of other conjugate pairs $(q^a(x),p_a(x))$. One can identify a one-parameter family of the physical observables $\mathcal{O}_T$ associated with the partial operator $\mathcal{O}$ such that the map $F_\tau:\mathcal{O}\mapsto\mathcal{O}_{T=\tau}$ for any $\tau$ is a homomorphism between the Poisson bracket of partial observables and a certain Dirac bracket (uniquely determined by the constraints and the choice of $T$) of physical observables. This implies that the reduced phase space approach is achievable, as finding the Hilbert space representation of $\mathcal{O}_T$ is as easy as that of $\mathcal{O}$. In the resulting Hilbert space, the physical observable $\mathcal{O}_{T=\tau}$ is interpreted as the observable $\mathcal{O}$ measured when the internal clock $T$ takes the value $\tau$, and (as heuristically $P$ acts as $-i\partial/\partial T$) the dynamics is generated by the evolution equation
\begin{equation}
\frac{d}{d\tau}\,\widehat{\mathcal{O}_\tau}
=-i \big[\hat{\mathsf{H}}, \widehat{\mathcal{O}_\tau}\big],
\end{equation}
where
\begin{equation}
\hat{\mathsf{H}} := \int d^3x\, \hat{\mathsf{h}}[q^a(x),p_a(x)]
\end{equation}
is called the \emph{physical Hamiltonian} in contrast to the Hamiltonian constraint as it corresponds to a nonvanishing physical observable that generates the genuine ``physical'' evolution.

Based on the model of Ref.\cite{Brown:1994py} with pressure-free dust, the reduced phase space approach to a quantum theory of GR has been formulated in the styles of LQG\cite{Giesel:2007wi,Giesel:2007wk} as well as of the Master constraint program and AQG\cite{Giesel:2007wn}. Based on the model of Ref.~\cite{Kuchar:1995xn} with a massless scalar field, the reduced phase space approach has also been constructed in the style of LQG\cite{Domagala:2010bm}. These studies have led to a new appealing direction of QG alongside the standard LQG.

\subsection{Off-shell closure of quantum constraints}\label{sec:off-shell closure}
In the Dirac approach of canonical quantization, quantizing a constrained system faithfully, one would expect that the corresponding quantum operators represent the same structure of the classical constraint algebra. That is, in the case of gravity, in accordance with the constraint algebra given by \eref{PB bw constraints}, there exits a vector space $\mathcal{V}$ of spin networks upon which, for all $\ket{\Psi}\in\mathcal{V}$, the quantum operators act as
\begin{subequations}
\begin{eqnarray}
\label{off-shell commutator a}
\hat{U}_{\varphi_1}\hat{U}_{\varphi_2}\ket{\Psi} &=& \hat{U}_{\varphi_1\circ\varphi_2}\ket{\Psi},\\
\label{off-shell commutator b}
\hat{U}_{\varphi}^\dag\, \hat{\calC}[N]\, \hat{U}_{\varphi}\ket{\Psi} &=& \hat{\calC}[\varphi^*N]\ket{\Psi},\\
\label{off-shell commutator c}
\big[\hat{\calC}[N],\hat{\calC}[M]\big]\ket{\Psi} &=& \hat{\calC}_\Diff[\,q^{ab}(N\partial_bM-M\partial_bN)]\ket{\Psi},
\end{eqnarray}
\end{subequations}
where we have disregard the Gauss constraint $\calC_\mathrm{G}=0$, which is trivially satisfied via intertwiners.
This, however,  is not the case in the standard formulation of LQG. In the diffeomorphism covariant regularization scheme as addressed in \sref{sec:quantum scalar constraint}, \eref{off-shell commutator a} is trivially satisfied and \eref{off-shell commutator b} corresponds to \eref{Re Re'}, but \eref{off-shell commutator c} on the other hand is replaced by (see \eref{action of C})
\begin{equation}\label{on-shell commutator c}
\oinner{\big[\hat{\calC}[N],\hat{\calC}[M]\big]\eta(\Psi)}{\Phi}
:= \lim_{\epsilon\rightarrow0} \oinner{\eta(\Psi)}{\big[\hat{\calC}_{R_\epsilon}[N],\hat{\calC}_{R_\epsilon}[M]\big] \Phi}
=0,
\end{equation}
for all $\Phi\in\mathcal{V}$, where $\eta$ is the linear map in \eref{map eta}. In this sense, the standard LQG is said to be anomaly-free only at the \emph{on-shell} level, but not \emph{off-shell}.

Although there is no obvious self-inconsistency, the fact that the quantum algebra is not off-shell closed arouses various worries. It was argued in Refs.~\cite{Gambini:1997bc,Lewandowski:1997ba} that quantum anomaly is hidden on-shell simply because the local scalar constraint $C$ in \eref{constraint C} is specifically of density weight 1. To see this heuristically, consider the local scalar constraint $C^{(k)}\sim\abs{q}^{(k-2)/2}F\tE\tE+\dots$ of density weight $k$ associated with the smearing function $N^{(1-k)}$ of density weight $1-k$. The scalar constraint in $d$ spatial dimensions ($d\ge2$) is given by $\calC[N]=\int_\Sigma d^dx N^{(1-k)}(x)C^{(k)}(x)$ and scaling analysis tells that the regulated classical scalar constraint $\calC_{R_\epsilon}[N]$ scales as $\propto \epsilon^d\epsilon^{-2d(k-2)/2}\epsilon^{-2}\epsilon^{-(d-1)}\epsilon^{-(d-1)} =\epsilon^{d(1-k)}$. In the special case of $k=1$, there is no explicit dependence on $\epsilon$ and consequently the right-hand side of \eref{on-shell commutator c} is trivialized (as long as $\epsilon$ becomes sufficiently small). In a sense, the quantum operator $\hat{\calC}[N]$ implemented in the standard approach fails to grasp the geometrical significance dictated by \eref{off-shell commutator c} and hence the quantum theory might not faithfully correspond to the starting classical theory (however, see Ref.~\cite{Thiemann:2010es} for a counterargument).

If one choose to construct the scalar constraint operator by starting with $C^{(k)}$ of higher density weight, one would has enough scaling factors of $\epsilon^{-1}$ to match the scaling of $\calC_\Diff[\vec{S}\,]$  on the right-hand side of \eref{off-shell commutator c} (as $\calC_\Diff[\vec{S}\,]$ involves spatial derivatives via $N\partial_bM-M\partial_bN$, which gives rises to at least one factor of $\epsilon^{-1}$) and be able to obtain an off-shell quantum representation for the constraint algebra. This ``off-shell closure'' approach indeed can be rigorously formulated for a generally covariant $U(1)^3$ gauge theory, which can be understood as the $G\rightarrow0$ limit of Euclidean gravity\cite{Tomlin:2012qz,Henderson:2012ie,Henderson:2012cu}. For Euclidean LQG, the new study of Ref.~\cite{Laddha:2014xsa} presented some evidence that, working with $k=4/3$, there exists such a scalar constraint operator well defined on a vector space $\mathcal{V}$, which turns out to be a suitable generalization of the Lewandowski-Marolf habitat\cite{Lewandowski:1997ba}.

In the off-shell closure approach, the geometrical interpretation of the action of the scalar constraint operator becomes more transparent, which should help to elucidate the quantum dynamics and the low-energy physics. Furthermore, as the requirement of being anomaly-free at the off-shell level is supposed to impose more restrictions on permissible regulators, the infinite multitude of quantization ambiguities could be enormously reduced or even uniquely fixed.

\subsection{Loop quantum gravity vs.\ spin foam theory}\label{sec:LQG vs spin foam}
The canonical (Hamiltonian) formalism of LQG and the covariant (sum-over-histories) formalism of the spin foam theory are closely related to each other, yet up to now spin foam models have not been systematically derived from the standard canonical theory of LQG. Over the past years, the spin foam theory in relation to the kinematics of LQG have been clearly established\cite{Livine:2007vk,Engle:2007uq,Engle:2007qf, Freidel:2007py,Engle:2007wy,Kaminski:2009fm}. For the dynamics, on the other hand, the two formalisms bear close resemblance to each other but it remains a challenging open problem to clearly derive the relation in the 4-dimensional theory. (In the 3-dimensional theory, a clear-cut relation between the canonical quantization of 3-dimensional gravity and spin foam models has been nicely established\cite{Noui:2004iy}.)

\subsection{Covariant loop quantum gravity}\label{sec:covariant LQG}
Even though the precise connection between LQG and the spin foam theory is still not completely clear, the merger of the canonical and covariant approaches has suggested a rather well-established theory formulated in a covariant formalism known as ``covariant loop quantum gravity'', which provides us a brand new perspective on LQG\cite{Rovelli:2010wq,Rovelli:2011eq,Rovelli:2013ht} (also see Ref.~\cite{Rovelli:2014book} for a comprehensive account). Since various quantization procedures (canonical, spin foam, etc.) all lead to the overlapped results, perhaps it is the time to ``leave the ladder behind'' (as advocated in Ref.~\cite{Rovelli:2010wq}) and take the merged theory of covariant LQG seriously as the \emph{defining} quantum theory of gravity instead of ``deriving'' it by quantizing the classical GR. The kinematics of covariant LQG is well defined and background independent, and the dynamics of it is given in terms of a simple vertex function, largely determined by locality, diffeomorphism invariance, and local Lorentz invariance.

In the framework of covariant LQG, transition amplitudes for given boundary states can be computed explicitly and then compared with the classical theory by the techniques of \emph{holomorphic coherent states} (see \sref{sec:coherent states} and references therein). Particularly, the 2-point function of the Euclidean theory over a flat spacetime has been computed to the first order in the vertex expansion, and has been shown to converge to the free graviton propagator of QG in the large distance limit\cite{Bianchi:2009ri}. The $n$-point functions of the Lorentzian theory have yet to be computed and compared with the vertex amplitude of conventional perturbative QG on Minkowski space.

\subsection{Spin foam cosmology}\label{sec:spin foam cosmology}
In the framework of covariant LQG, by choosing the holomorphic coherent states of geometry peaked on homogeneous isotropic metrics, it is possible to compute the transition amplitude in the vertex and graph expansions for the evolution of a homogenous isotropic universe. The calculation has been performed and the resulting amplitude in the classical limit appears to be consistent with the FLRW evolution\cite{Bianchi:2010zs}. Furthermore, if a cosmological constant term is added, the result matches the de Sitter solution in the classical limit\cite{Bianchi:2011ym}.

This approach starting from the full theory of covariant LQG and then taking an approximation of the vertex and graph expansions is called ``spin foam cosmology''.
By comparison, in LQC one studies the exact solutions in a symmetry-reduced quantum theory, while in spin foam cosmology one studies the approximate solutions in the full quantum theory. Casting LQC in an spin foam-like expansion has also been considered\cite{Ashtekar:2009dn,Ashtekar:2010ve,Campiglia:2010jw,Henderson:2010qd}. Spin foam cosmology and the spin foam-like expansion of LQC can be regarded as two converging approaches to the sum-over-histories formalism of quantum cosmology.

\subsection{Quantum reduced loop gravity}\label{sec:QRLG}
Quantum reduced loop gravity (QRLG) is a framework introduced for the quantum theory of a symmetry-reduced sector of GR, based on a projection from the kinematical Hilbert space of the full theory of LQG down to a subspace representing the proper arena for the symmetry-reduced sector. It was first proposed in Refs.~\cite{Alesci:2012md,Alesci:2013xd} for an inhomogeneous extension of
the Bianchi I cosmological model. This approach provides a direct link between the full theory and its cosmological sector and also sheds light on the relation between LQC and LQG.

It was later shown in Ref.~\cite{Alesci:2013xya} that the QRLG technique can be applied to broader cases beyond the cosmological context whenever the spatial metric can be gauge-fixed to a diagonal form. The technique projects the Hilbert space to the states based on the reduced graphs with Livine-Speziale coherent intertwiners\cite{Livine:2007vk}.
The framework of QRLG could simplify the analysis of the dynamics in the full theory and other issues, such as the coupling between quantum geometry and matter and the relation between canonical and covariant approaches.

\subsection{Cosmological perturbations in the Planck era}\label{sec:Planck era perturbations}
Cosmological inflation is a popular paradigm in modern cosmology (see \cref{ch:Sato}). The theory of inflation has successfully explained how quantum fluctuations sow the primordial seeds that grow into the large-scale structure at late times of our universe and made a number of predictions that have been confirmed by a range of observations (also see Chaps.~\ref{ch:Bucher} and \ref{ch:Davis}). In the standard inflationary scenario, cosmological perturbations are described by QFT on \emph{classical} cosmological spacetimes, and thus the domain of validity excludes the Planck era.

In a series of papers Refs.~\cite{Agullo:2012sh,Agullo:2012fc,Agullo:2013ai} (also see Ref.~\cite{Ashtekar:2013xka} for a nice summary), by studying the dynamics of quantum fields representing scalar and tensor perturbations on \emph{quantum} cosmological spacetimes using techniques from LQG, the inflationary paradigm is extended to a
self-consistent theory covering the epoch from the big bounce in the Planck era to the onset of slow-roll inflation. This pre-inflationary dynamics could yield deviations from the standard inflationary scenario and give rise to novel effects, such as a source for non-Gaussianity. These novel effects might catch a glimpse into the deep Planck regime of the early universe in future cosmological observations.

\subsection[\enskip Spherically symmetric loop gravity]{Spherically symmetric loop gravity}\label{sec:spherically symmetric loop gravity}
LQC is the loop quantum theory of \emph{minisuperspaces} in the sense that degrees of freedom are truncated to a \emph{finite} number by the symmetry of homogeneity (and possibly also isotropy). The next simplest symmetry-reduced framework is the spherically symmetric \emph{midisuperspace}, in which the spherical symmetry is imposed but inhomogeneity along the radial direction is retained, thus still giving rise to a field-theoretical system of \emph{infinite} degrees of freedom. Ashtekar's formalism for spherically symmetric midisuperspaces and its loop quantization have been studied and developed with various degrees of rigor\cite{Bojowald:2004af,Bojowald:2004ag,Bojowald:2005cb,Campiglia:2007pr, Chiou:2012pg}. The theory of spherically symmetric loop gravity provides an arena for testing important issues that are too difficult in the full theory of LQG and trivialized in minisuperspace (LQC) models. Particularly, the $SU(2)$ internal gauge is reduced to $U(1$) and the 3-dimensional diffeomorphism is reduced to 1-dimensional diffeomorphism (in the radial direction), making the constraint algebra much simpler yet nontrivial. The framework of spherically symmetric loop gravity can be used to study loop quantum geometry of spherically symmetric black holes (Schwarzschild black hole, spherical gravitational collapse, etc.). It is indicated that in the resulting quantum spacetime the black hole singularity is avoided\cite{Campiglia:2007pr,Chiou:2012pg, Gambini:2008dy,Gambini:2013ooa,Gambini:2013hna}, suggesting that the information loss problem could be resolved accordingly.

\subsection[\enskip Planck stars and black hole fireworks]{Planck stars and black hole fireworks}\label{sec:Planck stars and black hole fireworks}
One of the key insights obtained in LQC is that the quantum geometry of LQG gives rise to \emph{opposing} quantum gravitational force, which becomes appreciable when the matter density comes close to the Planck scale density $\sim1/(G^2\hbar)$. Based on the insight, recently, a new possible scenario of collapsing black holes has been suggested\cite{Rovelli:2014cta}: The gravitational collapse of a star does not lead to a singularity but to the quantum gravitational phase called a ``Planck star'', where the very huge gravitational attraction is counterbalanced by the opposing quantum force. Accordingly, a black hole hides a core of a Planck star, which is of the Planck scale density but can be much larger than the Planck scale size (depending on the initial mass of the collapsing star). As the black hole evaporates, the core remembers the initial mass and the final explosion can occur at macroscopic scale, thus providing a possible mechanism for recovery of information loss. More interestingly, the objects of Planck stars could produce detectable signals of quantum gravitational origin\cite{Rovelli:2014cta,Barrau:2014hda}.

Furthermore, the macroscopic remnant can develop into a white hole. As shown in a recent paper Ref.~\cite{Haggard:2014rza}, there is indeed a classical metric that satisfies the Einstein field equation outside a finite spacetime region where matter collapses into a black hole and then emerges from a white hole. Therefore, a black hole can quantum-tunnel into a white hole in the similar fashion that the wave packet representing a collapsing universe tunnels into a wave packet representing an expanding universe in LQC. A distant observer thus sees a dimming star, after a very long time, reemerge and burst out matter---a phenomenon called ``black hole fireworks''. Under the scenario of black hole fireworks, the arguments over the information paradox have to be drastically revised.

\subsection[\enskip Information loss problem]{Information loss problem}\label{sec:information loss problem}
It is widely expected that QG, once fully developed, should resolve several important problems in QG, particularly the singularity problem and the black hole information paradox. As both the cosmological and black hole singularities are indicated to be resolved by the LQG effects, it is strongly suggested that the information lost in the process of black hole evaporation should be recovered in a singularity-free scenario. In Ref.~\cite{Ashtekar:2008jd}, it was analyzed in detail for 2-dimensional black holes that quantum geometry effects indeed recover the information loss, primarily because the black hole singularity is resolved and consequently the quantum spacetime is sufficiently larger than the classical counterpart. For 4-dimensional collapsing black holes, the aforementioned scenarios of Planck stars and black hole fireworks in particular suggest a similar mechanism for information recovery.

The model of Ref.~\cite{Ashtekar:2008jd} provides a concrete example of avoiding the information loss problem under the ``remnant scenario'', in which black hole evaporation stops at some point, leaving a black hole remnant that is correlated with the Hawking radiation and allowing the combined state to remain pure. The remnant scenario generally implies the existence of a very long-lived remnant, which, unfortunately, may cause other problems (e.g.\ the problem of infinite remnant production\cite{Giddings:1993km}).

Meanwhile, a recent paper\cite{Bojowald:2014zla} investigated an example from LQG in which the singularity problem is solved but the information loss problem is made worse. It was argued that the aggravated information loss problem is likely to be a generic feature of LQG, therefore putting considerable conceptual pressure on the theory of LQG.

The information loss problem has been around for a long time. There is still no consensus over how the problem is to be resolved and it remains one of the focal points in the research of QG (see Sec.~10 of \cref{ch:Carlip} for more discussions).

\subsection[\enskip Quantum gravity phenomenology]{Quantum gravity phenomenology}\label{sec:quantum gravity phenomenology}
A legitimate quantum theory of gravity must make unambiguous predictions that in principle can be empirically tested by experiments or observations. Contrary to popular opinion, LQG does make definite predictions. For example, any measured physical area (such as the total cross-section of a scattering process) must be given by the discrete spectrum of area. (Also see Ref.~\cite{Major:2010qg} for the possible effects of angle quantization on scattering.) The true problem is that these predictions demand experiments or observations probing the Planck scale, which is believed to be far out of current reach. The impasse could be overcome soon, as the Planckian physics may not be completely unreachable after all by present and near-future technology.

As mentioned earlier, loop quantum effects in the pre-inflationary era could give rise to sufficient deviations from the standard inflationary scenario. Some research works have attempted to reveal possible observable footprints of these effects on the Cosmic Microwave Background along the lines of Refs.~\cite{Grain:2009kw,Barrau:2009fz,Mielczarek:2010bh,Grain:2010yv,Barrau:2010nd}.

Also as noted previously, the existence of Planck stars could produce detectable signals. The possible phenomenological consequences of Planck stars in gamma-ray bursts were studied in Ref.~\cite{Barrau:2014hda}.

Additionally, while the Lorentz invariance can be made manifest both in LQG and in the spin foam theory (see Ref.~\cite{Rovelli:2010ed} and references therein), many other models suggest violation of Lorentz invariance in the deep Planck regime, which will modify the Lorentzian energy-momentum dispersion relation at lower-energy scale. Observations of Ultra-High-Energy Cosmic Rays---the most energetic particles ever observed---have thus far put strong constraints on deviations from the Lorentzian dispersion relation. Future investigation with improved precision will either impose even stronger constraints or unveil breakdown of the Lorentz invariance, thus enabling theorists to discriminate between different models of LQG and the spin foam theory. See Refs.~\cite{Smolin:2005re,Liberati:2011bp} for more discussions.

\subsection[\enskip Supersymmetry and other dimensions]{Supersymmetry and other dimensions}\label{sec:SUSY and other dimensions}
It is a celebrated virtue that LQG does not require any hypothetical ingredients, yet it has always been desirable to incorporate supersymmetry and extra dimensions at one's disposal for the sake of both theoretical interest and comparison with other QG approaches (string theory in particular). Early attempts can be found e.g.\ in Refs.~\cite{Jacobson:1987cj,Matschull:1993hy,Freidel:1999rr}. Recently, a series of papers\cite{Bodendorfer:2011nv,Bodendorfer:2011nw,Bodendorfer:2011nx,Bodendorfer:2011ny, Bodendorfer:2011pa, Bodendorfer:2011pb,Bodendorfer:2011pc,Bodendorfer:2011hs} have been devoted to rigorous formulation of the loop quantum theory for \emph{supergravity} and for \emph{all dimensions}. Some progress has been made toward calculating black hole entropy from loop quantum theory in higher dimensions\cite{Bodendorfer:2013jba,Bodendorfer:2013sja}. It would be very instructive to compare these results to those obtained by string theory inspired approaches.

\newpage

\phantomsection
\addcontentsline{toc}{section}{Acknowledgements} % show Acknowledgements in TOC
\section*{Acknowledgements}
The author would like to thank Wei-Tou Ni for warmly inviting and encouraging him to write this article, and is deeply grateful to Steven Carlip, Friedrich Hehl, and Chun-Yen Lin for their detailed comments and suggestions on the manuscript, which have been extremely helpful and have led to many improvements. Deep gratitude goes to Chih-Wei Chang for having inspiring discussions with the author on quantum gravity and providing him with a comfortable office at CCMS.
This article was written over the time when the author was supported in part by the Ministry of Science and Technology of Taiwan under the Grants, No.\ 101-2112-M-002-027-MY3 and No.\ 101-2112-M-003-002-MY3.

%\bibliographystyle{ws-rv-van}
%\bibliography{ws-rv-sample}
%\blankpage
%%\printindex[aindx]                 % to print author index
%\printindex                         % to print subject index
\end{document}